\documentclass[pdflatex,sn-mathphys-num,review
]{sn-jnl}

\usepackage{graphicx}%
\usepackage{multirow}%
\usepackage{amsmath,amssymb,amsfonts}%
\usepackage{amsthm}%
\usepackage{mathrsfs}%
\usepackage[title]{appendix}%
\usepackage{xcolor}%
\usepackage{textcomp}%
\usepackage{manyfoot}%
\usepackage{booktabs}%
\usepackage{multirow}
\usepackage[makeroom]{cancel}
\usepackage{siunitx}
\usepackage[toc,acronym]{glossaries}

\makeglossaries

\makeglossaries
\newacronym{AP}{AP}{Action Potential}
\newacronym{AFM}{AFM}{Atomic Force Microscope}
\newacronym{APTD}{APTD}{Action Potential's Time Derivative}
\newacronym{ATP}{ATP}{Adenozine Triphosphate}
\newacronym{CNS}{CNS}{Central Nervous System}
\newacronym{EM}{EM}{electromagnetic}
\newacronym{GHK}{GHK}{Goldman-Hodgkin-Katz}
\newacronym{HBP}{HBP}{Human Brain Project}
\newacronym{HH}{HH}{Hodgkin and Huxley}
\newacronym{HW}{HW}{hardware}
\newacronym{PID}{PID}{Proportional-Integral-Derivative} 
\newacronym{PSP}{PSP}{Post-Synaptic Potential}
\newacronym{AIS}{AIS}{Axon Initial Segment}
\newacronym{AI}{AI}{Artificial Intelligence}
\newacronym{AIMC}{AIMC}{Analog In-Memory Computing}

\newglossaryentry{setpoint}
{
	name={setpoint},
	description=\supplementbox
	{A \gls{PID} controller (Proportional-Integral-Derivative) maintains a process variable (PV)—the membrane potential—at a specific, desired target value known as the setpoint (SP).}
}
\usepackage[
]{changes}
\definechangesauthor[name=V\'egh, color=blue]{VJ}
\definechangesauthor[name=VJA, color=violet]{VJA}

\begin{document}

\title
{ \centering{The unified non-disciplinary model}\\of the operation of neurons}


\author*{\fnm{J\'anos} \sur{V\'egh}, 0000-0002-3247-7810}\email{Vegh.Janos@gmail.com}
%
%
\affil{\orgname{Kalimános Bt}, \city{Debrecen}, \postcode{4032}, \country{Hungary}}



\abstract{
The conflicting disciplinary theories of neuronal operation are reviewed, and a unified, non-disciplinary synthesis is introduced. The theory explains all known experimental observations without contradictions.
Although the first principles of science are identical in the inanimate and life sciences, the abstractions and approximations that underlie concrete laws differ, so the actual laws depend on the environment.
We challenge the universality of protein mechanisms as well as the disciplinary attempts to describe neuronal operation.
By using physically and biologically plausible approximations, a non-disciplinary cooperation of thermodynamics and electricity 
perfectly describes processes in living matter. 
Biology works on the boundary of the
continuous and discrete views of physics, and in some cases,
those views must be bridged.
Most importantly, we explicitly consider the discrete nature 
of biological currents, that their charge carriers are slow and they repel
each other. That feature, overlooked by both mentioned theories, 
enables us to interpret the electrical and thermodynamic behavior of important 
biological objects and processes. The speed of biological currents changes 
by several orders of magnitude during operation. They keep the electrolyte 
in continuous movement, making some microscopic regions of the volume 
responsible for the observed action potential and the resting potential.
The cross-disciplinary model enables us to derive
the correct electrical equations
and the thermodynamic description of the action potential. In a way 
demonstrating the need for the non-disciplinary discussion,
we derive the thermodynamic Carnot cycle of \gls{AP} from
measuring electrical parameters, calculating its energy consumption and efficiency, interpreting the neuron's partially reversible operation, and
the long-sought solution to the heat emission/absorption issue.
We defy the claim that science cannot describe life: the underlying physics constrains biological phenomena, including cognitive ones.
}



\maketitle
\added{
The table of contents is present only for the comfort of reviewers}
\tableofcontents

\section{Introduction}

The \textit{dynamic} operation of individual neurons, their connections, and higher-level organizations, connections,
the brain with its information processing capability, and finally, the mind with its conscience and behavior,
are still among the big mysteries of science: \textit{at which point
	{the non-living matter becomes a living one}} and
\textit{at which point {the living matter becomes intelligent}} and conscious. Moreover, whether and how science can handle all this stuff. 
In neuron models, dynamics refers to how the neuron's state variables (primarily membrane potential and ion channel properties) change over time in response to internal processes and external inputs. However, unlike in our approach, the usual processes use a time parameter in empirical or mathematical functions; see, for example,~\cite{NeuralDynamicsGertsner:2014}. We use a genuine dynamics,
based on physics, in the sense that Newton introduced his laws of motion to describe
response of neurons.

The "great journey into the unknown [understanding the brain]"~\cite{BrainInitiative10:2024} must begin at a much lower level: revisiting the fundamental phenomena, disciplines, laws, interactions, abstractions, omissions, and testing methods of science; furthermore, applying them to the brain's neurons. Research must build on classical science, but be reinterpreted for living matter.
There is no independent 'life science', there is only science. It is based on the same 'first principles' but using different abstractions and approximations for living and non-living matters, and having the appropriate relations between them.
We arrived at the boundaries
of classical science disciplines and are moving now through terra incognita. "There is a
clear need for a tighter and more carefully managed integration and realignment of the work"~\cite{HBP_1st_Review:2015}.
Without (re)aligning the knowledge elements along the first principles, the "integration between data and knowledge from different disciplines", lacks integration.
However, even the review of the \gls{HBP} summarized that 
"\gls{HBP} is not developing with the expected level of integration and the project
controls in place are not adequate to achieve this aim."~\cite{HBP_1st_Review:2015}. We desperately need a "new understanding"~\cite{HBP_1st_Review:2015}.

The hype in the newly launched projects is excessive, but
they typically do not target understanding the scientific base
of neuronal operation.
For example, the newly (at the end of 2025) launched “\url{https://brainminds.jp/en}{~Brain/MINDS 2.0}” program in Japan was launched
using only mathematical models, without targeting the understanding of the underlying physical processes. 
"A biological model is often understood to be simply a diagram depicting the interrelationships of various (sub)systems in a process, whereas a physical model is expected to be a theoretical description of a process involving a number of equations of motion stemming from the first principles (if possible), testable against a range of tunable experimental conditions. It must lead to a quantitative prediction and not simply reproduce already known results"~\cite{MolecularBiophysics:2003}.

Our model, although it targets describing biological operation of neurons, is undoubtedly a physical model, in the above sense.
We provide a holistic, multi-level view of neuronal operation and describe (at least semi-)quantitatively the actual physical processes; this way, 
we achieve the promised 'new understanding' (and the need for rewriting books in neuroscience). Although the classical neuron model aims to be an electrical model, it violates fundamental laws of electricity and conservation laws; furthermore, it cannot interpret phenomena from other disciplines (thermodynamics, mechanics, optics). The thermodynamic model(s), on the other hand, cannot explain the electrical features of the observed processes.
We prove that no compromise is needed: the pressure and the voltage
measurements observe the same physical process by different disciplinary means. Biophysics and neuroscience forgot that the slow neural currents consist of ions, which repel each other in a structured confined volume that contains objects with static charge; furthermore, that ions are charged and have mass, which are abstractions from different classical disciplines. 

As Feynman R. P~\cite{FeynmanThinking:1980} wrote, "the separation of fields [of science] \dots is merely a human convenience, and an unnatural thing.
Nature is not interested in our separations, and \textit{many of the interesting phenomena bridge the gaps between fields}". "When we go to investigate \textit{we should not pre-decide what it is we are trying to do} except to find out more about it. \dots The first principle is that you must not fool yourself, and you are the easiest person to fool."
Physiology, despite \gls{HH}'s and Feynman's warning, pre-decided that only the science discipline 'electricity' is authorized to describe neurons' operation.
However, electricity has no place for the mass and speed of its charge carriers. After seven decades, it should be realized, admitted, and fixed.

We agree that "Biological laws are much more difficult to find than physical ones"~\cite{VollmerLimitsOfBiology:1995}.
Research already provided the detailed description of a correct ‘net electrical’ neuron model~\cite{VeghNeuronAlgorithms:2025}, the laws of motion for describing biology~\cite{VeghNon-ordinaryLaws:2025}, and now provides a correct non-disciplinary (unified and scientifically complete electrical/thermodynamic) model.
The deeper reason for the apparent conflict between the two competing theoretical descriptions is science's disciplinarity, as we discuss it separately in Appendix~\ref{sec:Physics-Fundamental}, since it is somewhat philosophical and touches not only the fundamental concepts of physics but also the way of scientific thinking. Shortly, the disciplinary separation proved to be successful for
most phenomena of inanimate science. Electrolytes are an exception. Thermodynamics (due to excluding a long-range interaction between its particles), excludes electric charge (and related quantities) from
the quantities it studies. Electricity (due to implying an 'electron cloud' as a charge transmission method) similarly excludes the possibility that long-range interaction may exist between its discrete charge carriers; that is, 
that they may have thermodynamic features based on the Vlasov equations.
This way, electricity excludes mass (and related quantities, such as heat) from
the quantities it studies. Neither of the two disciplines, alone,
can completely describe electrolytes (and so: biological systems): some of the system's quantities 
remain outside the scope of the given discipline. The measurements 
are not complete: some quantities are not measured, so 'miracles' (such as heat emission and absorption)
can occur. Nature does not respect science's disciplinary separation.
The disciplinary force analysis does not show a driving force, so some "new force or whatnot"~\cite{Schrodinger:1992} must be assumed (in physiology, "protein mechanisms"), in contrast with science's principles.  Our non-disciplinary approach, however, can successfully 
describe the operation of a neuron without miracles.

The structure of the paper is as follows.
Section~\ref{Physics-StateOfTheArt} presents the state-of-the-art of theoretical neuron modeling. A strongly simplified 
discussion in section~\ref{sec:Single_Nutshell}, without mathematical, physical, and electrical details, presents the unified model. It also derives how the separate disciplines 
electricity and thermodynamics can be bridged for explaining neuronal operation.
Electricity of ions and its application to the case of neurons is the subject of section~\ref{sec:Physics-segmented}, which includes the specifics of
ion currents, the electrical features and operation of biological objects, including the correct explanation of the origin of the resting potential; all without magic 
protein mechanisms. Section~\ref{sec:Physics-FormingPotential} presents the correct physical mechanisms that 
define and maintain the
membrane's potential in resting and transient states. The importance of the action potential deserved a separate section \ref{sec:ActionPotential},
where the physical processes, their correct "net electrical" model, with the related mathematics, are described; again without physically impossible cooperating ion channels and protein mechanisms. It also explains the observation that the accuracy of timing is about two orders of magnitude better than the one that the length of \gls{AP} provides. The discussion culminates in section~\ref{sec:Physics-Thermodynamics}, which on the one side derives 
the thermodynamic parameters describing \gls{AP} and their long sought calculation method (including the principle of the thermodynamic operation of neuron, its energy consumption and efficiency), on the other provides evidence that only the cooperation
of the competing disciplines can succeed: the thermodynamic parameters can
be calculated by measuring parameters of electrical processes.

\section{State-of-the-art\label{Physics-StateOfTheArt}}

Today, two major branches of disciplinary theories compete 
for being applied to describe the neuronal operation \cite{CriticalElectricity:2018,ThermodynamicAPDrukarch:2022,PerspectivesNerveSignalPropagation:2024,ComparisonHHandSoliton:2010,HH_Potential_Controversies_2017,PiezoelectricNeuron:2025}. Those cited references also compare those theories.
The 2-decade-old Heimburg-Jackson thermodynamic theory is trying to break in alongside or displace the 7-decade-old all-electric Hodgkin-Huxley theory. The two theories apply one of the scientific disciplines developed for inanimate nature, unchanged, to living nature. There are unexplained observations for each of the two theories, and neither can describe the biological neuron's functioning without contradictions. 
In a disciplinary view, there is no chance of the two theories fusing. Although they can describe a wider range of phenomena together, the conditions under which they are applicable exclude their joint use. Moreover, to conceal contradictions and fill gaps, biophysics assumes (typically unspecified protein) mechanisms that contradict the fundamental principles of physics, and some of them even the disciplinary laws.

The electrical view essentially stems from the (Nobel Prize-winning) work of Hodgkin and Huxley \cite{HodgkinHuxley:1952}. They could not make a perfect job, mainly due to the lack of
discoveries made several decades later, including ours, about handling
the finite speed of the biological ionic currents (and, due to that, the dynamic features), which drastically change
their conclusions. In contrast with their \textit{empirical description}, which delivered \textit{mathematically formulated measured observations}, our discussion, although it follows essentially
the same principles, reinterprets and pinpoints the used fundamental terms, sets up a physical model, and \textit{explains} the physically underpinned processes. "However, the theory could not explain the physical
phenomena such as reversible heat changes, density changes,
and geometrical changes observed in the experiments" \cite{ComparisonHHandSoliton:2010}. Given that
our (non-disciplinary) model natively connects charge and mass of ions~\cite{VeghNon-ordinaryLaws:2025}, it can explain the missing phenomena.
The thermodynamic view roots in the (possibly Nobel-prize-winning)
idea of Heimburg and Jackson \cite{SolitonPropagation:2005}. They proposed that the action potential is essentially a pressure wave (a soliton). ``However, there are several other questions that this has to answer like ion flow involvement in nerve signal propagation as stated by the \gls{HH} model and also the faster propagation in myelinated nerves than in unmyelinated'' \cite{ComparisonHHandSoliton:2010}.
If we consider that the presence of ions in the neuron's closed volume means an unbalanced force component exerting only on ions (but not on the vibrating neutral molecules)
and the ion flow means charge propagation in a membrane tube where the tube's specific capacity depends on the thickness of the axonal membrane's myelin sheath (aka the distance across the membrane surfaces), the faster propagation is not mysterious anymore.

The excellent textbook's statement that "the resting membrane potential results
from the separation of charge across the
cell membrane"~\cite{PrinciplesNeuralScience:2013} is only half the truth; we tell the second half in section~\ref{sec:Physics-segmented}. 
We refute their statement that "
resting ion channels
establish and maintain the resting potential"~\cite{PrinciplesNeuralScience:2013}, page 126.
The ion channels are passive players. The interplay of 
the lipid condenser and the electrolytes on both sides
of the permeable membrane establishes the resting potential, and ionic gradients maintain it.
The textbook separates the neuron membrane's states into resting and transient states. 
It introduces resting and transient ion channels,
but does not introduce that the different physical processes in those states need different models. 
As we show in section~\ref{sec:Physics-FormingPotential}, \textit{two different physical mechanisms
operate in the resting and the transient states}, so two different models are needed for describing the same components.

A primary reason neurophysiology has been misguided is that it has in mind a static picture of the dynamic life of neurons.
It started from the static
inanimate anatomic microscopic pictures, continued
with the static clamping investigations (where
the negative feedback of an external electric circuit eliminates the natural gradients) and underpinning the
theory by analyzing frozen (inanimate) samples
by electron microscopes.
Those investigations revealed a tremendous amount of crucial details,
but they obscured the idea
that those samples are just snapshots of a permanently
changing system.
Neuroscience projected the mechanisms of the static resting state to the dynamic transient state and attempted
to understand the transient state 
as perturbations~\cite{PerturbationNeuralComputation:2002} to the  (entirely different) resting state.
Of course, that attempt required more and more ad hoc non-scientific assumptions.
We agree that "Understanding and predicting molecular responses
in single cells upon chemical, genetic or mechanical perturbations is a core question in biology."~\cite{PerturbationSingleCell:2023}
However, it is not possible without understanding that the transient state is a state on its own right,
with different physical mechanisms, rather than a perturbation of the entirely different resting state.

Due to the lack of the needed expertize in physics,
biophysics introduced the idea of the "whatnot"~\cite{Schrodinger:1992} protein mechanisms.
Claims such as "pumps [by using protein mechanisms] \dots transport ions \textit{against their
electrical and chemical gradients}"~\cite{PrinciplesNeuralScience:2013}, page 101, 
or "the flux of ions through ion channels is passive,
requiring \textit{no expenditure of metabolic energy} by the
channels."~\cite{PrinciplesNeuralScience:2013}, page 107, are as scientific as in the pre-Newtonian age,
there was a notion related to Aristotelian philosophy. This concept gave credence to the “theory” that the celestial spheres and, later on, the planets were pushed or moved by celestial movers, celestial intelligences, or angels against physical forces. Life science does not have its laws of motion, in the sense as Newton's Laws in physics, at least at cellular level.

The book~\cite{PrinciplesNeuralScience:2013} asks the central questions, "How do ionic [i.e., electrical and chemical] gradients contribute to the resting membrane potential? What prevents
the ionic gradients from dissipating by diffusion of
ions across the membrane through the resting channels?"
However, it leaves them essentially open by giving only a qualitative answer, discussing only membrane permeability, without explaining how the resting potential is created.
We demonstrate in section~\ref{sec:Physics-FormingPotential} that classical methods of electricity enable us to calculate the membrane's potential resulting from charge separation and polarization, why and how its resting value is created, furthermore, how it is regulated.
We make some physically plausible assumptions to provide numerical figures.

Our results underpin that "the crucial system in biology isn't a molecule or a molecular class whatsoever, but the interface created by biomolecules in water"~\cite{LivingSystemPhysics:2021}. We add that mainly
physical processes at various speeds (instead of molecular classes such as proteins) and ionic gradients (instead of protein mechanisms) control the processes.
More precisely, inseparable thermodynamic and electrical processes set up and maintain the potential established by the electrical and thermodynamic backbone defined by the lipid/protein structure of the cell and the ionic solutions it contains. It is one of the frequent cases when 
one effect implements a functionality (the backbone closely defines the frames of the operation), and the other (in this case, combined thermodynamics and electricity) corrects it when it implements a complex operational (dynamical) functionality:
"stimulated phase transitions enable the phase-dependent processes to replace each other ... one process to build and the other
to correct"~\cite{BiologicalConservationLaw:2017}.

\section[Operation in a nutshell]{Operation in a nutshell\label{sec:Single_Nutshell}}

We have Schrödinger's warning in mind when constructing the {physical picture
about neurons}, that we must be very careful. \textit{We must not use the
'ordinary' laws of physics derived for non-living matter without revisiting them} because "\textit{the construction is different from anything we have yet tested in the physical laboratory}"~\cite{Schrodinger:1992}. The discussion requires a deep knowledge in science,
so in this section, we remain on the surface and provide a holistic picture of the model.
The deeper layers are explained in later sections. 
Appendix~\ref{sec:Physics-Fundamental} details the reasons.

The fundamental principles are the same, but we must use different 
{approximations
and abstractions} for the different disciplines of science and for living matter. (Notice that when defining the subject and laws of mechanics and thermodynamics,
we used the approximation that electric charge does not play 
any role; similarly, in electricity, mass plays no role.
Neither of those approximations is valid for ions; so \textit{those disciplines, alone, cannot be used to describe electrolytes; consequently, to describe processes in biology}.)
For this reason, it may be more correct to call those laws governing living matter 'non-disciplinary' rather than 'non-ordinary' laws.

We must also include the discoveries from the past seven decades,
among others, that a component \gls{AIS} can be found between the neuron's membrane and the axon. It is observed that \gls{AIS} has a crucial role in creating the \gls{AP}, but
that role has not been integrated into the existing theories.
As a consequence, in the transient state, a neuron shall be modeled
as a serial (i.e., differentiator-type) {electrical oscillator}, which has gated current 
inputs (instead of a parallel oscillator 
without gating its inputs, which is a good model only in the static view of the resting state) and provides a timed output. We also explicitly add that the charge carriers, the neuron 
works with, are \textit{ions} (instead of electrons) and that the signal propagation
mechanism is \textit{bioelectric} (instead of purely an electrical
or mechanical wave or "protein mechanism"), in many cases even without having free carriers in the respective volume. (We do not consider the
ideas that the associated mechanical, optical, etc. changes cause electric phenomena. They reverse the true causality: they do not have 
a triggering cause; the view inherited from the narrative biology.)
Electrochemical operation,
especially that of {semipermeable membranes}, requires special attention: 
charge carriers are "created" inside the biological matter.
Omitting the clear physical background results in performing
{wrong measurements} and introducing {wrong concepts} into biology.
The conclusions of the model are in perfect agreement with the experimental results from the cited publications, including classic publications and textbooks, although
in some cases their corresponding conclusions must be revisited, since the papers used wrong models. 

It is hard to compare our cross-disciplinary model to the different variations of the existing classical disciplinary models because they lack some concepts our model uses (or replace them with ad-hoc assumptions). They cannot consider
ions' charge and mass simultaneously; correspondingly, the mathematical formulas exclude either the charge or the mass of ions, so they cannot describe the complex case of living matter. It is also hard to validate it by some experiments
since those experiments are designed along the concepts 
of the classical disciplinary theories.
 For example, they do not include \gls{AIS}; the electrical models 
work with "fast" currents and the thermodynamic one quite without currents. Bioelectrical operation,
especially that of {semipermeable membranes}, requires special attention: 
charge carriers may be "created" inside the biological matter;
a fact omitted by all existing theories.
(We do not consider the
ideas that the associated mechanical, optical, etc. changes cause electric phenomena. They reverse the true causality: they do not have 
a triggering cause; a view inherited from the narrative biology.)
Unlike our model, neither of the classical theories can describe all observed neuronal operation without contradictions.

\subsection{Model's components\label{sec:Physics-ModelComponents}}

In the (highly simplified) non-disciplinary physical model (the 'abstract physical neuron', as we call it), a neuron is an elastic insulating sphere with two lipid layers (membrane), a similar output tube (axon), and an incompressible, partially ionized electrolyte fluid with different compositions and concentrations inside and outside.
They contain ions, simple chemical molecules, and complex biological formations.
These components may have gradients and move at continuously changing speeds that can vary by several orders of magnitude during operation. The lipid layer is partially permeable (controlled or uncontrolled) for the components above (channels, mainly ion channels).

The membrane represents a confined space in which ions behave differently from in an infinite space; furthermore, it significantly alters the properties of the few-nanometer-thick layer near the membrane.
The elastic wall of the membrane is covered with ion layers, and therefore, a potential difference is created there (the resting potential).
In this layer, the concentration and potential change significantly with the distance from the membrane due to interactions between the membrane's ion layers and the electrolyte.
Ions sent by other neurons can enter the internal volume through controlled inputs (synapses).
Between the membrane and the axon, there is a large amount of uncontrolled ion channels (\gls{AIS}). The conductance of the channels in the membrane's wall
is about 50 to 100 times smaller~\cite{ActionPotentialGenerationNatrium:2008,AIS_Updated_Viewpoint:2018} than that in the \gls{AIS}.

\subsection{Model's operation\label{sec:Physics-ModelOperation}}

The membrane essentially functions as a \gls{PID} control circuit, but it operates on slow ion currents.
Under a small external influence, a neuron remains in its resting state and regulates itself through non-controlled ion channels in the wall of its membrane.
If the external influence exceeds a threshold, the controlled channels suddenly release a large amount of ions into the internal space, switching the neuron into a transient state.
The membrane potential changes suddenly (within a few dozen nanoseconds), and, due to the mutual repulsion of ions, the pressure acting on the membrane wall and the ion concentration in the $\approx$\SI{1}{\nano\meter} layer near the membrane also increase significantly.

Since the \textit{same physical action generates the voltage gradient and the concentration gradient} (and, consequently, the pressure), they are inseparable and proportional to each other.
Generating a voltage change (by clamping, direct current, magnetic pulse) or a pressure change (by ultrasound, mechanical, shock waves) on the membrane causes the other quantity to change. Inputting ions (by synaptic input, followed by rush-in) generates both changes.
The control circuit tries to restore the resting state, during which the mentioned quantities decrease proportionally.
The transient state of the neuron can, in principle, be described by any of the mentioned entities, but, as a practical matter, \textit{the effects of changes in the other quantity must also be taken into account}: \textit{no single discipline, alone, can describe the changes}.
The elastic membrane implements an almost critically damped vibration, but the "vibrating mass" and the "spring force" are constantly changing.
In an electrical resonant circuit, the capacitor's voltage and capacitance change due to the finite amount of charge stored on it.

\textit{The pressure wave caused by a force shock and the voltage wave caused by a voltage shock describe the same effect}: the sudden increase of ions in the layer in question.
In both cases, the other effect must be taken into account to some extent: an infinitesimal movement of an ion also involves changes in both pressure and voltage.
Changes in pressure and charge may alter the membrane's thickness (and therefore its capacity), and they can cause a structural change (a phase change, "melting"; density; size) in the lipid chains that make up the membrane.

In the case of a pressure wave, the speed is naturally finite (mass must be moved), but the laws must be adapted to the case of delivering charged particles (that repel each other).
For the electrical wave, the laws tacitly assumed instantaneous interaction of massless carriers; so, they must be adapted to slow ion currents.
It is simpler to describe the generation of the action potential as an electrical oscillation (although slow currents' propagation must be accounted for), and its propagation as a pressure wave (although this only represents well the wave front: the \gls{AP} lasts as long as there is non-equilibrium charge in the neuron: the repulsion between the ions pushes the ions out of the closed space, so the intensity of the wave continuously decays).
Hyperpolarization is described in the electrical view as a capacitive current in the sequential oscillator circuit, in the thermodynamic view as a suction effect occurring in the opposite phase of the membrane. 
In both cases, however, it must be taken into account that charge and mass are inseparable in ions, and that their cross-disciplinary effects significantly reduce the applicability of the disciplinary laws.

In simple words, a neuron combines different disciplines when generating an \gls{AP}. In the thermodynamic view, it works like a two-tact biological internal combustion engine, and can be described theoretically as a special Carnot-type thermodynamic engine. 
In the electrical view, it operates as a serial $RC$ circuit 
(that, of course, has a significant capacitive current in the second phase of operation). The synaptic input currents operate the ignition. The influx of $Na^+$ ions acts like an explosion, producing a large pressure impulse due to electrical repulsion between ions.
The neuron's volume remains practically unchanged (an iso-volume process) during the ignition phase. The extension phase is practically limited to the layer near the membrane: ions move slowly.
The process is similar to the generation of a sound wave, with negligible material transport.

The force analysis shows that the thermodynamic force, generating a pressure wave, 
causes a strongly damped longitudinal vibration, while the changed potential adds
a potential-dependent longitudinal force component for moving the ions toward the \gls{AIS}; as long as there are excess ions in the neuron.
The resulting force moves the ions with variable speed out of the neuron through the only available output channel: the axon. 
The process is entirely in line with the principles of science,
but the ion, with its mass and charge, combines two disciplinary forces. After summing the two disciplinary forces, one can use Newton's Laws to describe the ion's motion.
Notice, however, that the electrical charges move at a
speed, which is the bargain between the mechanical wave and the electrical interaction.
So, the truth is halfway between the disciplinary explanations:
the charge moves under the effect of the resultant force
that derives from the vibrating mechanical force exerted on the mass
and the electrical repulsion force exerted on the charge
of the ions. 

In the first stroke, the pressure impulse invests energy into suddenly compressing the elastic membrane and starting a compression wave (a soliton); in the electric view, it suddenly charges the condenser.
The two waves are synchronized, see Fig.~\ref{fig:HeimburgElascticity}. In the second stroke, a strongly varying combined thermoelectrical/mechanical force (rather than some fixed-voltage magic battery) pumps out ions through the ion channels in the \gls{AIS}, which "resists" it, thereby generating a potential wave known as \gls{AP}; well described by electricity. 
Electricity interprets the backward current as capacitive current.
Physiology observes hyperpolarization 
and explains that the outward current in the channel consists of $Na^+$ ions,
and when the resultant potential changes its direction, 
the ions change to $K^+$; a kind of alchemy. Alternatively, 
it hypothesizes intelligent ion channels that synchronize 
their operation to the operation of distant ion channels regarding the type and amount of ions
they deliver, without any evidence for the existence of 
a communication channel or a control mechanism. Another alternative is that the two kinds of positive ions
travel against each other, without a driving force and without their charge acting on each other.

The pressure wave is well described in the thermodynamic view.
The compression
produces heat, as evidenced by a rise in the temperature. The expansion consumes heat, as evidenced by a decrease in temperature. These are well-understood processes in thermodynamics.
However, they are still unexplained in biophysics, although they were observed
seven decades ago~\cite{HeatProductionNeuron:1958}.
The elastic membrane functions as a damped oscillator, which, in its negative amplitude phase, reverses the direction of electrolyte flow (and, consequently, the direction of the current carried by the ions).
Correspondingly, one can observe, in addition to the electrical potential wave, mechanical, optical, and other features changing. However, thermodynamics cannot account for
any electrical change, given that electric charge is
outside the region of its interest.
In the discipline of electricity, there is no direct way
to interpret heat emission and absorption (see section~\ref{sec:Physics-HeatAbsorption}), given that the
heat is outside the notions used in electricity.

The shocking conclusion in~\cite{MechanicalBrainPulses:2018}, that "brain cells communicate
with mechanical pulses, not electrical signals", more precisely sounds that mechanical pulses (pressure waves) deliver the charge carriers
of the electrical signals, in close interaction with the repulsion of the delivered charged carriers. In Equation~(\ref{eq:IonicForces}),
we also must consider the mechanical force delivered by the pressure wave.
Furthermore, the nervous pulse transmission over the axons
must be revisited. The classical mechanism based on the coordinated action
of ion channels is undoubtedly wrong. The enormous pressure, the incompressible electrical fluid,
and the repulsion of the delivered particles suggest
an alternative and more reasonable mechanism.

\subsection[Forces]{Forces exerted on 
	 ions\label{sec:Physics-ForcesMembrane}}
\begin{figure}
	\centering
	\includegraphics{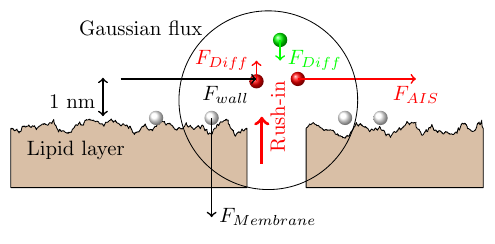}
	\caption{The forces exerting on ions at the moment when the rush-in finished.
		\label{fig:Physics-MembraneForces}}
\end{figure}

Figure~\ref{fig:Physics-MembraneForces} shows the forces exerted on ions 
near the internal surface of the membrane 
at the exit of an ion channel
in the moment next to that
$Na^+$ ions rush into the internal segment of the neuron.
The white (likely mainly $K^+$ with some $Na^+$) ions are sitting on the surface, and the charges on the opposite side of the membrane ($F_{Membrane}$) firmly fix them. They are not "free charge carriers": they provide the resting potential. They are at rest because of the counterforce exerted by the 
lipid layer.

At the beginning of the \gls{AP}, $Na^+$ ions rush into the intracellular segment. The accelerating field across the membrane's surfaces moves them at the corresponding Stokes-Einstein speed through the ion channel.
When they reach the ion channel's exit,
in the absence of the enormous field, the
ions slow down to the speed corresponding to the
local field in the electrolyte; practically, in a short distance, they "stop".
This way, a large amount (about $10^7$; about the amount of the free charge carriers in the segment) of ions appear suddenly
in the layer near the surface;
that is, the charge near the ion channel's exit port
suddenly increases. (In the previous moment, the ions present in the volume segment are free to move, but the resultant force of the thermodynamic and the electrical forces keeps them in place
in a balanced state.)
Since the total electric flux through any closed surface (Gaussian surface) is directly proportional to the enclosed electric charge, the local electric field suddenly increases compared to its environment. Hence, the ions attempt to move along the local gradient. 
There are constraints, however.

In the resting state, most of the ions in that local volume are $K^+$ ions.
The rushed-in ions suddenly increase the local $Na^+$
concentration that practically kicks out the $K^+$ ions,
which help to block further ion input.
Their appearance has two different effects
in the directions perpendicular to and parallel with
the membrane's surface.
In the perpendicular direction,
relatively small $F_{Diff}$ forces move the ions
toward the former 'resting state' (the forces are composed
of the local thermodynamic and electrical forces).
In the direction of the ion channel (where they arrived from), they face an electric field of opposite direction.
Given that the fields are low, the speed of the in-place ions
is low (much above the diffusion speed but much below the rush-in speed), a significant ion enrichment in the near-membrane region occurs.
Notice that the ions move under the resultant force
due to the thermodynamic and electrical gradients, so the
$Na^+$ ions move toward the bulk, the $K^+$ ions
toward the membrane (due to the different thermodynamic gradients). This movement, however, is slow
compared to the speed of the gradient change of an \gls{AP} process.

In the horizontal direction, only the ions in the 
"{dynamic layer}" participate in the game.
Three forces exerted on the ions are: 
\begin{itemize}
	\item the increased electrical field pushes them 
	toward the \gls{AIS}
	\item the increased local $Na^+$ concentration (thermodynamic plus electrical field) pushes the ions toward the \gls{AIS}
	\item the injected ions repel each other, so they arrive at the membrane's wall, and they push the 
	ions toward the \gls{AIS} (a constraint force)
\end{itemize}

Based on the electric fields (the forces acting on a charge
are proportional to the local field), one can estimate the
order of magnitude of the forces in Fig.~\ref{fig:Physics-MembraneForces} (not proportional). 
We can assume that around the peak voltage of the \gls{AP} is \SI{100}{\milli\volt}, that means across the \SI{50}{\micro\meter} \gls{AIS}
a $\frac{10^{-1}}{50\times 10^{-6}}=$\SI{2}{\kilo\volt\per\meter} electric field. That is, one can safely assume that the force
perpendicular to the membrane's surface (even if it decreases
during generating an \gls{AP} 
to the half compared to its value in the resting state) is large enough 
to keep the ions fixed on the surface. Therefore, the 
composition and the amount of charge on the surface can be considered 
as constant, and the electric (and the related other) changes shall be
attributed to the charges in the {dynamic layer}.

All this happens in the {layer of thickness of 1~nm} near the
surface of the membrane, as observed by physiology. For the ions, the only way out is to leave the membrane through the
\gls{AIS}. The \gls{AIS}, with its large number of ion channels,
can be modeled as a resistor and the ion current generates a potential on it. As described, the potential changes its value and direction. Correspondingly, ions flow through the \gls{AIS}
initially outward, later inward, then again outward. There are no separated and cooperating $K^+$ and $Na^+$ channels with precisely timed activation. Instead, the mechanisms discussed above are valid. The \gls{AIS} current consists mainly of $Na^+$ with a small amount of $K^+$. 
As discussed, new $Na^+$ ions from the outside space flow into
the intracellular space. Although most of the $Na^+$ flows out through the \gls{AIS}, the leftover diffuses into the bulk and must be pumped out.
This fact is why Na/K pumps are needed.

At any given position,
the Gauss flux of a volume continuously changes as the amount of
charges in the given volume changes due to the vibration of the electrolyte, plus the mentioned forces. The electric field and
the potential are proportional to the flux and so, to the charge, and so, to the current.
The speed of the charge wave
depends only on the medium (i.e., constant), and
the amplitude is simply an exponential discharge.
The observer can measure the effect.

Due to turbulence in the material flow, the "white" ions may also move toward the bulk for a short distance before reaching the dynamic layer.
That is, all positive ions may move out through the \gls{AIS},
without selectivity. 
As discussed in section~\ref{sec:Physics-BackgroundLogistics},
the hydrolyzis of \gls{ATP} produces "fresh" ions near the 
membrane and the local field moves them toward the membrane,
so the neuron can restore its '\gls{setpoint}' as described in 
section~\ref{Physics-ControlTheory}. That means the "downhill" movement of ions across the membrane through ion channels does not violate the conservation of energy.
However, claims such as "\textit{Ion channels cannot be coupled to an energy source} to perform active transport,
so the transport that they mediate is always passive ('downhill')" \cite{MolecularBiology:2002}, are wrong.
The surfaces of the membrane serve as a kind of 
ion- and energy buffer, and the continuous material transfer
(the background logistics of life) performs
refilling the buffer.

\subsection[The Magic Connection]{Connecting electricity and thermodynamics\label{sec:Physics-MagicConnection}}

The injected ions suddenly act on the elastic membrane as an "impulse force", so the membrane (and the incompressible electrolyte it contains) performs a damped oscillation.
Given that the resultant of the above-mentioned  electrical
thermodynamic and constraint forces points toward the \gls{AIS}, the ions will move with the corresponding Stokes-Einstein speed toward the \gls{AIS}.
The electrical and thermodynamic forces would essentially produce a simple discharge, with only one current direction.  However, $F_{Wall}$ is an oscillating force that changes its direction and may be larger than the other forces.
The
damped oscillation has positive and negative phases,
thus increasing and decreasing the electrical force.
In a thermodynamic view, the membrane "sucks back"
the electrolyte containing the ions. In the electrical view,
the direction of the ion stream (aka current) changes to the opposite, causing the 'capacitive current' to change
the sign of the resultant current (the neuronal membrane
can be modeled as a condenser).  In a physiological view, 
polarization, depolarization, and hyperpolarization take place. As the model clearly suggests, the ions can be likely 
 $Na^+$ (and some $K^+$), the question is only their proportion.

Essentially, the Gauss flux connects electricity and thermodynamics. The electrical field is proportional with the number of charges in a  Gaussian volume and so is the pressure with the number particles in the volume; so, both are proportional with the number of ions in the volume. (In a strict sense, due to the finite speed of the ions, this statement is valid only in equilibrium state). By using the value of the force acting on a unit charge
in the field across the membrane is
\begin{equation}
	F_{Na^+}= 10^7 *1.60217663* 10^{-19} [V] [C] = 1.6*10^{-12}\ [N] \label{eq:UnitForce}
\end{equation}
we can estimate how the pressure of the neural cell increases due to the rush-in changes at the beginning of the
\gls{AP}. 
As evidence shows, the local potential at the internal surface of the membrane is in the range of \SI{100}{\milli\volt} in the resting state and increases by $\Delta U=$ \SI{100}{\milli\volt}  in the transition state. This increase means a change in the force acting on an ion (see Eq.~(\ref{eq:UnitForce})) by  $1.6*10^{-12}\ [N]$.
When we assume $10^7$ rush-in ions and unchanged cell size, the total force acting on the membrane increases by $1.6*10^{-5}\ [N]$.
This change in force means on the neuron's $8*10^{-9}\ [m^2]$ surface (see~\cite{JohnstonWuNeurophysiology:1995}, page~12) a pressure change 
\begin{equation}
	\Delta P = \frac{1.6*10^{-5}\ [N]}{8*10^{-9}\ [m^2]}
	= 2*10^{3}\quad \biggl[\frac{N}{m^2}\biggr]
	\label{eq:CellPressureChange}
\end{equation}
\noindent The pressure can be calculated using electrical parameters; i.e., in a cross-disciplinary way.
By assuming that the ions are distributed evenly on the
neuron's surface, their average distance is about \SI{30}{\nano\meter}. That is, their distance on the same size
of the membrane is about 6~times larger than the distance across the membrane. This way, the parallel force is about 40~times smaller than the perpendicular one (furthermore, the parallel force is almost perfectly balanced by the neighboring ions).
That means, we can use the approximation that the perpendicular force firmly keeps the ions on the surface.

\subsubsection{Resistance in neurons\label{sec:Physics-ResistanceNeuron}}

As charge carriers, ions differ in many features from electrons, tacitly assumed in physical electricity. Their conduction mechanisms
are entirely different, and they are not necessarily present 
in the discussed medium (say, when the medium is surrounded by charged surface, the field moves free ions out of the volume). They may be produced by biological mechanisms, 
and they are slow (have a couple of $m/s$ speed, creating the illusion
of 'delayed current').
Disciplines physiology and electricity provide an atomic level description, but unfortunately they use the same word for their concepts, although
their meaning can be entirely different. 

In solids, as the Drude model describes, 
electrons constantly bounce among heavier, stationary crystal ions, that make up the structure of the material. With each collision, though, the electron is deflected in a random direction with a velocity that is much larger than the velocity gained by the electric field. The net result is that electrons take a zigzag path due to the collisions, but they generally drift in a direction along the electric field. Important, that \textit{the collisions are inelastic}, so the electrons lose (most of) their energy. The energy is absorbed by the solid's gridpoints, which later is released in form of heat. \textit{The process is irreversible}.

In biological matter, so called ion channels
are used for ion transfer. Here diffusion and electromigration takes place, at by orders of magnitude different speeds under the effect of an electric field, see Equ.(\ref{eq:StokesEinsteinSpeeddV}). The resistance 
is interpreted as the interaction of charge carriers (this time, their mass is relevant), which are transported by a driving force, with the neutral particles in the material.
In solid state, the volume considered as "infinitely large" (that is, the field is homogeneous
and the propagation is isotropic), and the charge carriers are uniform; that is, only one type of interaction is present.
In biological objects, the volume is more or less limited (is finite), charged objects may be present that make the propagation non-isotropic and the charge carriers are non-uniform that introduces thermodynamic interaction in addition to the electrical interaction.

Ions, under an electrical/thermodynamic field (and maybe mechanical constraints) accelerate to the Stokes-Einstein speed (see Eq.(\ref{eq:StokesSpeed})), and move against a friction in the electrolyte.
The ions suffer (essentially elastic) collisions with neutral molecules and transfer their momentum (and so: energy) to them. 
From the point of view of ions, the process is energy loss and momentum loss.
From the point of view of neutral molecules, the process is energy gain and momentum gain.
Given that the collisions are elastic, no energy dissipates;
so, in principle, \textit{the process is reversible}.
Recall that the momentum loss results in momentum gain of the neutral molecules. This way, the electrical force
accelerates also the nearby neutral molecules; furthermore, their momentum (using an elastic membrane) can fully contribute to the mechanical force shock that triggers the \gls{AP}).
Here comes to light an important difference between the 'infinite space' approximation and the closed volume of biology. In a free space, from the point of view of the ion, the 
energy would dissipate, despite the elastic collisions.
 If the acceleration occurs in a narrow tube 
such as an ion channel, the wall prevents gaining momentum in the 
direction perpendicular to the direction of acceleration.

The 
conduction mechanism of ion channels provides another nice example for the cooperation of
science's disciplines behind the scenes. The energy lost by ions due to the Stokes-Einstein friction is not 
dissipated as in the case when electrons collide with fixed
gridpoints in a solid. Instead, the neutral atoms and molecules
they collide with on their way, receive an impulse (and kinetic energy)
in the direction of the accelerated ion.
This way, \textit{indirectly, also the neutral atoms and molecules are accelerated by the electric field across the neuron's membrane}. They also contribute to the force shock,
and they represent the overwhelming majority of the 
energy/impulse required to issue an \gls{AP}.

\subsubsection{Mechanical consequences\label{sec:Physics-MechanicalWave}}
{ 
	The pressure of electrical origin
	enables one to compare theoretical pressure and force values
	with the experimental ones. 
	That pressure on a $10\times10~\mu m$ tooltip is converted to a force
	$2*10^{3}*100*10^{-12} = 200~pN$ force. The measured force value~\cite{MechanicalPropertiesNerves:2025} is about $600~pN$,
	so our estimation is in the correct range. 
	The measured pressure value is~\cite{PressureChangeActionPotentialMeasured:1980} $5\ \frac{dyn}{cm^2} = 0.5\ \frac{N}{m^2}$, which is an average value
	for some period. If we assume a $1~Hz$ frequency, and a $1~ms$
	period for the duration of the \gls{AP} peak, it means $1*10^{3}\, \bigl[\frac{N}{m^2}\bigr]$.
	So, the measured mechanical change values~\cite{NeuronalDeformation:2020} do not contradict our hypothesis that the increased pressure increases the neuron's size via its elasticity.
	The peak of swelling was shown to coincide fairly accurately with the peak of the action potential (i.e., when the concentration reaches its peak value; see also Fig.~\ref{fig:HeimburgElascticity})~\cite{RapidMechanicalBiphaseActionPotentialMeasured:1981}.	
}

The consequences of the changes in the electrical charge are large enough to explain why 
pressure wave and other mechanical changes~\cite{MechanicalWaves:2015} also start at the beginning of the
\gls{AP},
and other (such as optical, density) changes 
are accompanied by it. Nature invests energy also 
in the conventional way, a term $\Delta V\times P$, into the thermodynamics of neural operation; not only in the form of storing energy in the 
changed electrical field. Thermodynamics and electricity, not to mention elasticity, must not be 
separated when discussing the neuronal energy budget.

Notice an important difference. The acceleration of an ion is unbelievably large. It would be sufficiently large to keep the potential at the same value along the membrane,
provided that the ions must follow a small change, such as due to the
"leaking" current through the
\gls{AIS}.
Similarly, the ions can 'instantly' follow quick changes such as
a square wave gradient.
However, if many similar ions are ahead, their repulsion and the viscosity decrease
the acceleration, and the ion travels only at a few $m/s$ speed.
The force decays quickly. The ions start to move
'instantly', but the charge carriers can move only with a limited speed,
much below the interaction speed of
\gls{EM}
interactions. The effect can propagate only with that lower speed; which effect calls for interpretation by the soliton theory.
However, in the case of an axon, there exists a mechanical constraint that
the ions cannot spread through the wall.

\subsubsection[Energy]{Energy relations\label{sec:Physics-EnergyRelation}}

We can estimate the work done when the rush-in charge
extends the neuron, as $P\times\Delta V$. We consider an average value $P$, and we calculate the change of the volume as neuron's surface $10^{-8}\ m^2$ multiplied by 
the change of the radius $\Delta r=10^{-9}\ [m]$\cite{NeuronalDeformation:2020} multiplied by the pressure $P$, that gives $E_{mech}=2*10^{-14}\ [J]$.
Another way to estimate the work the constant electric field across the membrane performs on the ion when moving through the ion channel is $1.6{\times} 10^{-12}\times 5\times 10^{-9}=$ 
\SI{1e{-20}}{\newton\meter}, which amounts to \SI{1e{-13}}{\joule} total for $10^7$ ions.
However, a large fraction of that energy is lost
due to friction. The rest is transferred by collisions
to the neutral molecules, which matches perfectly the value
derived another way round and the experimental data~\cite{EnergyNeuralCommunication:2021,NeuralEnergyConsumption:2017} measured by direct energy consumption.

We can also estimate the electrical energy $E_{electr}=\frac{1}{2} C*(\Delta V)^2 = 1.4*10^{-12}\ [J]$, see also~\cite{EnergyNeuralCommunication:2021}.
To estimate the time required to change the pressure, one can assume that the collision force of a $Na^+$ ion provides the pressure (the shock wave), and this can be calculated as $F = m\frac{\Delta v}{\Delta t}$. (Given that Eq.~(\ref{eq:UnitForce})  provides the force, and one must assume that the ion loses its 
$10^4~to~10^5\frac{m}{s}$ speed, one concludes that it generates a shock wave in $\Delta t = 10^{-8}~to~10^{-7}~seconds$.) By assuming that $E_{electr}$ is the kinetic energy of $10^7$ electrons, one derives the speed
of ions $10^3\ \frac{m}{s}$. This result underpins that
the force accelerating the ions comprises a thermodynamic force
in addition to the electrical one, resulting in a much higher
speed and shorter passage time. These types of calculations 
cannot be performed in a simplified disciplinary approach.

In line with the experimentally measured energy consumption~\cite{EnergyNeuralCommunication:2021,NeuralEnergyConsumption:2017} and our calculation in section~\ref{sec:Physics-DerivingThermodynamics},
the \gls{AP} cycle consumes energy in the range of $10^{-7}\ [J]$. 
The five orders of magnitude between the total and the electrical energy
seems to underpin that in addition to the $10^{-3}$ concentration of ions,
only a small fraction of the ions (the ones in the layers near the membrane) in the volume participate in the electric activity of issuing an \gls{AP}, while the entire volume in the mechanical activity.
The seven orders of magnitude between the total and the non-ionic thermodynamic energy suggests that
classical thermodynamics is not valid for ions.
That is, practically all the energy is invested as elastic energy, but the elasticity modulus of the membrane is exceptionally high.
The primary reason for neuronal operation is the appearance of new ions in the vicinity of the membrane, but this creates a secondary, vast mechanical shock wave. By considering the electrical repulsion force between particles in a slow current, which are entirely omitted in the classic theory, we open the way for explaining the observed thermodynamic and mechanical changes. Instead of alternative disciplinary theories, cooperation between the classical disciplines is needed.

\subsubsection[Speeds]{Speeds
	\label{sec:Physics-UnifiedSpeed}}

{ 
	A way to estimate the ions' rush-in speed is 
	that one assumes that the Stokes-Einstein speed describes
	the ions' speed (Fig.~3 of~\cite{HodgkinHuxley:1952}, quantitatively underpins the hypothesis, see~\cite{VeghStokesEinstein:2025}), both in axons and ion channels. If so, 
	the ratio of the potential gradients equals the ratio of the 
	speeds of ions. One can estimate that the $100~mV$ \gls{AP} on the 
	$50~\mu m$ \gls{AIS} generates a $2*10^4~\frac{V}{m}$ electrical gradient and 
	\cite{HodgkinHuxley:1952} measured $20~\frac{m}{s}$ for the speed of the \gls{AP}. By assuming that the electrical field across the membrane is $10^7~\frac{V}{m}$, one can estimate the speed at the exit of the ion channels as $10^4~\frac{m}{s}$.
}

\subsubsection[Electrical temperature]{Electrical temperature\label{sec:Physics-UnifiedElectricalTemperature}}

For thermodynamic distributions, one can interpret the "temperature of individual particles" with mean kinetic energy $E$ as $T=\frac{2*E}{3*k_B}$ (where $k_B$ is the Boltzmann constant and $T$ is the thermodynamic temperature of the
bulk quantity of the substance); that is, the temperature is directly proportional to the average kinetic energy.
The energy arrives at the neuron in the form of $10^7$ $Na^+$ ions,
and the average energy a single ion conveys is $2.5*10^{-14}~[J]$. One can assume that most (99\%) of that energy is consumed for moving
the ion against the Stokes-Einstein force in the viscous medium, 
so after exiting the membrane, the speed of the ions is $\sqrt{\frac{2*2.5*10^{-16}}{3.82*10^{-26}}}\approx 10^5~[m/s]$, 
which is well above the $\approx~500~m/s$ average speed.

The ions on the two sides of the membrane are in the state with temperature $T$ before and after generating the \gls{AP}. When the $Na^+$ ions rush into the intracellular space, they gain energy through electrostatic acceleration by the membrane's electrical potential, so the same ions appear on the intracellular side as having more energy, i.e., slightly higher temperature. The temperature is more than two orders of magnitude higher. However, the proportion of those "hot" ions is about six orders of magnitude lower, so one can expect a temperature change of up to a millikelvin, depending on the measuring conditions.
In the second phase of the \gls{AP}, those more energetic ions that provide excess local potential above the resting potential, leave the membrane's proximity through the \gls{AIS},
so the temperature decreases in the second phase to its original value. 
Although it is hard to quantify the effect, it provides another explanation of the general heat production and adsorption (the thermodynamic interpretation is described in section~\ref{sec:Physics-HeatAbsorption}); 
providing one more example that cross-disciplinary science instead one of its disciplines must be used.
In the original publication on solitons (essentially pressure waves)~\cite{SolitonPropagation:2005}, a wide range of measured and theoretical data is discussed. They are in line with our conclusions, among others, $80\ \mu K$ measured temperature change.
Furthermore, the electrical energy is about two orders of magnitude smaller than the elastic energy (it is called the soliton energy).

\subsubsection{Resting vs. leakage current\label{sec:Single-RestingCurrent}}
Although through the 
{thickness of the ion layer $\Delta z$}, see section~\ref{sec:Physics-OneSegment}, the ion's water-related behavior can influence the surface charge density,
one can conclude that when only one type of ions is dissolved in the liquid, approximately the same
concentration difference between the segments produces approximately 
the same "thermodynamic electric field" that can counterbalance the 
\index{electric field}
electric field of the biological condenser. One can notice that in the classical measurement by Hodgkin-Huxley, the sum of concentrations of $K^+$ and $Na^+$ in the cytoplasm is 450,  while the concentrations of $Cl^-$ outside is 560; practically equal, as our analysis derived it must be. The deviations are well within the uncertainty of the 
measured values; furthermore, might draw attention to
the quantification of the different ion concentrations.

\gls{HH} did a complete measurement, and derived \textit{precise} 
measured data. However, their measurement was not \textit{accurate} because 
their model was wrong.
In a balanced state, no 
current flows, so no voltage is generated.
For this reason, they assumed that a permanent leakage current $I_L$
was flowing on the membrane, and on its resistance, it generated
a $42.5\ mV$ voltage~\cite{CompanionGuideHodgkinHuxley:2022}.
In \gls{HH}'s picture,
the question \textit{what type of ions constitute the current} remains open. 
It was claimed that the only permeability pathways open at rest are $g_{leak}$ and $g_K$. It follows that at the resting membrane potential, the leak current equals the potassium current. That is, at rest, two currents flow (without driving force in a balanced state), one consisting of $K^+$ ions and another of one $leak^{+/-}$ ions (?) (depending on the direction, it could be positive or negative),
and neither of them changes the concentration neither on the departure nor on the arrival segment, neither the concentration of the other ions; defying also Nernst's law.
Also no problem, that the $K^+$ current is assumed  to flow
through the membrane, and its generated voltage (decrease)
is measured on the \gls{AIS}.  

As discussed,
 the voltage generated across the membrane results from charging rather than a voltage drop due to a permanently flowing current, so no dissipation occurs.
The idea of leakage current was wrong. The voltage 
\gls{HH} measured~\cite{CompanionGuideHodgkinHuxley:2022} is correct, but
-- as our model correctly explains -- it was \textit{generated by charge separation instead of a current through a distributed resistor}. (In \gls{HH}'s picture, the question what was the driving force of that invisible current that generated that voltage, remained unanswered.)
The charge-up voltage is permanently present while ionic segments are on the membrane's sides, even when no current flows. There are
alternative methods of producing a voltage difference across
the membrane's surface; with and without current. \gls{HH} made the wrong choice.
Measuring the energy consumption of neuronal operation~\cite{EnergyNeuralCommunication:2021} confirms that no resting current,
in the sense as \gls{HH} used the notion, exists.

\subsubsection[Heat absorption]{Heat emission/absorption\label{sec:Physics-HeatAbsorption}}
The lack of leakage current 
not only validates our model, but also solves the long-standing mystery of reversible heat release during the \gls{AP}, which is
\index{heat absorption}
inconsistent with the \gls{HH} model.
The issue here is that the dissipation was attributed to the "leakage current."
\index{leakage current}
In the framework of the classic
theory, Ohmic currents ﬂow through resistors that dissipate heat due to
friction, no matter in which direction ($W=I^2\times R$) the ion currents ﬂow.
As Hodgkin wrote: "Hill and his colleagues found~\cite{HeatProductionNeuron:1958} that an initial phase of heat liberation [of the \gls{AP}] was followed by one of
heat absorption. [...] a net cooling on open-circuit was totally unexpected
and has so far received no satisfactory explanation."~\cite{HodgkinConduction:1964}
"All authors came to similar conclusions: during
the \gls{AP}, \textit{no signiﬁcant net heat is produced}. Transient heat releases are mostly reabsorbed in the second phase of the \gls{AP}. \dots The ﬁnding of a reversible heat release during the action potential of
nerves is a striking and very fundamental fact. It is inconsistent with the
\gls{HH} model.
The physics underlying the nervous impulse
must rather be based on reversible processes".~\cite{ThermalBiophysics:2007}
Although the energy to build a potential on the membrane and to maintain it needs biological energy consumption, the process of generating an action potential is, in principle, mostly reversible.
Pressure changes accompany the electrical process and naturally account for the observed temperature changes.
The existence of "cooling", alone, undermines the credibility of the classic theory~\cite{HEIMBURGReversibleHeatProduction:2021}.

Our discussion on the energy consumption of the nervous impulse underpins that the complex physical process comprises mostly reversible disciplinary processes:
most of the energy of \gls{AP} is stored as reversible elastic potential energy. Only about 1\% of the energy of \gls{AP} is in the form of electrical energy (Only the $10^{-3}$
fragment of the electrolyte moves against the damped Stokes-Einstein friction force), so only an unmeasurable fragment of the energy dissipates: "no signiﬁcant net heat is produced"~\cite{ThermalBiophysics:2007}. 
(Alternatively, it may be hypothesized that the positive $K^+$ ions, while traveling ahead of the positive $Na^+$ ions under the effect of a potential-free space in the second phase of \gls{AP}, not only turn
with their positive charge the offset voltage to negative, but on their way, the $K^+$ ions collect the heat dissipated by the  $Na^+$ ions. "Laws of science cannot describe life".)

\subsection{Background logistics\label{sec:Physics-BackgroundLogistics}}
The neurons use a quasi-hidden background energy production and a state-restoration process that operates in parallel with neuronal function at a quasi-permanent speed.
The \gls{ATP} is produced and delivered to the spot by 
independent subsystems. The \gls{ATP}'s delivery inside neuron, similarly to that of the ions,
obeys its component gradient. When the neuron's membrane (through the \gls{AIS}, resting ion channels, and pumps) 'loses' ions,
their local gradients in the vicinity of the membrane decrease.
The interesting processes (ionization, ion absorption, and so on)
occur near the membrane.
The processes enable producing 'fresh' ions from the available neutral molecules by hydrolysis to restore their concentration; then
new \gls{ATP} diffuses in the place of the 'used' ones. 
Those ions build up on the membrane, increasing its electrical potential. That energy is a kind of potential energy and enables forwarding ions to the other segment (their presence slightly increases
the potential there), while part of the energy is dissipated as the potential moves ions in the viscous fluid.  

As discussed in section~\ref{sec:Physics-TwoSegments}, outside
the lipid bilayer, electric and concentration gradients exist.
They enable the ions to move, but since their speeds are proportional
to the resultant gradients, at different locations, the speeds of 
the macroscopic movement of ions, as well as the individual speeds
of ions along their paths, can change by several orders of magnitude,
from the 
{drift speed} to the 
{potential-accelerated speed}, while they travel
a nanometer-long path under nanosecond periods.
For example, a $Na^+$ ion is in rest (has the 
{drift speed}) on the high-concentration side
when the ion channel opens; accelerates to a
{potential-accelerated speed}; then slows down to the 
{drift speed} again on the low-concentration side. 
Even its velocity components in directions parallel and perpendicular to the membrane's surface can differ by orders of magnitude when fields are exerted on them in the mentioned directions.
Describing the ion-gradient-related movement needs special care,
since the driving force is different for the different chemical elements, and the movement of ions in close vicinity to each other
(such as in ion channels) strongly affect each other.
Measuring those speeds and fields in nanometer distance 
is a real challenge.

\section{Membrane's electricity\label{sec:Physics-segmented}}

The importance of the subject was correctly estimated decades ago~\cite{UpdateHodgkinHuxley:1990}:
"For the past forty years, our understanding and our methods
of studying the biophysics of excitable membranes have been significantly
influenced by the landmark work of \gls{HH}. Their work has
had a far-reaching impact on many different life science sub-disciplines where
concepts of cell biology have come to be important.
The ionic currents and electrical signals generated by
neuronal membranes are of obvious importance in the nervous system."
To fix the flawed and incomplete models and hypotheses has a similar importance.

Concerning the origin of the membrane's resting potential, we disagree that "the resting membrane potential is the result of the [low intensity]
passive flux of individual ion species through several
classes of resting channels"~\cite{PrinciplesNeuralScience:2013}.
The claim was already question-marked whether the membrane permeability is a primary property for the 
membrane potential generation, for a review, see~\cite{MembranePotentialPermeability:2024}.
This fallacy is rooted in the hypothesis and the measurements by Hodgkin and Huxley~\cite{CompanionGuideHodgkinHuxley:2022}, as discussed in section~\ref{sec:Single-RestingCurrent}.
Their numeric figure was correct, but they mistakenly interpreted 
(due to erroneously projecting the mechanism in the resting state to the transient state) the voltage they measured as being produced by a "leakage current" on the membrane's parallel resistance. Furthermore, their interpretation
leads to causality issues~\cite{HH_Potential_Controversies_2017}.

It was already stated that the membrane's resting potential is, of course, of electrostatic origin~\cite{MembranePotentialAbsorption:2014,OriginMembranePotential:2018,Hodgkin-HuxleyAdsorption:2021,MembranePotentialPermeability:2024,MembranePotentialKeyFunction:2023}.
Paper~\cite{OriginMembranePotential:2017} stated correctly "that separation of charges produces a potential" and derived the thermodynamic part in two different ways, but did not account for the electrical potential. 
However, as discussed below, it is only a small part of the potential; the rest is contributed by the polarization of the electrolyte's dipoles.

\textit{Biology represents a complex case where the phenomena cannot be reasonably reduced to a single discipline}, as classic inanimate science does when deriving its laws for "material points", as A.~Einstein coined, abstracted to either mass or charge, but not both. Moreover, \textit{the closed volume of the neurons drastically differs from the infinite space.
The charges attract and repel each other, but since they experience the counterforce 
of the limiting surface (the membrane), a static mechanical pressure and a shock wave are also created, which 
naturally connect the electrical and mechanical effects, as well as the electrical, mechanical, and thermodynamic theories of operation.}
We must discuss them in a non-disciplinary way because the participating ions belong simultaneously to the disciplines of electricity and thermodynamics; furthermore, we must cross the borderline between particle-based and continuous points of view 
of nature~\cite{VeghNonOrdinaryLawsForLife:2025}.
Moreover, the features of living matter may change while being measured 
in physics laboratories as usual:
"the construction is different from anything we have yet tested in the physical laboratory"~\cite{Schrodinger:1992}.

When the balanced state is perturbed, the system attempts to find a 
new balanced state, using temporal processes.
In this scenario, the 'carrier' - the ion - can be influenced by two interactions.
We discussed elsewhere that those phenomena in living matter 
need deriving and applying 'non-ordinary' laws of science~\cite{VeghNonOrdinaryLawsForLife:2025}, and the time course of those processes can be described by the laws of motion of biology. Fundamentally, due to the mixing of interaction speeds, \textit{considering only one of the interactions (the disciplinary way) leads to incorrect conclusions and theoretical interpretations}.

\subsection{Condenser\label{sec:Physics-Condenser}}

Cell biochemistry sufficiently well describes the lipid bilayers that constitute the membrane~\cite{OriginOfLifeMicelles:2021,ReproductionLife:2023}.
The negative and positive ends of the lipids that constitute the membrane, in an electrolyte solution, attract and bind ions, forming a charged layer on the membrane surfaces. In this way, the membrane in electrolytes acts as a condenser, comprising two charged sheets on its surfaces, with ill-defined parameters. Their boundaries, thicknesses, and compositions depend on the operating state (unlike conducting plates with well-defined sizes and boundaries). This condenser is not ideal: the charged sheet comprises ions with attached aqueous constituents from the electrolyte, so its thickness is far from the ideal zero thickness of classical electricity. Furthermore, in its wall, it inseparably comprises ion channels (from an electrical point of view: resistors), and the condenser plates are in polarizable electrolyte solutions. 

Cell science does not establish the membrane's electrical phenomena; instead, it attempts to elucidate complex protein activities that explain charge transfer, resting and transient potentials, and explains the entire \textit{biological operation using mathematical formulas without a realistic physical background}, by using believed protein mechanisms. Electrophysiology combines currents of electrons and ions, and \textit{uses Ohm’s Law for its admittedly non-Ohmic systems}. It introduces foreign (clamping) currents into biological systems, attributing their effects to changes in membrane conductance without explaining how \textit{voltage and current can be independent of charge carriers}.

\subsection{Electrolytes\label{sec:Physics-ElectrolytesConductivity}}	
Specific chemical substances can naturally hold either a positive or negative electrical charge and react to their micro-, macro-, and electrical/mechanical/thermodynamic environments.
A molecule has internal electrical forces that keep its ions in place, so it has two charge centers (dipoles).
When another dipole or a macroscopic external electrical field (which can be of electrical, magnetic, mechanical, or chemical origin) appears near the molecule, its perturbing effect can affect the relation of the ions to each other.
The two charge centers initially increase their distance (the molecule polarizes). When that disturbance is strong enough, the ions can entirely separate (the molecule ionizes). The local electrical field fluctuates, so their state is dynamic: molecules dissociate, and free ions recombine. Furthermore, the process also plays an important role in the energy supply of neurons: \gls{ATP} hydrolyzes molecules, and the electrical field near the membrane moves ions to the "condenser plates", thereby accumulating "potential energy". (The claim that the ions pass through ion channels by using the "downhill method"
does not use energy, is wrong: they are accelerated by the condenser's field and consume energy in that way.)  Depending on their environment, substances can exist in base, ionized, and polarized states (and exhibit corresponding behaviors).

(In physics, the term 'polarization' is reserved for the redistribution of charges in multiply charged systems without generating an external current. Changes in the membrane's potential difference are a consequence of complete charge separation (and the movement of slow charges), even though, unfortunately, physiology refers to it as polarization. As discussed below, true polarization is also present, which confuses the description. External currents contribute to the changes of potential.)

In biological cells, in a resting state, a small (approximately $10^{-3}$) fraction of the molecules dissociates (i.e., ions separate from their counterparts with opposite charges) and moves freely within the volume.
In other words, the electrolyte liquid can conduct electricity (under the effect of an electrical or thermodynamic field) because the positive and negative ions are mobile. 
The remaining molecules can be more or less polarized, allowing the possibility of generating an internal electrical field in the solution. 
In a transition state, all components (including ions, molecules, and \gls{ATP}) have gradients, and they move towards their balanced state.

\subsection[Thermodynamic field]{Deriving the "thermodynamic electrical field"\label{sec:Physics-ElectrolyteInhomogeneity}}

In a segmented electrolyte, electrical charges are typically globally balanced; however, they may become locally unbalanced due to physical factors.
The Nernst-Planck electrodiffusion equation in a balanced state
\begin{equation}
	\frac{d}{dz}V_{m}(z)=-\frac{RT}{q*F}\frac{1}{C_{k}(z)}\frac{d}{dz}C_{k}(z)\label{eq:NernstPlanck}
\end{equation}
describes the mutual dependence of the \emph{spatial gradients} of the
electrical and thermodynamic fields on each other. In good textbooks (see, for example,~\cite{KochBiophysics:1999},
Eq (11.28)), its derivation is exhaustively detailed. In the
equation, $z$ is the spatial variable across the direction of the
changed invasion parameter, $R$ is the gas constant, $F$ is the
Faraday's constant, $T$ is the temperature, $q$ the valence
of the ion, ${V_m(z)}$ the potential, and ${C_k(z)}$ the concentration
of the chemical ion. The two sides are the derivatives of the equation
\begin{equation}
	V_{m} =\frac{RT}{q*F}\ln{\biggl(\frac{C_{k}^{ext}}{C_{k}^{int}}\biggr)}\label{eq:Nernst1}
\end{equation}

\noindent known as \textit{Nernst equation}.  In other words, Eq.(\ref{eq:NernstPlanck}) and Eq.(\ref{eq:Nernst1}) are the differential and integral formulations of the same knowledge. The limits of the integration are chosen arbitrarily (by choosing the reference concentration $C_{k}^{int}$).
Hence, the derived potential inherently comprises an additive term (a potential difference), so they must not be compared directly if they use a different reference $C_{k}^{int}$
(furthermore, it is nonsense to combine  quantities from the intracellular and extracellular sides additively,
whether concentrations, mobilities, or Nernst voltages, as it happens in the \gls{GHK} equation; see section~\ref{sec:Goldman-Hodgkin-Katz}).
The derivatives are spatial derivatives; the temporal derivatives (needed for describing the time course) of the concentration(s) and voltage are derived in~\cite{VeghNonOrdinaryLawsForLife:2025}. 
In 'ordinary' physics, where the charge and mass are independent,
changing one side changes the other in the opposite direction.
In 'non-ordinary' physics~\cite{VeghNon-ordinaryLaws:2025}, the two differentials must change in the same direction,
given that ions' charge and mass cannot be separated 
(BTW: this is why Eq.~(\ref{eq:Nernst1}) results in a negative potential difference).
(The derivation in 'ordinary physics', such as ~\cite{KochBiophysics:1999},
Eq.(11.30) is wrong, at least because of two reasons: the same speed is assumed for mass and charge transport; furthermore, the used partial derivatives are not independent; the details are discussed throughout the paper. It is also a question, whether the low amounts of material transport are differentiable at all: recall that a similar problem led to the birth of quantum mechanics.)

\begin{figure}
	\includegraphics[width=.8\columnwidth]{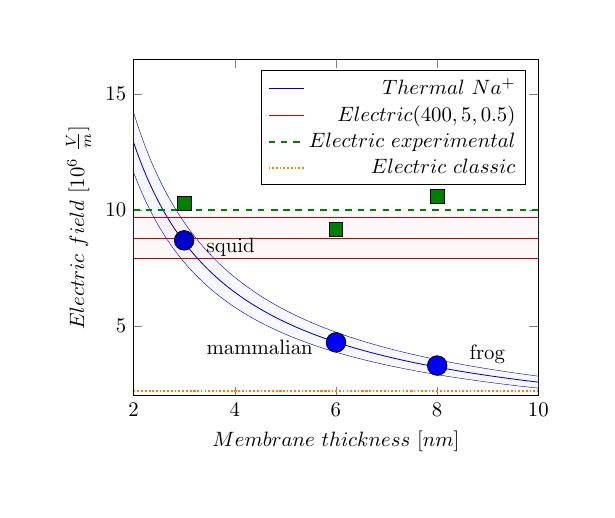}
	\caption{
		The thermodynamic "electrical field" depends on the membrane's thickness; see Eq.(\ref{eq:NernstPlanckThermal2}). The electrical gradient depends on the charged layer's concentration and thickness; see Eq.(\ref{eq:ElectricGradient}). The dots mark measured data (from Table 2.1 in~\cite{JohnstonWuNeurophysiology:1995}) shown in Table \ref{Tab:SummaryTable}.
		\label{fig:NernstPlanckThermalWidth}
	}
\end{figure}

It is a common fallacy in physiology that the applied potential $U$ generates
a current. Instead, its gradient ($E=U/d$) has the effect, where $d$ is the distance, for example, between the clamping point and the membrane or the \gls{AIS}.
We assume that, in a balanced state, the concentrations on the two sides of the membrane are $C_{k}^{ext}$ and $C_{k}^{in}$, respectively; furthermore we assume that \[\frac{dC_{k}}{dz}\approx\frac{C_{k}^{ext}-C_{k}^{in}}{d}\]
To simplify the math writing, we neglect the change caused by considering 
the difference (maybe we should apply a near-unity factor); we use the larger concentration magnitude instead. Using the constants and $T=300$, furthermore, that $U=E*d$, we can derive the "equivalent thermodynamic electrical field" for the case of a permeable ion channel in the membrane

\begin{equation}
	E_{thermal}^{C_{k}}=-\frac{RT}{q*F}\frac{1}{C_{k}^{ext}}\frac{C_{k}^{ext}-C_{k}^{in}}{d} \approx -\frac{RT}{q*F}\frac{1}{d}
	\label{eq:NernstPlanckThermal1}
\end{equation}
\noindent
It does not depend on concentration, but rather is linear in temperature and inversely proportional to membrane thickness (given in $nm$). Furthermore, it is valid separately for all $C_k$
\begin{equation}
	E_{thermal}^{C_k}(d)=-2.585*10^{-2}\frac{1}{d}\quad \biggl[ \frac{V}{m} \biggr] \label{eq:NernstPlanckThermal2}
\end{equation}
Figure~\ref{fig:NernstPlanckThermalWidth} depicts the field's dependence on the membrane's width and how it compares
to the charge-generated electrical field.
The figure illustrates the setpoint, the boundaries of the changes, and the interesting phenomena that occur within the shaded cross-section. The figure is an illustration, but its figures are not far from the true ones. For a $5\ nm$ membrane thickness, it results in $5.17*10^{6}\ \frac{V}{m}$, which compares well to the value provided by Eq.~(\ref{eq:ElectricGradient}).
(Notice that the membrane's thickness may change 
during the action potential, see~\cite{HeimburgPhysikOfNerves:2009}: 
the rush-in extra charge compresses the membrane. Similarly, the concentration of the electrolyte in proximity to the membrane 
also changes.) Correspondingly,
\begin{equation}
	U_{thermal}^{C_k}(5)=-25.85\quad \biggl[ {mV} \biggr] \label{eq:NernstPlanckThermal2A}
\end{equation}

In simple words, the Nernst-Planck equation states that the changes in
concentrations of ions create changes in the electrical field (and
vice versa), and in a stationary state, they remain unchanged.
Notice that in the case of an ion mixture, \textit{joint charge-generated electrical fields exist, and they are equal, per ion, to the concentration-generated "electrical fields"}, see section~\ref{sec:PHYSICS_RESTINGSTATE}. 

In the steady state, some other forces also contribute to
the mentioned forces. To discuss how nature restores the steady state when a microscopic change happens in a
balanced state of a biological solution, we write the well-known Nernst-Planck
equation (see Eq.(\ref{eq:NernstPlanck})) in a slightly extended form:
\begin{equation}
	0 = \underbrace{ e_{el}*E_{Gap}^{Total}(c,\Delta z)}_{Electrical~force} +\underbrace{e_{el}*E_{thermal}^{\textcolor{red}{C_k}}(d)}_{Thermodynamic~force} \quad\underbrace{\bigl(+F_{ext}(z)\bigr)}_{Constraint}\quad\underbrace{\bigl(+F_{Transp}(z)\bigr)}_{Transport~force}\label{eq:NernstPlanckExtended}
\end{equation}
We multiplied the usual two terms by the elementary charge, so its terms are expressed as forces, plus we added a transport-related external force (its role is discussed below).
Furthermore, changing one force triggers a corresponding counterforce and/or causes the ion to move in a viscous fluid.
(An excellent summary of the possible biology-related forces is given in \cite{QuantitativeBiophysicalForces:2014}, with a lot of cited references.) 
This equation, expressing Newton's first law, has no place for effects such as "Pumps that maintain
ion gradients \dots transport ions \textit{against their
electrical and chemical gradients}"~\cite{PrinciplesNeuralScience:2013}, page 101. 
See also section~\ref{sec:Physics-IonPumps}.
Biophysics still did not provide
either the origin of the 'new force' or the reason why Newton's laws are 
not valid for ions in living matter, which undermines the credibility of
those protein-based mechanisms.

Expressed differently
\begin{equation}
	-F_{constraint} = Eq.(\ref{eq:ElectricGradient}) + Eq.(\ref{eq:NernstPlanckThermal2}) + Eq.(\ref{eq:StokesEinsteinSpeeddV}) + External
	\label{eq:IonicForces}
\end{equation}
\noindent
That means when describing an ionic transfer process, \textit{we must not separate the electrical current from the mass transfer} (in other words, to discuss electrical or thermodynamic models in a disciplinary way): they happen simultaneously and mutually trigger each other. Notice that \textit{the thermodynamic term is ion-specific while the electrical term is not}. To be entirely balanced, the system must be balanced to all elements. This way, changing one concentration implicitly changes all other concentrations and the electrical field.

Notice the special role of the constraint force by the closed volume represented by the membrane's wall. 
The thermodynamic contribution also means a pressure from the colliding particles, and the electrical repulsion exerts a force on the membrane's wall. Those disciplinary forces act simultaneously and inseparably. 

We call attention to the statistical nature of the derived 
thermodynamic force. It is valid as long as the 
conditions of the Nernst-Planck equation are valid; furthermore, it can be used with the same limitations.
That is, one cannot use it for describing the movement of
individual ions, but it can describe the charge/mass transfer
represented by a statistical population of ions. 

An interesting option is when the counterforce combines two disciplines, see section~\ref{sec:Physics-MagicConnection}.
When $Na^+$ ions rush-in at the beginning into the membrane,
they produce a huge electrical and thermodynamic gradient simultaneously.
The repulsion between the ions creates a huge mechanical pressure
that presses the elastic membrane. The counterforce starts a mechanical 
shock wave (soliton)~\cite{SolitonPropagation:2005}, that is simultaneously
an electrical potential wave that is measured as \gls{AP}. Only a cross-disciplinary discussion (non-ordinary laws) enables understanding of neuronal operation.  
That means that when describing an ionic transfer process, \textit{we must not separate the electrical current from the mass transfer}: they happen simultaneously and mutually trigger each other. Notice that the thermodynamic term is ion-specific while the electrical term is not. To be entirely balanced, the system must be balanced to all elements. This way, changing one concentration implicitly changes all other concentrations and the electrical field.

It was a colossal mistake to introduce equivalent circuits with their fixed-voltage generators. It forces one to assume that the 
conductances of the players (membrane, synapses, \gls{AIS}) 
change 
without any reason, and prevents understanding how the competition between thermodynamic and electrical processes governs neuronal operation.
It leads, among others, to attributing conductance change to membranes, which are simple isolators with no charge carriers that can implement
charge transfer, see section~\ref{sec:IonicCurrent}; this way
attributing the change of an electrical entity to the biological material.
This assumption neglects the driving force that the mentioned forces provide. 
Instead, it attributes the magic ability to the ion channels that they
can change their transmission ability as the actual situation requires.

In balanced states, no transport occurs, so the transport force cancels, and the mobility has no role. The counterforce adapts to the situation.
That force may be a mechanical one:
the ions sitting on the surface of the membrane press the
surface due to the attractive force of ions, and the membrane mechanically provides the needed 
counterforce. 
If the boundary of the segments is not freely penetrable, the counterforce equals the difference between 
those two forces. This way, no force acts on the ions; the two gradients persist. When ions can move freely between segments, they will continue to move until they establish a concentration gradient (a thermodynamic force) for that ion.

\subsection{One segment\label{sec:Physics-OneSegment}}

For the discussion below, we assume that the segment has a two-dimensional surface boundary, and we discuss the gradients along a line perpendicular
to that plane's surface.
We neglect the electrokinetic and surface-related electrical effects connected to membranes and nanopores~\cite{OriginMembranePotential:2018} and focus
on the balanced state of membranes in ionic solutions.
We compose the segments from such layers (sheets), $(x,y)$ 
parallel plates, and describe the gradients in the direction of $z$.
We introduce the concept of a 'thin physical layer' parallel to the membrane, characterized by a finite width. 

In the calculations below, we require the 'surface charge density' (interpreted as an 'infinitely thin layer' in physics, expressed in $C * m^{-2}$).
We can derive (assuming singly-charged ions) the \textit{volume charge density}
\begin{equation}
	\sigma_{V}(c) = c*N_A * e_{el} = c*6.023*10^{20} * 1.602 * 10^{-19}  = 96.5 * c\quad \biggl[\frac{C}{  mM  * m^3}\biggr]\label{eq:Physics-VolumetricChargeDensity}
\end{equation}
\noindent where $N_A$ is the number of ions in a $millimol$
and $e_{el}$ is the elementary charge. Notice that all ions contribute 
to the charge density, so simply $c=\sum_k C_k$.

Here, we run into a conflict between the 'infinitely thin'
layer of physics and the biologically implemented 
(finitely) 'thin physical layer', so we use a thickness parameter $\Delta z$. (It is an empirical parameter rather than a theoretical fittin funtion. As soon as experiments can derive its value, it can be replaced with a measured one.)
By assuming an arbitrary 'physical layer thickness' $\Delta z$, we can calculate the  \textit{surface charge density} 
\begin{equation}
	\sigma_{A} (c,\Delta z)= 96.5*c*\Delta z \quad \biggl[\frac{C }{m^2}\frac{1}{mM * nm}\biggr] 
	\label{eq:Physics-SurfaceChargeDensity}
\end{equation}

We interpreted classical physics' notions from the theory of 'continuous' electricity,
interpreted for 'infinite' cases, for the finite world of biological objects  
and the quantized atomic world. Our idea is similar to Boltzmann's for connecting the continuous medium to particles.
When ions are present in an electrolyte having concentration $c$ 
in a 'conducting layer' of thickness $\Delta z$ on an isolating surface, they produce a field
\begin{equation}
	E_z(c,\Delta z) 
	= \frac{\sigma_A(c,\Delta z)}{\epsilon_o} =	
	10.89*10^3 * c  * \Delta z \quad 
	\biggl[ \frac{V}{m} \frac{1}{nm*mM} \biggr]
	\label{eq:Physics-ElectricFieldFromSigma}
\end{equation}
\noindent
\begin{figure}
	\includegraphics[width=.8\columnwidth]{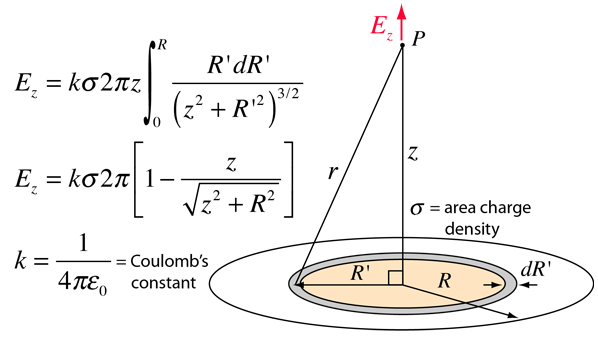}
	
	\caption{\href{http://hyperphysics.phy-astr.gsu.edu/hbase/}{The electrical field of a disc of charge} can be found by superposing the point charge fields of infinitesimal charge elements. This field can be facilitated by summing the fields of charged rings.\label{fig:Physics-PointCharge} “From Hyperphysics by Rod Nave, Georgia State University". \copyright hyperphysics.phy-astr.gsu.edu/hbase/
	}
\end{figure}

\subsection{Two segments\label{sec:Physics-TwoSegments}}

Now, we split the volume into two segments and compose them from 'thin physical layers' (sheets) with different potentials (and concentrations).
We consider the immediate environment of the neuron's membrane
as adjacent parallel $(x,y)$ plates and find the electrical field's $z$ component of those plates in the points
as shown in Fig.~\ref{fig:Physics-PointCharge}.
In classic electricity, the charge remains on the surface of the conducting layer, and the distribution is homogeneous (although the charge is concentrated on the surface and the electrical field is zero inside the conducting layer). Hence, the electrical field contributions cancel each other.
In electrolyte segments, (a tiny fraction of) the molecules decompose into an ionized state (dissociate), and the created ions interact with a bounding
membrane using a not entirely understood mechanism~\cite{MembraneBindingDipole:2025}. The rest of the dipole molecules become polarized but do not dissociate, forming virtual charges that allow the electrolytes to contain electrical fields. Although not explicitly stated, we assume that electrical double layers are present near the membrane, and ions from the dispersion medium are adsorbed onto the particle's surface, depending on the medium's chemical properties. 

Correspondingly, the theory non-equivocally explains the potential near the membrane surface; see the discussion and review in~\cite{SurfacePotential:1985}, mainly their Figure~1.
That model considers a symmetrical condenser, that is, both plates
contain charges of the same sign.
The Donnan function is a (discontinuous) step-like potential.
Its continuous version introduces a mathematically smooth,  non-physical function at the boundary of the charged layer.
One must describe a physically discontinuous layer (electrolyte, ion layer on the membrane's surface, and isolating membrane) with a single continuous function and gradients (potential and concentration) with physical meaning.

The behavior near the boundary layers is problematic, primarily because
the layers themselves are ill-defined.
Physically, they are not ideally plain and well-defined surfaces; the layer thicknesses are limited by the size of the ions/molecules and their associated ions; the layers do not necessarily
satisfy the ideal requirements of the respective physical laws.
Mathematically, a single 'universal' function that spans 
all regions surely cannot accurately describe the behavior
of the physical model. So, we chose a piecewise 
description where the pieces seamlessly fit at the boundaries. We assume that between the surfaces of the membrane, classical
electricity describes the electrical field; the charged layer and the electrolyte jointly produce the electrical field; we have a uniformly charged layer from separated charges on the surface and a polarized electrolyte outside them. 

Unlike in the classical theory, the electrolyte segment next to the surface layer contains dipole molecules
that have more or less balanced charges (the polarization also depends on the external electrical field), so they have much less mobility than the dissociated ions: their size and mass are bigger, and their charge is a fraction of that of the ions.  Their virtual charge becomes apparent only in the appropriate environment, yet it generates an electrical field similar to that of a real charge. The final reasons for creating virtual charges are real charges and/or external fields, so an electrical field due to virtual charges can also exist
inside dielectric matter.
We assume the membrane is transparent
for the electrical interaction (the electrical field affects the ions
in the other segment on the other side of the membrane) but not for
their masses (mechanically separate the segments). 
The membrane does not need to generate any counterforce (except for collisions).

\subsubsection[Charge layers]{Charge layers in segments\label{sec:Physics-LayersInSegment}}

As we show, the finite thickness will result in
a lack of balance (introducing inhomogeneity and creating a voltage and concentration gradient) proximal to the surfaces of the membrane, even if the concentrations on the two sides are the same. Changing the
bulk concentration or potential in one of the segments creates a corresponding
gradient across the separating membrane, increasing the inhomogeneity proximal to the membrane.
The ions will experience an extra force due to the gradient; however, the mechanical counterforce of the membrane will keep them back in the segment. The \emph{concentration and
	potential, inseparably and having the same time course}, will change
across the two sides of the membrane, just because of the gap's physical
features the membrane represents. 

We consider that the segment is composed of electrically conducting discs (the ions are free to move on the surface), and the charged discs' contribution to the 
electrical field at point $P$ (see Fig.~ \ref{fig:Physics-PointCharge}) can be calculated as known from the 
theory of electricity. Due to the symmetry in the $z$ direction, in a homogeneous solution, the resulting electrical field in the plane perpendicular to the $z$ direction is zero, as we have contributions of equal magnitude with opposite signs.

Notice that here we used the abstraction of an infinitely thin "charged sheet", and there is a step-like
gradient in the electrical field in the gap. The physical reality is that the charge is carried by ions, which simultaneously carry mass. In equilibrium, the forces due to the voltage and concentration gradients must be counterbalanced by an external force.
Interference between scientific disciplines can also manifest here. We know simultaneously that at the boundaries of electrolytes, different interfaces, including electrically neutral electrical double layers, can be formed by only partly known 
processes~\cite{ElectricDoubleLayer:2023}. The presence of those structures makes it hard to draw quantitative conclusions.

\subsubsection[Physical condenser]{Simple conducting sheet (a physical condenser)\label{sec:Physics-SimpleSheet}}

First, we consider a condenser that obeys 
the laws of classical electricity and has 
the geometrical size of a biological neuron.
We apply the physics terms to that pseudo-biological object
and test whether the values we derive are consistent with physiological phenomena. A tiny (according to~\cite{JohnstonWuNeurophysiology:1995}, page~12, about~$2*10^{-3}$) portion of the positive ions leaves their negative
counterions behind and forms a thin positively charged ion layer on the surface of the membrane (of course, the same happens on the opposite side with negative ions). Experience shows that -- also in biology -- two skinny layers of ions are formed on the two sides of the membrane of finite width, 
{see the caption of figure (Fig.~11.22 in~\cite{MolecularBiology:2002})}:
"The ions that give rise to the membrane potential lie
in a thin ($< 1\ nm$) surface layer close to the membrane"
(As shown below, these ions give rise only to part of the
potential, and the dipoles in the electrolyte contribute the rest).
A sub-nanometer-thick conducting layer covers the two surfaces at the top of a $ 5~nm$-thick insulator. 
So, we model the cell with two finite-width "conductive sheet" layers.
These layers represent an electrical condenser (so we can calculate the internal electrical field between the plates) and counterbalance each other's electrical field outside the condenser.
We may assume that the solid surface represents the needed
counterforce to keep the charges at rest.

We cannot expect geometrically flat surfaces, as some biological objects of up to $1~nm$ in size are present on the surface.
So, we must assume
a physical 'uniformly charged sheet' with an average thickness $\Delta z=0.5~nm$ on the surface.
For the sake of simplicity, we assume a step-like function
for the electrical field. It is zero in the segment outside 
the "conducting sheets" (the 'bulk' portions) and inside the sheets.  Between the sheets, it jumps to the value of the electrical field of a charged sheet with surface density $\sigma_A$ (see Eq.(\ref{eq:Physics-SurfaceChargeDensity})).
(In this picture, we see a jump in the value of the electrical field
on the two sides of the physical layer. The physical layer emulates the well-defined boundary
on the side opposite to that proximal to the membrane. Here, we still assume that a mechanical counterforce due to the surface's roughness keeps the charges inside the charged layer.)

We assume that all unbalanced (dissociated) ions are in the layer,
and the rest contains no dissociated ions; furthermore, 
the ion concentration is unchanged in the segment.
In the case of a $400~mM$ solution and a $0.5~nm$ 'thin physical layer'
Eq.(\ref{eq:Physics-ElectricFieldFromSigma}) evaluates to an electrical field 
\[E^{Gap}_{Classic}(400,0.5)=0.22*10^7 \biggl[\frac{V}{m}\biggr]\]
and at a $5\ nm$ membrane thickness, the voltage on a classical (non-dielectric) condenser across the plates becomes
\begin{equation}
	U^{Gap}_{Classic}(400,0.5,5)= 
	E^{Gap}_{Classic}(400,0.5)*10^{-3}* d  = 10.84\ \bigl[mV\bigr]
\end{equation}
\noindent
Our simple model seems to be
a strong oversimplification, and explains why a classical condenser cannot explain the voltage in neuronal condenser.   The number of uncompensated ions needed for the cell, using the method in~\cite{JohnstonWuNeurophysiology:1995}, is $0.88*10^7$.
The calculated values are about a factor of $5$ lower than the experimentally derived values.
"An
electrical potential difference about $50-100\ mV$ ... exists across
a plasma membrane only about $5\ nm$ thick, so that the resulting
voltage gradient is about $100,000\ V/cm$"~\cite{MolecularBiology:2002}. "The number of uncompensated ions needed for the cell is $4.7*10^7$"~\cite{JohnstonWuNeurophysiology:1995}. 
We are in the correct order of magnitude, but we arbitrarily assumed a layer thickness, a non-dielectric medium, and a physically unreasonable step-like electrical field. 
The deviation from the expected values suggests that dipoles in the electrolyte bulk significantly alter the achievable electrical field and potential across the plates.
\begin{figure}
	\includegraphics[width=.8\columnwidth]{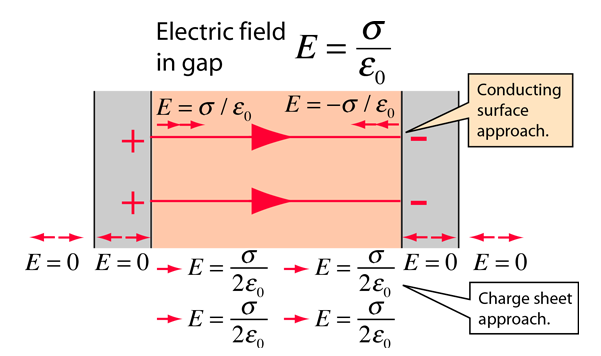}
	
	\caption{
		The oppositely charged parallel surfaces of the membrane are treated like  conducting plates (infinite planes, neglecting fringing),
		then one can use Gauss's law to calculate the electrical field between the plates. Presuming the plates are at equilibrium with zero electrical fields inside the conductors, the result from a charged conducting surface can be used. “from Hyperphysics by Rod Nave, Georgia State University".  \label{fig:Physics-Capacitor} \copyright 
		\href{http://hyperphysics.phy-astr.gsu.edu/hbase/}{http://hyperphysics.phy-astr.gsu.edu/hbase/}
		}
\end{figure}

A usual parallel plate condenser (shown in Fig.~\ref{fig:Physics-Capacitor}) can be derived with such a model.
We have two infinite conducting sheets and 
an insulator layer between them. The magnitude of the opposite charges on the two plates must be the same as in the classical picture.
\textit{Given that the amount of surface charge densities is concluded from the ionic concentration in the segment (in a balanced state), the 
sum of ionic concentrations of the neuron must be the same in the segments.}
A constant electrical field is present inside the membrane (across the plates of the condenser), and there is zero electrical field inside the parallel conducting plates and outside the condenser.
In this ideal picture, the charges are aligned along the boundary between two (infinitely thin) conducting layers and cannot move.
The attractive force between them and the opposite charges on the other plate 
keeps them fixed in the direction of $z$, and the repulsive force
between the charges with the same sign, the surface density in the $(x,y)$ plane remains uniform: the infinitely thin plates are equipotential.
This picture illustrates the equilibrium state of charges, consisting of infinitely small non-dissociating charge carriers and perfectly smooth surfaces.

\subsubsection[Biological condenser]{Condenser in dielectric material\label{sec:Physics-CondenserEffect}}

In its integral form, 
{Gauss's flux theorem} states that the flux of the electrical field out of an arbitrary closed surface is proportional to the electrical charge enclosed by the surface, irrespective of how that charge is distributed.
The surface layer represents a steep potential and concentration gradient.
Above, we assumed that the counterforce that keeps the charges in their place
against their electrical attraction on the membrane's surface is a mechanical force: the ions cannot pass through the membrane.
However, such a counterforce does not exist on the side toward the 'bulk' part of the segment. 
The charges on the plates do not generate an electrical field
toward the bulk, but the concentration they represent does, according to the Nernst-Planck equation, provided there are charge carriers in the bulk. 
We hypothesize that virtual 
electrical charges exist in the electrolytes, and their field provides the missing electrical field.

When explaining the effect of dielectricity, we must explicitly consider the duality of ions that they obey laws of electricity and thermodynamics \textit{simultaneously}; furthermore, the complexity of the electrical structure of the solution and that the charge carriers have finite size (see "electron size" vs. "dipole size").
According to the theory of electricity, the free charge carriers (the dissociated ions) are located on the surfaces where they form powerfully charged thin layers (condenser plates) in the segments separated by a membrane on the two surfaces of the membrane; two proximal charge layers on the surfaces of the membrane.
We assume that those ions behave as point charges, i.e., 
due to the attraction from the opposite charges on the opposite plate,
they do not produce an electrical field toward
the side of the bulk of the electrolyte layers.
In the classical picture, those layers represent step-like gradients in the electrical field along the $z$ axis, and the constraint that ions must not enter the membrane provides the necessary counterforce.

From a thermodynamic point of view, a driving force acts
as long as a concentration gradient between neighboring layers exists. 
From an electrical point of view, the layers are simple parallel-plate condensers
that produce no electrical field outside their closed volume
and are connected serially; plus, the top layer has free ions. The electrical field is proportional to the bulk concentration in the layer.
Changes in the local electrical gradient may also alter the degree of dissociation and polarization, producing a graded local electrical field. 
These two driving forces have opposite directions
and in a balanced state, the same magnitude.
The counterforce acts in a way that constrains the molecules to stay in their layers; it adapts
(also compensates for the different interaction speeds)
while the gradients are changing. It presses the proximal layer against the membrane and the neighboring layers against each other.

Due to the membrane's finite width and surface roughness, the conducting layer is also finite. 
The electrical field is homogeneous within the layers. The rest of that electrical field toward the bulk layer directs the dipoles proximal to the neighboring dipole layer, and their polarization increases; in other words, they "produce" an electrical field.
The final effect is that a dipole layer is attracted to
the charged layer, and it forms another layer that shows 
a charged layer toward the bulk. 
The process repeats and results in 
a decaying electrical field based on the distance from the membrane's surface.

In our model, we build the volume of the electrolyte
from thin layers of directed dipole molecules (the size of molecules limits the thinness $\Delta z$) in the volume, separated into two segments with electrolytes by a membrane with
a finite thickness~$d$ (that is, a point's distance in the electrolyte from the 
two disks will be $z$ and $z+d$, respectively). 
The contributing potential of an infinitesimal volume on the axis at point $z$ due to the charged sheet is
\[
dU(c,\Delta z,z) = 
\underbrace{E^{Gap}_{Classic}(c,\Delta z)*z}_{Potential~ from~sheet}*\frac{1}{z^2}
=  
E^{Gap}_{Classic}(c,\Delta z)\frac{1}{z}
\]
The total potential due to all elements $dz$ from the left side is (in the mathematical formulas below, we express the distance in units of $\frac{z}{d}$, so here $z$ is dimensionless).
\begin{equation}
	U_{left}(c,\Delta z,d)=E^{Gap}_{Classic}(c,\Delta z)*d\int_{-\infty}^0\frac{1}{z} dz =\\ E^{Gap}_{Classic}(c,\Delta z)*d \biggl[\ln | z | \biggr]_{0}^{\infty}\label{eq:Physics-InternalPotentialL}
\end{equation}

In the case of a single-segment volume, within the segment, a similar potential with an opposite sign is generated by the charges on the neighboring sides of the considered charged sheet,
counterbalances the potential described by Eq.(\ref{eq:Physics-InternalPotentialL}). However, when a membrane with thickness $d$ separates the segments (with no charges in the gap), the potential on the right side will be
\begin{equation}
	U_{right}(c,\Delta z,d)
	= E^{Gap}_{Classic}(c,\Delta z) * d *\biggl[\ln | z+1 | \biggr]_{1}^{\infty}\label{eq:Physics-InternalPotentialR}
\end{equation}
\noindent
We can write the electrical field in the form
\begin{equation}
	E(c,\Delta z,d)=
	E^{Gap}_{Classic}(c,\Delta z)*ln\biggl(\frac{z}{z+1}\biggr)
	\label{eq:ElectricFieldOutside}
\end{equation}
That is, the gap sets the potential difference across the membrane to
	\begin{equation}
	U(c,\Delta z,d)=
	E^{Gap}_{Classic}(c,\Delta z) *d *\biggl(\biggl[\ln | z | \biggr]_{-\infty}^{0}-\biggl[\ln | z+1 | \biggr]_{1}^{\infty}\biggr)\label{eq:Physics-InternalPotentialFunction}
\end{equation}
We use the approximation that $\ln(\infty)\approx \ln(1+\infty)$,
and we arrive at that
\begin{equation}
	U(c,\Delta z,d)=E^{Gap}_{Classic}(c,\Delta z)*d*\biggl[ln\biggl(\frac{z}{z+1}\biggr)\biggr]_{0}^{1}\label{eq:Physics-Internal1 }
\end{equation}
\noindent describes the potential across the plates; that is,
the classic potential shall be multiplied by the result of the integral. It was experimentally confirmed~\cite{DonnanPotentialExperiental:2022} that the potential 
of the membrane is proportional to the logarithm of the distance from the membrane.
 
Given that the same number of charged ions must be present
on the two sides of the membrane, their resultant surface density $\sigma_A$ must be the same. 
In the classic measurement by \gls{HH}, the sum of the concentrations of $K^+$ and $Na^+$ in the cytoplasm and extracellular fluid  (in mM) is 380 and 390, respectively, while the  concentration of the $Cl^-$ 
are 450 and 460; practically all equal, as our analysis has derived they should be. The deviations are well within the uncertainty of the 
measured values.  
Notice that the effect is purely electrostatic, resulting in an asymmetric ion distribution; in that balanced state, no permeability is needed. If the membrane is (slightly) permeable, ions will
move across the membrane until equilibrium is reached.
The resulting potential difference is directly proportional to the concentration difference.

\begin{figure}
	\includegraphics[width=1\textwidth]{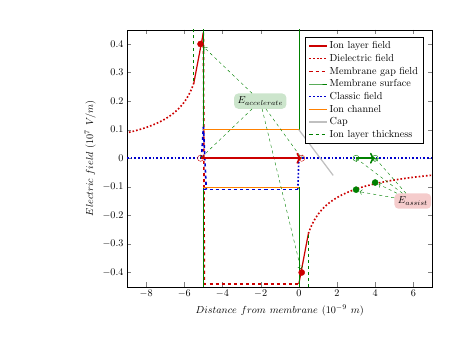}
	\caption{The neuronal membrane's \emph{electrical field} in the function
		of the distance from the membrane's surface.
		The thickness of the 'atomic layers' proximal to the membrane's surfaces
		are also shown.
		The electrical field is linear inside the uniformly charged atomic ion layer on the surface and is described by Eq.(\ref{eq:ElectricFieldOutside}) in the electrolyte.
		For visibility, a layer thickness $0.5\ nm$ is assumed.
		For illustration, a simple capped ion channel in the membrane's wall
		is also displayed. \label{fig:The-membrane's-extra-gradient}}
\end{figure}

The classical condenser has an electrical field, as the blue dashed line shows in Fig.~\ref{fig:The-membrane's-extra-gradient}. No field is outside the condenser or inside the condenser plates (see Fig.~\ref{fig:Physics-Capacitor}). There is a sudden jump in the field on the conductor's surface (the armatures) and a constant field between the plates.
(The charged layer can be ideally thin if electrons are sitting on the surface, but is of finite thickness on the opposite side where ions are on the surface.  This latter effect is not discussed in classical electricity.)

The biological condenser behaves differently; see the
red diagram line. An electrolyte (instead of an insulator) outside the condenser significantly alters the field's structure. The electrical field on the surface and between the plates changes by a factor of 3-4. Outside the plates, the polarization creates a field that changes logarithmically. An ion layer with finite thickness is built near the membrane's surface. It is uniformly charged, so the field inside it is linear. At the internal surface of the condenser, it takes the field value calculated for the gap; on the other side of the layer, it takes the logarithmic field value at that position. The dashed line represents the gap field, the continuous line represents the charged layer's field, and the dotted line represents the electrolyte's field. We observe the "correspondence principle": discrete and continuous fields join seamlessly; furthermore, as the electrolyte's polarizability decreases toward zero, the field approaches the classical value.

\subsection{Bridging the discrete and continuous views\label{sec:Physics-BridgingElectricity}}
We assumed that the function can be interpreted for regions
$(-\infty,0)$ and $(1,+\infty)$. However, we have arrived at the boundary between particles and continuous electricity.
We use a trick similar to the one Boltzmann used in his famous equation, except that we calculate the number of charge carriers from
geometrical instead of statistical assumptions.
The minimal thickness of layers is limited, at least by the size of ions.  Furthermore, the membrane's surface is not flat; there are structures (mainly proteins/lipids) with sizes of up to $1~nm$, so it is likely 
realistic to consider a layer thickness of $0.5~nm$ as we did above
(on our mathematical scale $0.1=\frac{0.5~nm}{5~nm}$), for which a multiplier of $1.7$ between the classic and the electrolytic fields exists.
A similar calculation using a layer thickness of $0.1~nm$ (on our mathematical scale, 0.02) results in a multiplier of $3.2$;
so we use a multiplier of value 3 (that corresponds to a layer thickness of $0.125~nm$ (the radius of a $Na^+$ ion is $0.116~nm$)). (This arbitrarily chosen value can be replaced with a measured value, as soon as it will be available.)
Correspondingly, we assume an electrical field 
\begin{equation}
	E^{Gap}_{Dielectric}(c,\Delta z) = 3*E^{Gap}_{Classic}(c,\Delta z)  \quad 
	\biggl[ \frac{V}{m} \frac{1}{nm*mM} \biggr]
\end{equation}
\noindent
due to the dielectric properties of the segment.
This contribution arises from the dielectricity in the segment (for simplicity, we have neglected the relative permittivity). It is to be added to the value of the 
classic contribution, the 
field generated by the conducting plates, so the 
final multiplier is $4$. 
(The experience
is known in technical electricity: the 
{electrolytic capacitors}
\textit{achieve several times higher charge storage capacity by using 
	{pseudocapacitance}}.
Using "roughened anode foil" increases the electrolyte thickness, and the roughening provides the necessary mechanical support. 
A thicker electrolyte layer wraps the condenser plates, and the dipoles in the thicker electrolyte provide an additional charge storage facility.)

Correspondingly, 

\begin{equation}
E_{Gap}^{Total}(400,0.5)= 4*E^{Gap}_{Classic}(400,0.5) = 8.8*10^6 \bigl[\frac{V}{m}\bigr] \label{eq:ElectricGradient}
\end{equation}
\noindent and at a $5\ nm$ membrane thickness, the voltage across the plates becomes
\begin{equation}
	U_{Gap}^{Total}(400,0.5,5)= E_{Gap}^{Total}(400,0.5) * (5*{10^{-9}}) = 45\ \bigl[mV\bigr]\label{eq:UGapTotal400}
\end{equation}
\noindent
"The plasma
membrane of all cells, including nerve cells, is approximately 6 to 8 nm thick and consists of a mosaic of lipids
and proteins."~\cite{PrinciplesNeuralScience:2013}, page 71.
Using such a value for the membrane's thickness may yield a thickness up to 60\% higher. 
 Furthermore, the usual concentration is 
also up to 10\% higher.
Hodgkin, in 1964, measured molarity values in squid axons for ions $K^+$, $Na^+$, and $Cl^-$, (400,50,40-150) inside and (20,440,560) outside, and they provided potential values 
$55-75~mV$~\cite{JohnstonWuNeurophysiology:1995}.
Given that we used a plausible but ad hoc "charged layer thickness"
and gap distance, we cannot expect a better agreement.
However, our qualitative conclusions remain valid and can be adjusted
quantitatively to the results of dedicated measurements.

We have the boundary conditions that we know the electrical field at the membrane's surface through the volume charge density, and the Nernst-Planck equation delivers the change
of the electrical field due to the change in concentration. 
The physical constraint is that the layer thickness
cannot be infinitely small. If we use a parameter for layer thickness equal to that of the surface layer, 
the calculated electrical field distribution will be
scalable.

\subsection{The dynamic layers\label{sec:Physics-DynamicLayer}}
This layer's importance and operation are in
line with the Earth's atmosphere. Its features drastically deviate
from the features of the bulks on its two sides. It is separated by
a sharp contour on one side and an ill-defined border on the other;
its volume is far from being homogeneous. Gravity keeps it in place, and it is at rest. However, sometimes,
for some periods, also other (thermodynamic and electrical) forces are evoked
inside it and lead to transient changes, moving huge masses with high speeds
inside it. Their thickness and mass are negligible compared to those
of the bulks on their two sides, and we can describe the bulks without
considering their density, mass, size, etc. Still, this thin layer is
responsible for the weather.
It can temporarily absorb
products of slow processes (water evaporation) and deliver masses of
high density (much above its density, such as water, sand, etc.) to
continental distances, creating the illusion that it temporarily stores that matter.
Minor changes (natural ones, such as a slight difference in air temperature, and artificial ones, such as injecting condensation nuclei in clouds)
can result in enormous changes. We can even imagine volcanic eruptions
as semipermeable gates for material with apparently random operation
and distribution of the injected material.

To describe those complex and continuous phenomena at least approximately,
we must separate them into stages. We can describe the stages approximately using omissions, approximations, and
abstractions, usually considering
only one dominant phenomenon. The described phenomena are interrelated
in a very complex way and depend on different parameters. To some
point, we can describe that thin layer using a static picture and
provide an empirical description of its processes, even though 
we can give some limited validity mathematical descriptions for those
stages. However, we understand that for describing the time course of the transition
(contrasting with step-like stage changes) between those well-defined
stages of the atmosphere, we need a \emph{dynamic description} to
discover the \emph{laws of motion} governing the processes.

The same is true of neuronal membranes and neuronal function.
Now, we are at the point where their decades-old static description
is insufficient. We need
to derive the corresponding laws of motion to describe the neuron's dynamic behavior. We need a meticulous and
unusual analysis to derive them. 
In a neuron, in the abstraction science uses, we put together only
an ionic solution, a semipermeable membrane, and currents that enter and leave
it. All these belong to non-living matter.
As experienced, \textit{at some combination of their parameters and gradients, qualitatively
different phenomena happen, which, in the abstraction biology uses,
are called signs of life};
our system starts to belong to living matter.
Given that the approximations, the derived abstractions,
and the mathematical formalisms describing them are different for
the two cases, \emph{it looks like we have two different, only loosely
	bound worlds}. We realize we have arrived at the boundary of non-living
and living matters, and we must go back to the \emph{first principles
	of science} to clarify where their boundary is. However, by using our approach, we may defy that "the
emergence of life cannot be predicted by the laws of physics"~\cite{ConservationOfInformation:2021}.
Our artificial duck looks, quacks, and swims like a duck.

\subsection{Ionic currents in a cell\label{sec:IonicCurrent}}

On a microscopic scale, a charge generates a potential field that acts on other charges. It is a common fallacy in physiology that a "potential field travels": neither potential nor current changes without a physical movement of charges. In a conducting wire filled with ions (the solids have a different current transfer mechanism), there are free charges; their
number per unit volume is given by $n$, and $q$ is the amount of
charge on each carrier. If the conductor has a cross-section of $A$,
in the length $dx$ of the wire, we have charge $dQ=q*n*A*dz$. If the charges move with a macroscopic speed $v=\frac{dz}{dt}$, 
at the macroscopic level, we define the current $I$ as the charge moved per unit of time as
\begin{equation}
	I=\frac{dQ}{dt}=q*n*A*v\label{eq:DriftCurrent}
\end{equation}
The effect was hypothesized at the very beginning: "it seems difficult to escape the conclusion that the changes
in ionic permeability depend on the movement of some component of the
membrane which behaves as though it had a large charge or dipole moment."
"it is necessary to suppose
that there are more carriers and that they react or move more slowly"~\cite{HodgkinHuxley:1952}. The charged ions, attracted by the opposite charges on the other plate and repelled toward the single exit point \gls{AIS}, really move slowly and behave as if they were a component of the membrane.

Notice that if any of the factors is zero, the macroscopic current
$I$ is zero. \textit{Microscopic carriers must be present in the volume},
and have charge, the cross section must not be zero, and the charge
carriers must move with a potential-assisted speed, which needs an
electrical field generated by the system, see Eq.(\ref{eq:IonicForces}).
When there is no constraint force, the thermodynamic and electrical forces will produce an effective force that moves the ions.
That is, the ions can move without an external electrical field, as experienced
in biological systems. 
One of the fundamental mistakes by \gls{HH} was neglecting the Coulomb force in ion-electrical interactions and assuming that diffusion processes or ideal electrical batteries create the electrical potential.

In biological systems, where ions represent the "current", it is either a native current (without an external potential) or an artificially injected current or potential generating a current.
This way, the current is consistently accompanied by a change in concentration gradient, as the moving ions represent both mass and charge transfer simultaneously (also, the repulsion leads to a "skin effect": in electrical wires and axons, the charge travels near
the outer surface).
The potential and current are
connected through the features of the medium (material) that hosts
our measurement. 

According to Stokes' Law, to move a spherical object
with radius $a$ in a fluid having dynamic viscosity $\eta$, we need
a force
\begin{equation}
	F_{d}=6*\pi*\eta*a*v\label{eq:StokesForceOnIon}
\end{equation}

\noindent (drag force) acting on it. A (microscopical) electrical force
field $\frac{dV}{dz}$ inside the wire would accelerate the charge
carriers continuously
\begin{equation}
	F_{electric}=k* E(z) * q
\end{equation}
\noindent with a constant speed $v$. It is not the \emph{drift} speed because of the
electrical repulsion, it is a \emph{potential-assisted} or \emph{potential-accelerated} speed that can
be by orders of magnitude higher.  
The medium in which the charge moves shows a (macroscopic,
speed-dependent) counterforce $F_{d}$, which in steady state equals
$F_{electric}$, that is :
\begin{equation}
	I=\frac{k*q^{2}*A}{6*\pi*\eta*a}*n*E(z)
	\label{eq:StokesCurrent}
\end{equation}

We extend the original idea by using
\[
E(z)  = E_{electric} + E_{thermal} + E_{external}\quad in\ electrolytes
\]
that is, for living matter
\begin{align}
	E(z) & = \frac{\Delta V}{\Delta z} & in\ linear\ potential\\
	& = \frac{dV}{dz} & in\ graded\ potential\\
	& = E_{electric} + E_{(thermal)}  (+ E_{(external)}) & in\ electrolytes
\end{align}

\textit{The amount of current in a wire is not only influenced by the electric
	force field (specific resistance) but also by the number of charge
	carriers $n$.} While the latter is commonly considered constant and
part of the former, this is not necessarily the case for biological
systems with electrically active structures inside. The medium's internal
structure ("the construction) introduces significant modifications. \textit{Applying an electric
	field to a biological wire can generate varying amounts of current as the number
	of charge carriers changes due to biological effects.} For axons, we use a single-degree-of-freedom
system, a viscous damping model, so the \textit{ions will move with
	a field-dependent constant velocity in the electrical field}; so it takes time while they appear on the membrane.
The activity of potential-controlled ion channels in its wall may
change $n$ in various ways; furthermore, that change can result in
'delayed' currents during measurement, for example, in clamping, as physiological measurements witness.

If we have a concentration
$C(z)$, in the volume $A*dz$, we have $dQ=C(z)*A*dx*q$
charge, resulting in another expression for the current

\begin{equation}
	I=\frac{C_k(z)*A*dz*q}{dt}=C_k(z)*A*q*v\label{eq:ConcentrationCurrent}
\end{equation}
By combining equations (\ref{eq:StokesCurrent}) and (\ref{eq:ConcentrationCurrent}): 
\hypertarget{eq:StokesLawSpeed}{ }
\begin{equation}
	v(z)=\frac{k*q}{C_k(z)*\eta*6*\pi*a}*E(z)\label{eq:StokesSpeed}
\end{equation}

The higher the resulting space derivative (gradient) and the fewer ions that
can share the task of providing a current, the higher the speed. We hypothesized (it needs
a detailed simulation) that in the case of this charged fluid, the
electrical repulsion plays the role of 'viscosity'~\cite{VeghNonOrdinaryLawsForLife:2025}. The higher the charge
density, the stronger the force equalizing the potential; so $\eta$
is the lower, the higher the charge density (proportional to $C_{k}$).
For the sake of simplicity, we assume that the speed is proportional
to the space gradient of the local concentration and potential. Recall that \emph{our equations
	refer to local gradients only. The electrical gradient can propagate
	only with the speed of the concentration gradient}, given that only
the chemically moved ions can mediate the electrical field~\cite{VeghNonOrdinaryLawsForLife:2025}. \emph{The
	lower interaction speed limits the other interaction speed if the
	interactions generate each other.}

The dependence of the diffusion coefficient on the viscosity can be modeled by the Stokes-Einstein relation:
\index{Stokes-Einstein relation}
\begin{equation}
	\label{eq:StokesEinstein}
	D = \frac{k*T}{6*\pi*\eta*a}
\end{equation}
so we can express the speed with the diffusion coefficient $D$ and temperature $T$
\begin{equation}
	v(z) = \frac{D}{T}*\frac{q}{C_k(z)}*E(z)\label{eq:StokesEinsteinSpeeddV}
\end{equation}
We introduce \textit{a time-dependent speed gradient
	$v(z,t)$
	which plays a vital role in the processes that occur in living matter}.

\subsection{Imitating slow currents\label{sec:SlowCurrent}}
At the dawn of finding methods for describing neuronal operation,
\gls{HH}
published high-precision measurements~\cite{HodgkinHuxley:1952}
enabling detailed testing of theories explaining the observed physiological
behavior. \emph{Their good physical model that }``movement
of any charged particle in the membrane should contribute to the total
current'' \emph{lacked considering the mutual repulsion and finite speed} of those particles (they are about 50,000 times heavier and a million times slower than the electrons, and they move in a closed space segment). Furthermore, 
they have started from the commonly used wrong assumption that conductance
is a primary electrical entity.
They could "find equations which describe the conductances with
reasonable accuracy and are sufficiently simple for theoretical calculation of the \gls{AP} and refractory period".
The "reasonable accuracy" and the need for a simple calculation method (performed on a mechanical calculator) obscured
the fact that the laws for electrons are not the same as the laws derived for ions.
However, good textbooks  warn about that "These equations describe
the process of excitation primarily in terms of changes
in membrane conductance and current. \textit{They tell little
about the mechanisms that activate or inactivate channels} in response to changes in membrane potential or
selectivity for specific ions."~\cite{PrinciplesNeuralScience:2013}, page~162.
It is at least hard to interpret that "The trace was obtained by blocking
the $K^+$ channels with tetraethylammonium and subtracting the
leakage and capacitive currents electronically". It subtracts 
the leakage current that, as we discuss in section~\ref{sec:Single-RestingCurrent}, does not exist; the capacitive current corresponding to the wrong oscillator model; after "blocking the channels", see section~\ref{sec:BlockingChannels}.

As discussed in connection with Eq.~(\ref{eq:DriftCurrent}),
the current depends linearly on the number of charge carriers $n$.
The physical processes changing the number of charge carriers
also define the intensity of the currents, providing a way to imitate
slow currents by fast currents (by using current generators using special function shapes), where the spatiotemporal time course of the physical process of producing ions in the system modulates the number of charge carriers.

\begin{figure}
	\includegraphics[width=0.75\textwidth]{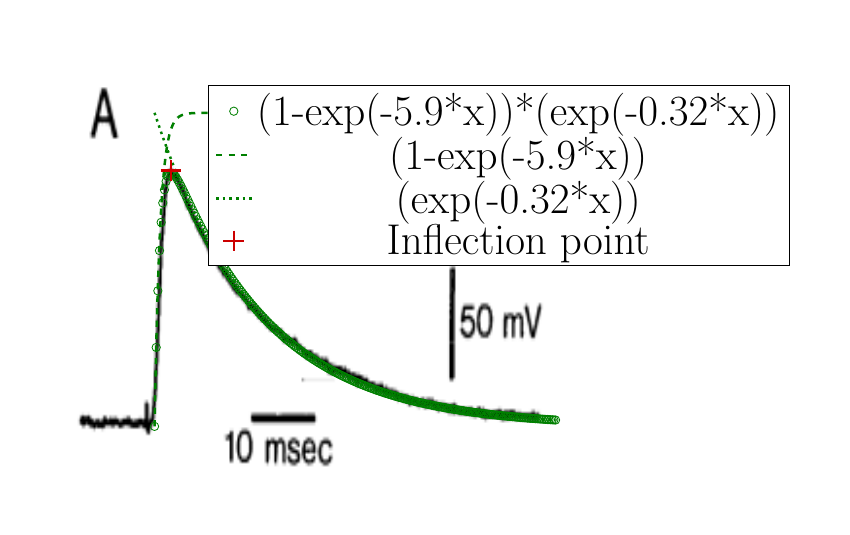}
	\caption{The \gls{PSP} shape measured by~\cite{SynapticTransmissionMason:1991} and fitted by function forms
		given by  Eq.(\ref{eq:BioCurrentForm}).
		\label{fig:MasonPSP} }	
\end{figure}

As an example, we discuss the case of slow current
traveling in an axon.
The ions travel a finite distance with
a finite speed, so there must be a delay between their entry and exit times. The case of switching a clamping voltage
on, to some measure, is analogous to the arrival of a spike. Initially, the axon contains
no ions. 
In the classical picture, the axonal current flows into the membrane with
capacity $C_{m}$ and increases the membrane's voltage $V_{m}$
with a time constant 
\begin{equation}
	\frac{dV_{m}}{dt}=-\frac{1}{C_{m}}I_{axon};\ V_{m}(t)=\frac{I_{wall}*(1-e^{-\alpha*t})}{C_{m}}\label{eq:MembraneCurrent}
\end{equation}
\noindent that generates a change in the membrane current
\begin{equation}
	\frac{dI_{m}}{dt}=\frac{1}{R_{m}}\frac{dV_{m}}{dt};\ I_{m}^{on}(t)=g_{m}(V)V_{m}(t)\label{eq:TimeDependentConductance}
\end{equation}
\noindent where $g_{m}=\frac{1}{R_{m}}$ is the conductance of the
membrane. That is, the measurable current equals the product of the conductance
and the clamping voltage. Equs.\emph{(3)-(5) in~\cite{HodgkinHuxley:1952}}
express this relation.
If we assume that the axonal current is "slow",
we naturally conclude that Ohm's Law is correct and valid also for
biology: \textit{the conductance/resistance is constant, but the number of charge carriers changes}, see section~\ref{sec:IonicCurrent}. However, as we discuss in section~\ref{sec:Physics-GatingLayers}, the voltage $V_M$ is not constant.  \emph{If one assumes that the axonal current
is ``fast'', we arrive at the wrong conclusion that the conductance
is voltage- or time-dependent.} \textit{Due to the wrong idea of using equivalent circuit with constant voltages, the actual change of membrane voltage in the product in~Eq.(\ref{eq:TimeDependentConductance}) is misinterpreted as 
the change of the corresponding conductance.} Again, biology projects the static picture constructed for the resting state
to the dynamic operation in the transient state that some magic effect works against the fundamental principles and laws of science and changes the conduction of the membrane. It is sticking to an ad-hoc assumption against the principles of
electricity it (apparently) uses.

The \textit{slow} input
currents (although due to slightly differing
physical reasons for the different current types) are described by
the following analytic form valid for \textit{fast} currents:
\begin{equation}
	I_{in}=I_{o}*(1-\exp(-\frac{1}{\alpha}*t))*exp(-\frac{1}{\beta}*t)\label{eq:BioCurrentForm}
\end{equation}

\noindent where $I_{o}$ is the current amplitude per input, and $\alpha$
and $\beta$ are timing constants for current in- and outflow, respectively.
They are (due to geometrical reasons) approximately similar for the
synaptic inputs and differ from the constants valid for the rush-in
current. To implement such an analog circuit with a conventional electronic
circuit, voltage-gated current injectors (with time constants around
1~msec) are needed.
For example, see in Fig~\ref{fig:MasonPSP} the \gls{PSP} measured function shape published in~\cite{SynapticTransmissionMason:1991}, fitted with our theoretical function Eq.(\ref{eq:BioCurrentForm}). The corresponding voltage derivative is as follows:

\begin{equation}
	\frac{dV_{in}}{dt}=\frac{I_{o}}{C}*\biggl(\frac{1}{\alpha}*exp(-\frac{1}{\alpha}*t-\frac{1}{\beta}*t)-\frac{1}{\beta}*exp(-\frac{1}{\beta}*t)*exp(1-exp(-\frac{1}{\alpha}*t))\biggr)\label{eq:PSPderivative}
\end{equation}

\subsection{Effects of membrane's electricity\label{sec:Physics-MembranesElectricity}}

The new and reinterpreted concepts fundamentally change
the operation of neuronal components.

\subsubsection{Ion transport in cells\label{sec:Physics-IonTransportInCells}}

In balanced states, 
the transport force cancels, and the mobility has no role (when speaking about mobility provided by protein mechanisms, one should derive the corresponding contribution to Eq.~(\ref{eq:IonicForces})). The counterforce adapts to the situation.
That force may be mechanical:
the ions sitting on the surface of the membrane press the
surface due to the attractive force of ions on the opposite side, and the membrane mechanically provides the needed 
counterforce. 

When ions can move freely between the segments, they continue to move until they produce a concentration gradient (a thermodynamic force) for the given ion that counterbalances the electrical gradient (the electrical force), at which point the transport stops. However, in a system with composite electrolytes, no balanced state exists, since
the thermodynamic force depends 
on the chemical nature of ions, while the electrical force does not. A local resulting transport force is needed to reach the Stokes-Einstein speed (see Eq.~(\ref{eq:StokesEinsteinSpeeddV})) in a viscous fluid.
In a quasi-balanced state, for ions having only slightly different reversal potentials, the driving forces are small.
Correspondingly, the currents (that is, the ion transports) are slow as observed in physiology.
Due to the limited resources, when one ion leaves or arrives at one segment
of the neuron, it changes the local potential on both sides which changes
the driving force exerting on the other ions. This way, 
the different reversal potentials, alone, keep up the illusion that the ion channel "pumps" ions in and out (this effect is misinterpreted as "mobility"), without needing some magic protein mechanisms.
Those ion channels with low-speed material transport 
are
called "ion pumps".
This effect has no role in forming the 'setpoint' of the resting potential (in contrast with as the equation \gls{GHK} believes), only in keeping the balance  at the point defined 
by the neuronal condenser as described in section~\ref{sec:TransientState}.

We must not forget that initially, the Nernst-Planck equation described
a transfer process (i.e., a dynamic equation), assuming the same speed for mass and
charge transfers. We used it only to describe the balanced state (i.e., a static equation at zero speed) and added the term describing external (such as mechanical constraint) force that
may be involved; it affects the process, but it does not belong to either of the respective fields. Commonly known mechanical constraints appear in a non-conducting layer called the membrane, where that counterforce prevents ions from penetrating the membrane layers.
A membrane is a perfect isolator, i.e., no charge carriers exist
between its two surfaces. (There may be ion channels built into the membrane that deliver ions.) The charge-up results from \textit{charge separation}, a mechanism distinct from \textit{polarization}, as typically described in most biology textbooks.

\subsubsection{Gating layers\label{sec:Physics-GatingLayers}}

The operation of the ion channel, alone, cannot explain why
the channel closes after a given number of ions have passed through the channel; that number is not (entirely) random.
The local behavior of the membrane's surface layers regulates the number of ions that pass through. 
The segments are no longer mechanically separated when the cap is
removed. The charged ions are enabled to rush into the lower concentration
segment. They experience an enormous accelerating gradient; that,
comparable to that of an electrostatic particle
accelerator, "snorts" the ions from the high-concentration side
into the low-concentration side and causes a process "like
a flee hopping in a breeze"~\cite{KochBiophysics:1999}. 
Consequently, "transport efficiency of ion channels
is $10^5$ times greater than the fastest rate of transport mediated
by any known carrier protein"~\cite{MolecularBiology:2002}. 
Having gates (caps) is needed only for the synchronized operation of the channels: opening and closing work
also in the absence of caps, as discussed by~\cite{HeimburgPhysikOfNerves:2009}.

The snorted ions "hop" into the layer from another layer.
At the beginning, with
their \emph{voltage-accelerated} speed, it could take less than $\frac{5*10^{-9}m}{10^{3}m/s}\ s$
to pass the channel (simulation~\cite{IonChannelSimulation:2016}
uses a $psec$ representative time interval).
In the end, they may
slow down to the \emph{voltage-assisted} level as the potential gradually
decreases (which is still $\frac{5*10^{-9}m}{10^{-1}m/s}\ s$), so
we can omit that time when calculating the charged layer formation
(the estimated speed is about $10^4~to~10^5\ m/s$).
Due to the enormous speed difference between the \emph{accelerated}
and \emph{assisted} speeds, the passage is practically instant.
On the high-concentration segment, only the ions in the
layer in the immediate vicinity of the entrance can feel the accelerating potential and move with the potential-accelerated speed.
The after-diffusion
with the \emph{potential-assisted} speed from the next neighboring
layer in the high potential segment is by orders of magnitude lower than the passage through the
hole with the \emph{potential-accelerated} speed. 

\begin{figure}
	\includegraphics[width=0.65\textwidth]{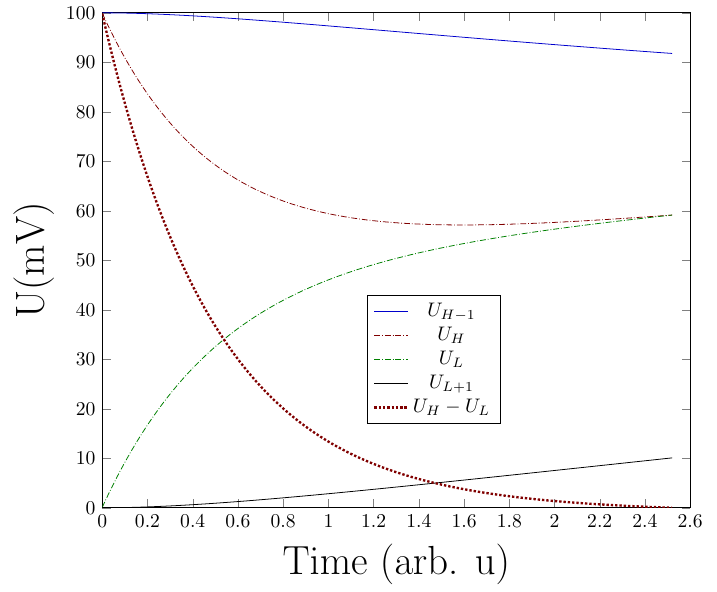}
	\caption{The potential values connected to the operation of ion channels. The arbitrary values of the transfer speeds (for visibility, instead of the real values) are $\alpha = 1$; 
		$\beta  =.1$; $\gamma = 0.0001$; furthermore the time scale is also arbitrary.
		\label{fig:IonChannelRK} }	
\end{figure}

The accelerating potential around the channel's mouth gradually (but quickly) disappears (see Fig.~\ref{fig:IonChannelRK}) when
the particle exits the ion channel (see the green ion in the figure),
and the ion arrives at the bulk potential. It practically stops: it
can continue only with its \emph{potential-assisted }(later with\emph{
	drift})\emph{ speed, which is several orders of magnitude lower.}
However, the rest of the ions in the channel are still accelerated through the channel,
and somewhat later, they also land in the formerly low-concentration
layer, further increasing its potential and concentration. The passed-through
ions increase the local potential in the layer in the low-concentration
segment and decrease the local potential in the layer in the high-concentration
segment. Given that the after-diffusion speeds in the layers are limited,
"as ion concentrations are increased, the flux of ions through a
channel increases proportionally but then levels off (saturates) at
a maximum rate"~\cite{MolecularBiology:2002}.

Here, the effect of the finite resources explicitly appears: about $10^3$ ions are transferred 
per channel. These ions are snorted from one layer in the 
high-concentration segment into another layer in the low-concentration layer.
The driving force gradually decreases because ions leave the first layer
and appear in the second layer.
The potential-assisted speed to replace the leaving ions in the first segment from the bulk, as well as to diffuse out from the second segment without appropriate driving forces, is by orders of magnitude slower; therefore, we can approximate the process as one in which a gradually decreasing accelerating force drives the ions.
In a very short period, the potential difference between the layers
disappears (see Fig.~\ref{fig:IonChannelRK})
that prevents further ions
from entering the formerly high-concentration layer: the process
of transferring ions through the channel closes the door behind the needed
amount of ions. Now the gradient diminished, and the van der Waals force
can close the channel again.

The limited resources means here that the passing ions decrease the concentration and the electrical field
on the high-concentration side, and, conversely, increase them on the low-concentration side.
At the exit, the field continues to accelerate the ion. 
However, the increasing potential and concentration 
increasingly decelerate the passing ions, so they quickly brake them to the assisted speed. Then, the effect of the electrical field cancels, and the ions will move with their diffusion speed ("the ion stops").
The local potentials, connected to the operation of the ion channels,
act as an "electrical lock". 

The empirical description given by~\cite{SodiumChannelsGating:2023} is essentially correct:
"Voltage-gated sodium channels have two gates: an activating gate that is voltage-dependent and an inactivating gate that is time-dependent. The opening of the activating gate allows the influx of sodium and cell depolarization. \textit{The closing of the inactivation gate will stop the flow of sodium regardless of the status of the activation gate}. These two gates work in tandem to ensure that depolarization occurs in a controlled manner: after being open for a few milliseconds, the voltage-gated sodium channels will inactivate, stopping the flow of sodium, even in the presence of persistent stimulation. The channel will remain unable to open again until the cell repolarizes to a threshold voltage that varies depending on the cell type." We only add that the activating gate is mechanical and the inactivating gate is electrical, 
furthermore, that it is charge-dependent rather than time-dependent.

Moving a charged particle inside the ion channels (nanopores) in membranes is a complex phenomenon and its details are discussed in~\cite{OriginMembranePotential:2018}. They have studied theoretically and experimentally the
generation of membrane potential at zero current in nanoporous
membranes separating two reservoirs with different electrolyte concentrations. Although not specifically for our environment,
they confirmed our statement that the membrane potential is of electrostatic origin (exists at zero current) and its value is in the order of a couple of $mV$.
Their finding that "the corresponding
Donnan potentials appear at the pore entrance and exit, leading to \textit{a
	dramatic enhancement of membrane potential} in comparison with an uncharged dielectric membrane" inspires beginning further research 
in uncovering quantitatively the details of the operation of the dynamic layer governing neuronal operation; furthermore, moving toward 
studying non-equilibrium processes.

We want to describe the change of voltage due to
limited resources using an idea similar to the predator-prey model.
We have charged sheets (represented by ions in the electrolyte) with potential $U_H$ and $U_L$
on the two surfaces of the membrane, 
plus we consider 
the corresponding neighboring layers on their side toward the "bulk" of the segment. We assume the layers' capacities are constant and equal, so the change in charge is linearly proportional to the change in voltage.

We consider that a charge transfer happens from the layer with potential~$U_H$~to the layer with potential~$U_L$~
(that is, the \textit{same charge} is removed and added, respectively) with a high speed (we call it \textit{potential-accelerated} speed);
furthermore, that with a much lower speed 
(we call it \textit{potential-assisted} speed)
the charge in the neighboring layer increases and decreases, respectively, and those voltages are ($U_{H-1}$~and~$U_{L+1}$).

That is, we assume four voltages and the initial conditions
\[
U_{H-1} = U_{H} = U; \quad U_{L=1} = U_{L+1} = 0;
\]

We assume constant layer capacity and that the amount of transferred charge (or voltage) is proportional to the voltage difference between the layers.
\[
\frac{dU_{H}}{dt} =  -\alpha * (U_H-U_L) + \beta *(U_{H-1}-U_H)
\]
\[
\frac{dU_{L}}{dt} = +\alpha * (U_H-U_L) - \beta *(U_{L}-U_{L+1})
\]

Furthermore, we assume that the after-diffusion from and to the layers next to the proximal layers is negligible aside from the high speed of the charge exchange
between the proximal layers, that is
\[
\frac{dU_{H-1}}{dt} = -\gamma*(U_{H-1}-U_{H})
\]

\[
\frac{dU_{L+1}}{dt} = +\gamma*(U_{L}-U_{L+1})
\]

As Fig.~\ref{fig:IonChannelRK} depicts, voltages $U_H$ and $U_L$
quickly approach their balanced values. When they get equal, the driving force cancels. In the meantime, the voltages $U_{H-1}$ and $U_{L+1}$ tend to approach $U_H$ and $U_L$, respectively. The diffusional and energy-producing processes, depending on the local conditions, change the voltages 
on the two ends of the channel, creating the illusion that the channel opens and closes (ad-hoc constants are used to illustrate the concept). As~\cite{HeimburgPhysikOfNerves:2009} 
analyzed, \textit{this on/off behavior of currents can be observed  
	for natural and synthetic lipid membranes, without caps}.
Despite the slightly different experimental conditions, the amplitudes and typical time scales are similar.

\subsubsection{Ion channels\label{sec:Physics-PassingIonChannels}}

The commonly used picture about the operation of ion channels~\cite{HilleIonChannels:1999} is definitely wrong, because
\begin{itemize}
	\item the potential generated across the membrane is entirely neglected
	\item the ions have no driving force, and the hypothesized protein carriers are too slow~\cite{MolecularBiology:2002}
	\item the considered van der Waals force is too weak to be noticed by the ions
	\item even if the weak van der Waals force would work for a single ion,
	the next ion would be rejected by the strong Coulomb force due to the first ion
	\item the assumed force by the 'cation-attractive negative ends' destabilizes the ion path: as the deviation from the central path increases, so does the deviating driving force
\end{itemize}

 Although it is true that conformation changes occur, but they are the \textit{consequence} of the enormous electrical field of the passing ions, rather than the \textit{cause} of their passing.
 The wrong structural model also misguides their operation, such as voltage sensing or ion sensitivity.
We also explain how the wrong model, accompanied by using unphysical ad-hoc assumptions, could explain the correct observations.
Using more ad hoc assumptions, without reasoning, requires neglecting very fundamental principles of science, such as moving huge masses for conformation changes without a driving force, without energy consumption and a clear reason.
Furthermore, that the actions require time and communication between proteins ("the
opening and closing of gap junction channels  involve a concerted twisting and tilting of the six subunits that make up the channel"~\cite{PrinciplesNeuralScience:2013}, page~109). There are unfortunate guesses to find reasons
"The energy for gating may come from mechanical forces that
are passed to the channel through the cytoskeleton". 
The whole concept builds on mechanical/structural changes,
entirely forgeting the enormous field that acts on the ions.
For the discussion, see below.

The idea of ion channel is rather confusing. 
The observation that ion channels are a "water-filled pathway"~\cite{PrinciplesNeuralScience:2013}, page~101,
is not compatible with believing that they are filled with
proteins: "in all ion channels so far studied the channel protein
has two or more conformational states that are relatively stable"~\cite{PrinciplesNeuralScience:2013}, page~108.
In two adjacent sentences~\cite{PrinciplesNeuralScience:2013}, page~107, 
claims that "the direction for this flux is determined by the electrostatic and diffusional driving
forces across the membrane" but that movement (performing a work by those forces on the charge)  "requires no expenditure The research result that "\textit{in the absence of proteins}, synthetic lipid membranes can display quantized conduction events for ions that are virtually indistinguishable from those of protein channels"~\cite{RoleProteins:2013} challenges whether really proteins are behind the ion channels' operation.

The celebrated electron microscopic images are static: they capture only a snapshot of an ion channel and the ions (the ion channel is frozen). However, they cannot reveal how the transfer occurs. Our description is dynamic. 
In Fig.~\ref{fig:The-membrane's-extra-gradient},
an ion channel is
depicted in the middle of the figure with a diameter of about $3\ nm$ for visibility. It is a "water-filled pathway" (it does not contain ions because of the enormous electric field across the two ends; furthermore, protein structures that would be mechanical obstacles for the ions).
Furthermore, we assume that the ion's size and, correspondingly, the
thickness of the atomic layer in the electrolyte on the surface of
the membrane is approximately $0.1\ nm$ (although with their electrical double layers, they can also reach a size of $1\ nm$). For comparison, recall that an \gls{AFM} tip's size is typically about $10\ nm$, and the
size of the tip of the clamp pipette is in the range of $1,000\ nm$
and the size of the soma in the range of $10,000\ nm$.

Figure~\ref{fig:The-membrane's-extra-gradient} also hints at how ions can move in the proximity of the membrane. On the left side, the electrical field increases toward the membrane, so the ions cannot move in that direction without an initial force. The case is the same on the right side because of the opposite charge.
An ion must gain energy to get closer to the membrane, which means, at the same time, higher potential energy. At that position, it is attracted by the charges on the opposite side and kept back by the mechanical counterforce of the membrane plate on this side. This situation explains why the ions do not diffuse into the bulk region. 
The ion channels 
can work in continuous mode (the "resting ion channels") and 
impulse mode (the "transient ion channels" are gated by "caps"):
"Two types of ion channels—resting and gated—
have distinctive roles in neuronal signaling. Resting
channels are primarily important in maintaining the
resting membrane potential, the electrical potential
across the membrane in the absence of signaling. Some
types of \textit{resting channels are constitutively open and
are not gated} by changes in membrane voltage; other
types are gated by voltage but can open at the negative resting potential of neurons. \textit{Most voltage-gated
channels, in contrast, are closed when the membrane is
at rest and require membrane depolarization to open.}"~\cite{PrinciplesNeuralScience:2013}, page 126.
Closing ion channels is important for making ungraded layer potentials.

The resulting electrical field cannot accelerate an ion 
except when the ion is at the beginning of the open ion channel in the wall of the membrane. In that case, the mechanical counterforce is missing, and the field moves the ions along the "water-filled pathway". 
In the classical model, no mechanism is provided for how the ion pumps gain energy from \gls{ATP} and converts it into forwarding force. In our model, the potential gradient near the membrane, the \gls{ATP}, by hydrolysis, generates ions, which move
to the "plates" of the condenser, thus increasing its potential energy; see section~\ref{sec:Physics-BackgroundLogistics}.
In the transient state that voltage accelerates the ions to a high speed in the viscous liquid. The empty red circles show
an ion that traverses through the channel.
The figure shows the electrical fields on both sides of the channel at the beginning and end. (On the right side, the electrical field is negative, but so is the charge of the ions.)

The electrical acceleration sharply decreases after leaving the ion channel (see the right side). As shown by the difference in the vertical positions of the filled red and green balls, the difference in the electrical field values is huge across the membrane ($E_{accelerate}$) and considerable inside the dielectric
electrolyte ($E_{assist}$), as shown by the difference in the electrical field values of the green balls. The field still accelerates, but significantly less intensely than the one in the gap. Suppose we assume that the ion quickly accelerates to its Stokes-Einstein speed. In that case, the travel speeds in
those space regions are proportional to the difference of the fields
in the shown positions and the reciprocals of their travel times. 

The fundamental issue with describing neuron's operation is
the fallacy that
"The primary function of ion channels in neurons is
to generate transient electrical signals"~\cite{PrinciplesNeuralScience:2013}, page 109,
and they have no relation
with the mechanical (mainly pressure) phenomena.
As we discussed, the charge transmission mechanism 
(energy and impulse transmission in the narrow viscous "water-filled pathway" by the Stokes-Einstein friction)
generates mechanical signals. The latter are not only significant, but play a definitive role in signaling.
The cooperation of disciplines of science plays a fundamental role.

\subsubsection{Ion pumps\label{sec:Physics-IonPumps}}

The role of ion pumps must also be revisited.
"These pumps differ from ion channels in two
important details. First, whereas open ion channels
have a continuous water-filled pathway through which
ions flow unimpeded from one side of the membrane
to the other, each time a pump moves an ion, or a group
of a few ions, across the membrane, it must undergo a
series of conformational changes. As a result, the rate of
ion flow through pumps is 100 to 100,000 times slower
than through channels. Second, pumps that maintain
ion gradients use energy, often in the form of \gls{ATP}, to \textit{transport ions against their
	electrical and chemical gradients}. Such ion movements
are termed active transport."~\cite{PrinciplesNeuralScience:2013}, page 101.

It is not clear what a force can "transport ions against their
electrical and chemical gradients". It is undoubtedly "not on the ground that there is \textit{any ‘new force’} or whatnot"~\cite{Schrodinger:1992} present: the ions,
independently of whether they are in an inanimate or living environment,
can be moved only by electrical or thermodynamic forces,
instead of some magic protein mechanism (ion pumps). The resultant potential
(the sum of the electrical and thermodynamic potentials for the individual ions) moves the ions through the channels in the membrane.
In a typical mammalian cell, in the resting state, see Table~\ref{Tab:SummaryTable},
the sum of the electrical and thermodynamic potentials is $-32\ [mV]\ K^+$ ions and $+4\ [mV]\ Na^+$ ions.
The same values for squids are $-33\ [mV]\ K^+$ and $+13\ [mV]\ Na^+$. 
Correspondingly, in the resting state, the resulting force moves sodium ions out of the cell and potassium ions into the cell
(for more numerical examples see section~\ref{sec:AP-PhysicalProcess}).
The resultant potential difference actually implements
what a "Na-K pump" is supposed to do by using some 
hypothesized protein mechanisms.
Claims like "dissipation of ionic gradients is prevented by
the sodium-potassium pump ($Na^+-K^+$ pump), which
moves $Na^+$ and $K^+$ against their electrochemical gradients"~\cite{PrinciplesNeuralScience:2013}, page 139,
are wrong.  The  pumps works correctly: the transfer ions
according to the electrochemical gradients, not against them.

Claims such as "The flux of ions through ion channels is passive,
requiring no expenditure of metabolic energy by the
channels."~\cite{PrinciplesNeuralScience:2013}, page 107,
are wrong. The enormous electric field across the membrane
performs work on the ion, and the effect of the transmitted ion decreases the accelerating potential: some (white) ions 
are released, and those are replaced by ions produced by 
the hydrolysis of ATP as discussed throughout the paper. 
\textit{There is no "free passive transport"}. 
That energy is prepared "globally" (not at the proximity where it is used) and continuously (during the full life cycle of neuronal operation), mainly in the recovery period. As discussed in section~\ref{Physics-ControlTheory}, in the transient state,
to restore the condenser's voltage, those ions must be collected.

In our model, \textit{the ion pumps are ion channels which work in stationary
	rather than transient conditions}. In the resting state, the potential difference fluctuates around zero (due to the permanent slow material transport). In contrast, while at the beginning of the transient state, a sudden change around the entrance 
of the channel creates a large potential difference,
so an intense transport happens.
In the resting state, since the resultant forces exerted on the different ions differ, they permanently
pass through the membrane. As described, the passage changes 
the ionic gradients on the two sides of the membrane and
generates a slight "leakage current" and slightly fluctuating voltage.
In the resting state, the electrolyte layers have sufficient time for after-diffusion, allowing a graded ionic gradient to form near the membrane.
(The gradient changes in a fraction of a nanometer, so it is challenging to measure it.)
With reference to Fig.~\ref{fig:IonChannelRK}, the graded gradient
generates an electrical field on the ion channels, which is
by orders of magnitude lower than before rush-in.
Conversely, in the transient mode, the capped ion channels enable the creation of ungraded (step-like) gradients.
Consequently, the ion speed, furthermore, the charge and mass transport,
are by two orders of magnitude lower in the resting state than in the transient state, as observed, mainly because the
conductance of the ion channels used in the two states
differs by two orders of magnitude.

As Figure 6 of \cite{NeuralEnergyConsumption:2017} displays, 
the ratio of $K+$ and $Na+$ changes sharply during the transient state.
Theoretically, they assumed that the neuron “pumps 3 $Na^+$ ions out of the cell and two potassium ions in”; experimentally, they showed in their Fig. 6 that the ratio changes between 0.01 and 7.5.
According to our theory, the magnitude of the currents (and so the exchange ratio)
depends on the resulting potentials, so the ratio must change as the membrane potential changes during \gls{AP}.
The classical theory cannot explain this behavior and leads 
to exciting conclusions: 
"Furthermore, we analyzed energy properties of each ion channel and found that, under the two circumstances, power synchronization of ion channels and energy utilization ratio have significant differences. This is particularly true of \textit{the energy utilization ratio, which can rise to above 100\% during subthreshold activity}."~\cite{NeuralEnergyConsumption:2017}
(The simplest way to solve energy problems would be to use subthreshold excitations).
Our model says that (due to $Na^+$ rush-in) both the concentration and local membrane potential drastically change 
during the transient state. As long as the resulting potential is above 
$\approx 33\ mV$, the driving force acting on potassium ions is 
positive, i.e., the pump will move both ions out of the cell.
The composition of the output current depends on the 
input current.

\subsubsection{Maxwell demon in cells\label{sec:Physics-MaxwellDemon}}

After all, gated ion channels, see Fig.~(\ref{fig:The-membrane's-extra-gradient}), operate as Maxwell-demon'-like objects built into the separating membrane (from the point of view of the segments
and the observer).
In the classical theory, some power opens them, and they autonomously transfer ions
in a \emph{potential-accelerated} operating mode, and then that power
puts the cap back on the channel.
The segments are separated by a semipermeable membrane with ion channels capped on its surface. 
As long as the caps of the ion channels are closed 
and the ion concentrations on both sides of the membrane are the same, we do not observe any change in the state of the solutions.
Although the channels can stochastically open, close, and re-open,
they transmit charge quanta of varying degrees of definition.
Even the channels can
recognize the ions' chemical nature and transmit only a selected ion
type. The channels are passive during those processes, although the enormous
voltage gradient can rearrange their structure and change their behavior through that.
The demons also coordinate their actions using the layer containing charges
as a communication medium; their population maintains a well-defined macroscopic current across the membrane.
Physics can also explain the magic behind their operation.

\subsubsection{Gated ion channels in action\label{sec:Physics_GatedChannels}}

A gate is open when there is no mechanical obstacle against
material transport, and the local resulting potentials on the two sides of the membrane
in the near-membrane layers enable the transport of ions.
The cap is connected to the membrane only
at one point, so it cannot fly away, and also cannot close again as long as the charge 
on the surface is present. 
In the absence of charge, the cap undergoes random motion, and the short-range van der Waals force may eventually reposition it on the membrane, thereby closing the channel.
The voltage-sensing electrometer (see section~\ref{sec:Physics-VoltageSensing}) opens the channel, and the lack of 
charge on the surface enables it to close, but the closure is not immediate (the cap's mass is by orders of magnitude larger than the mass of an ion that can pass through the channel).
The fluctuation of the voltage gradient due to the slow
current in the layer in the proximity of the membrane near the ion
channel's exit opens, closes, and re-opens the channel in an apparently
stochastic way (actually, as the repulsion of charges due to the fluctuating current on the cap and
the membrane regulates it), as observed. 

\gls{HH} was right when "they postulated that a change in membrane
potential causes this charge to move across the electric field of the membrane, resulting in conformational
changes that open or close the channel"~\cite{MolecularBiology:2002}, page 163.
However, their followers changed the causality by claiming
that conformational changes (protein mechanisms) move the charges across the membrane, despite that synthetic lipid membranes can display "quantized conduction events for ions that are virtually indistinguishable from those of protein channels"~\cite{RoleProteins:2013}. The latter fact shows 
that not protein mechanism hide behind the observations, and that our electrical gating explanation in section~\ref{sec:Physics-GatingLayers} is correct.

When one cap is removed, the rushed-in
ions in the proximity of the channel's exit suddenly increase the
local potential (produce fast transient changes~\cite{KochElectricalPropertiesSpike:1983})
proximal to the spot centered at the exit in the layer on the membrane's surface.
The surface outside the spot remains at a lower potential, so the
ions in the layer start moving toward other channel exits, delivering
potential to those channel exits. Given that those channels are voltage-controlled,
they get open, and the process continues in an avalanche-like way~\cite{NeuronalAvalanches:2003}. The avalanche, as explained, needs a 
sufficiently large voltage gradient, which several
synaptic inputs can trigger if they sum up appropriately. Alternatively, a single spike with sufficiently steep front slope~\cite{LosonczyIntegrative:2006}
can be sufficient, providing a simple way to synchronize neuronal assemblies (and proving that \textit{not the voltage, but the voltage gradient, single or summed, controls the operation)}.

\subsubsection{Ion selectivity\label{sec:Physics-IonSelectivity}}

"Some types of cation-selective channels
allow the cations that are usually present in extracellular fluid -- $Na^+$, $K^+$, $Ca^{2+}$, and $Mg^{2+}$ -- to pass almost
indiscriminately. However, many other cation-
selective channels are permeable \textit{primarily} to a single type of ion, whether it is $Na^+$, $K^+$, or $Ca^{2+}$."
"The normal selectivity cannot be explained by
pore size, because $Na^{+}$ is smaller than $K^{+}$"~\cite{MolecularBiology:2002}, page 107.
The fact that ion selectivity changes with the membrane potential
is observed (under the name of changed specific permeability), but it remains unexplained.  "For example, the
negative resting potential of nerve cells is largely determined by a class of $K^+$ channels that are 100-fold more
permeable to $K^+$ than to $Na^+$. In contrast, during the
action potential a class of $Na^+$ channels is activated that
are 10- to 20-fold more permeable to $Na^+$ than to $K^+$.
Thus, a key to the great versatility of neuronal signaling is the regulated activation of different classes of ion
channels, each of which is selective for specific ions."~\cite{PrinciplesNeuralScience:2013}, page 101.
However, the question of why the permeability changes and how the channels are activated remains open. It is remarkable that \textit{an ion-specific channel is permeable also for other ions, only different transmission intensity is observed}.
Unfortunately, the measurements are incomplete: neither 
the time (say, relative to the last \gls{AP}), nor the neuron's state, nor the local potentials at the two ends of the channel (at the two dynamic layers), nor the location of the channel on the membrane are provided.

The above mechanism of channel passage can also help explain ion selectivity.  
The commonly used picture of the operation of selectivity filters
is surely wrong. The assumed mechanical operation of the pores
is too slow: the assumed structural change needs $10^{-8}\ s$
and the ions passage time is about $10^{-10}\ s$ (furthermore, it must be repeated about $10^{3}$ times per open period). If a wrong ion is caught,
it must be transported back to its departure side, through the right ions
(against their repulsion), and the right ions must retry.
Neither forward nor backward movement has an appropriate driving force in that picture (a 'whatnot' force). Furthermore, the transport capacity
is not sufficient for that kind of operation.
Not to mention that the "water-filled pathway" offers no
mechanism for selectivity.

In our model, as Eq.~(\ref{eq:NernstPlanckExtended}) states, the electrolyte separated by a permeable membrane is balanced
when the electrical and thermodynamic forces are equal.
The $E_{thermal}^{C_k}(d)$ "electrical field" depends on the chemical quality of the ion; the electrical field does not.
That means the resulting driving forces acting on the
different ions differ. While a resulting force is exerted on one ion and moves it to the other segment,
the other remains resting, or even moves in the opposite direction.
Furthermore, the rush-in of ions significantly increases the $Na^+$ concentration in the dynamic layer, naturally increasing
the number of $Na^+$ ions passing through the channel.
That increase alone creates the illusion of the increased $Na^+$ permeability. 
Of course, the ions moving across the segments change the field contributions,
indirectly affecting the balanced state of the other ions.
The permanently open ion channels participate passively in the process, and their apparent permeability changes with the applied fields. 
To provide channel penetrability data
is meaningful only if ion concentrations and membrane thickness data are also available.
The driving forces of the ions differ, and they produce, depending on the actual concentrations and local potential, the selectivity of the channels.
Of course, the driving force is a statistical quantity: it is not valid for the individual ions, but it is valid for the result
of a population of ion channels.

Here we interpret some claims from~\cite{PrinciplesNeuralScience:2013},  page~107.  Presumably, the neuron is in the resting state. The permeability (in our model, the resulting potential difference, i.e. the local concentrations inside and outside) strongly depends on the local conditions, especially the local streams near the membrane, in a nanometer range.
Without knowing any details, especially the local concentration gradients of the respective ions, it is hard to state anything definitely.
However, claims such as "some types of cation-selective channels
allow the cations that are usually present in extracellular fluid - $Na^+$, $K^+$, $Ca^{2+}$, and $Mg^{2+}$ - to pass almost indiscriminately" simply claims that the "water-filled pathway" observation is correct, and the fluctuating potential
moves all ions there and back, without discrimination.
Similarly, that "Most types of anion-selective channels are also highly
discriminating; they conduct only one physiological
ion, chloride ($Cl^-$)",
means that the fluctuating potential
moves the only available negative ion, practically constant 
driving force (concentration gradient). 
In a balanced state, inside and outside, persisting flows 
of the electrolyte may be formed. The ion that passes decreases the concentration of the respective ion, this way prepares
the after-diffusion of other ions. This process creates
the illusion of the "almost indiscriminate" ion passage.

"The net electrochemical driving force is
determined by two factors: the electrical potential difference across the membrane and the concentration
gradients of the permeant ions across the membrane."
According to our model, the third (maybe, most important) factor is the local dynamic potential in the layer
neighboring the membrane. To describe even the resting state,
dynamic concepts are needed.

"In some open channels the current varies linearly with driving force - that is, the
channels behave as simple resistors". Wrong. As said in the 
previous paragraph of the book, "the electrostatic and diffusional driving forces" move the ion, and only the "electrostatic force" is measured. Ion channels are not simple electric resistors; see also section~\ref{sec:Physics-ResistanceNeuron}.
"The rate of net ion flux (current) through a channel
depends on the concentration of the permeant ions in
the surrounding solution." If 'surrounding' means the
nanometer-size layer, correct. The concentration in that layer is not
identical with the concentration in the bulk. The resting current keeps the electrolyte in permanent moving, and the 
observation records that in the stationary (resting) state,
due to the fluctuating concentration (and voltage), ions pass
the ion channels.

\subsubsection{Blocking channels by drugs \label{sec:BlockingChannels}}

Drugs can strongly affect ion channel's operation.  When drugs are forcefully injected into a cell, correspondingly to the 
static view of physiology, the purpose is to replace
the electrolyte with another solution containing the drug. 
It is commonly believed that the drug somehow builds 
into the (wall? of the) ion-specific ion channels and
closes them, this way blocking the ion traffic across
those specific channels.
However, in fact, 
the original electrolyte is changed to another one,
and the ion compositions and concentrations change,
thereby the accelerating voltage across the membrane 
disappears.
The strong flow (compared to the weak flow
in the resting state) may remove
the surrounding ion layers from the surface,
where the electric processes take place; this way dismissing the permanent flows (apparently 'specific ion channels').
Since the working
conditions are destroyed, the \gls{AP} gets blocked.
However, it is not sure whether it is the effect of the drug
directly on the channels as the old model claims,
or removing the ions from the electrolyte
and thereby removing the gradients that move the ions across
the membrane as follows from our model.
The fact that the \gls{AP} is measured as the voltage drop
on the \gls{AIS} and the blocked potassium current is
expected to flow through the membrane, seems to support our model.
Depending on the conditions, restoring the cell state may require minutes.

\subsubsection[Delivering current]{Delivering current across the membrane\label{Electrodiffusion_DeliveringCurrentLayers}}
The passage is too quick to affect the bulk, given that the ions
can only use a \emph{potential-assisted} speed to reach distant places
in both segments. Again, the charge and mass conservation works: the
ions pass suddenly from the high-concentration side to the low-concentration
side, only from one layer to another.
The mentioned \emph{layers on the two sides will actively
	initiate and terminate the ion transfer through the ion channels,
	but the ions can only pass through an open channel.}
One layer saturates, and the
other empties. After a while, \emph{the source of ions will be exhausted}.
\emph{Those layers' existence suggests revisiting the idea of describing
	neuronal operation by two single potentials of the bulks on the two
	sides of the membrane}. The picture behind the equivalent circuits is definitely wrong, as it is based on the concept of "fast current".

Following their arrival, the driving force perpendicular to the membrane's voltage disappears, and the ions form a thin "hot spot" in the layer. 
The electrical repulsion acts in the direction parallel with the membrane's surface and
leads to distributing the ions (decreasing the gradient by distributing the charge locally) around
the channel's exit. However, the ions have a finite speed.
The ions saturate the layer on the membrane's surface with a time constant
between ($\frac{10^{-8}m}{10^{-1}m/s}\ s$) at the beginning
and $(\frac{10^{-8}m}{10^{-4}m/s}\ s)$
at the end of their arrival period (we assumed $10\ nm$ average distance between
ion channel exit positions on the membrane). We shall take the longer time,
so that we can expect a time constant for the saturation current around
the ion channel's exit in the order of $0.1\ ms$. When charging up
the membrane in an avalanche-like way, the ions must pass on average
a distance of about $0.05\ mm$ from its center to its farthest point,
so we expect a $0.5\ ms$ ($\frac{5*10^{-5}m}{10^{-1}m/s}\ s$) time
until the membrane's slow current charges up the membrane to its maximum
potential. The created charge must flow out from the farthest point
in the neuron membrane of size $0.1\ mm$ in time of order at or below
$1\ ms$ ($\frac{10^{-4}m}{10^{-1}m/s}$); see the length of the
$\frac{dV}{dt}$ pulse measured at the beginning of the 
\gls{AIS}
~\cite{BeanActionPotential:2007}, which time is prolonged
up to $10\ ms$ by the neuronal $RC$ circuit; the ions are slow when
the voltage on the 
\gls{AIS} is low, see Eq.~(\ref{eq:StokesSpeed}).
Assuming those distances and speeds, including the \emph{potential-assisted}
speed of the slow current, we are on a time scale matching the available
observations.
\index{current!across layers}

\subsubsection{Synaptic gating\label{sec:Physics-SynapticGating}}

The slow current also accounts for synaptic gating. The synaptic charge-up current flows into the membrane.
It causes transient changes~\cite{TransientResponses:2008,KochElectricalPropertiesSpike:1983} in the membrane's voltage, providing \textit{direct evidence that the membrane is not always equipotential. The ions on the membrane’s surface can propagate at a finite speed}.
Given that the ions in the axonal arbor must enter
the membrane against the actual membrane potential, the membrane potential acts as a valve. The potential stops the ion inflow to the membrane for the period while the membrane's voltage is above the threshold: it effectively inhibits further inflow through
all synapses. This behavior naturally explains the absolute refractory
period. After the membrane's voltage drops below the threshold value,
the ions can enter the membrane again (and the current intensity increases as the membrane's potential decreases), but they reach the
\gls{AIS}
later when in the meantime the membrane' voltage proceeded toward its hyperpolarized state; so they seem to appear dozens of microseconds later at the
\gls{AIS}, explaining the relative refractory period (as a delayed expansion of the absolute refractory period, with variable current intensity).
Of course, the length of the relative refractory period sensitively depends on the conduction velocity~\cite{APTemperatureDependenceRefractory:2001}.

\subsubsection{Voltage sensing\label{sec:Physics-VoltageSensing}}

"Voltage sensing by ion channels is the key event enabling the generation
and propagation of electrical activity in excitable cells."\cite{VoltageSensingIonChannel:2019}
The mechanism by which voltage-gating of channels works remained a mystery; one of the worst consequences of the \gls{HH} theory is that it separates the potential from ions and their current.
It is not
easy to investigate it experimentally: "the structural basis of
voltage gating is uncertain because the resting state exists only
at deeply negative membrane potentials"~\cite{VoltageGatedChannelStructure:2019}.
Usually, a "sliding helix" (structural)
model is assumed. The last chance is a "whatnot" protein mechanism, again.

Under certain conditions, an ion channel can be opened in only one direction and
only for a limited period, and this allows the membrane to become semipermeable.
\index{membrane!semipermeable}
We imagine an ion channel as a simple hole (a cylinder) between the
high and low-concentration segments with a cap on its top (on the
side of the low-concentration segment), in line with the "water-filled pathway" model of biophysics. At the spots
where the ion channels are located,
the ions cannot penetrate the
membrane until the cap is removed/lifted
(the channel gets open). Unlike the original Maxwell demon, our demon does not have
information in advance about which particle should be transmitted:
\index{Maxwell-demon}
it is passive in selecting the particle. 
(Passive here means that no biologically-produced local energy is used:
the electrical potential energy from the voltage difference across the membrane moves the ions to the other side of the membrane.)

In our model, a
voltage-controlled ion channel gets opened and closed due to purely electrostatic
reasons. It works as a two-plate simple
nano-scale electrometer (of type quadrant%
), similar to the ones used to measure small electrical potential
between charged elements (e.g., plates or fine quartz fibers).
\index{electrometer}
\index{ion channel}
Given that the membrane and the cap in their resting state are isolators,
no electrical repulsion evokes between them, and the adhesion
 sticks them firmly together, representing a permanent force. The van der Waals force is inversely
proportional to the squared distance between the dipoles in the cap and the membrane, respectively, and is linearly proportional to the perimeter of the channel.

However, when a slow ion current
flows into the surface layer in the proximity of the cap,
charges appear in the layer proximal to the membrane;
the membrane and the cap get covered by a fragile electrical skin. The charges locate forming a constant density on the surface.
The charge on the cap is proportional to the area of the cap and similarly inversely proportional to the squared distance between the cap and the membrane. 
\index{voltage gradient}
The local gradient of the slow ion current  generates a local voltage gradient, and the force acting on the cap is proportional to the
product of the voltage gradient and the area of the cap.
Given that the cap is slightly elevated, the repulsion
force may have a component in the direction of lifting the cap.
Since the van der Waals force has a fixed size, when the electrical repulsion exceeds it at a critical voltage gradient value, the channel opens.

\subsection[Fallacies about biological electricity]{Fallacies about biological electricity\label{sec:Physics-Fallacies}}

\subsubsection{Modelling}

It was a colossal mistake to forget that biological currents consist of ions, each with its electrostatic force.
It led to fallacies, among others, including delayed and leakage currents (see Section~\ref{sec:Single-RestingCurrent}) and heat absorption (see Section~\ref{sec:Physics-HeatAbsorption}). Their finite
(Stokes-Einstein) speed, see section~\ref{sec:IonicCurrent}, due to the ions' mass and the different conduction mechanism, explains the spatiotemporal behavior of living matter~\cite{RoleOfInformationTransferSpeed:2022} (contrasted to Newton's 'instant interaction').

It was another colossal mistake to introduce equivalent circuits with constant-value voltage generators. It forces one to assume that the 
conductances of the biological objects (membrane, synapses, \gls{AIS}) change 
without any reason, and prevents understanding how the competition between thermodynamic and electrical processes governs neuronal operation.
It leads, among others, to attributing conductance change to membranes, which are simple isolators with no charge carriers that could implement
charge transfer, see section~\ref{sec:IonicCurrent}; this way
attributing the change of an electrical entity to the biological material.
This assumption neglects that a driving force is needed to
move ions (\textit{denies Newton's laws of motion}; furthermore, energy and charge conservation). 
Instead, it attributes the magic ability to the ion channels that they
can change their transmission ability as the actual situation requires.

\subsubsection{Measuring\label{sec:Physics-Measuring}}

The classic textbook~\cite{JohnstonWuNeurophysiology:1995}, appendix~A, correctly warned
(especially when working with concepts of an electrical theory of 
neuronal operation): 
"An intuitive
grasp of concepts of electricity and electrical circuits is helpful for understanding some of the basic theory in cellular neurophysiology. \dots Under ideal
circumstances, the physical act of measuring a neurophysiological event
would have no effect on the electrical signal of interest. Unfortunately,
this is seldom the case in neurophysiology."

One of the bad examples is grasping the electrical characteristics
by clamping (we are not speaking against the method, but emphasizing
the need for its correct interpretation).
The textbook also provided the correct interpretation of the
measured quantities:
"In describing the resistance (or Ohm's law) properties of neurons, one
either \textit{injects current} and measures the resulting change in voltage (a current clamp) or changes the voltage to different values and \textit{measures the current necessary to hold (or clamp) the voltage} to these values (a voltage clamp)."

As discussed in section~\ref{sec:AP-PhysicalProcess}, physiology does not realize that, \textit{in resting and transient states, different physical processes take place and correspondingly, different circuit models shall be used}. In the resting state, the low-intensity ion current is "shunted" by the ion channels in the membrane's wall;
moreover, the "parallel $RC$" model works sufficiently well.
Since the current is slow, it is not simply distributed,
but only the "overflow" reaches the \gls{AIS}.
However, as the current intensity exceeds the ion-conductance capacity of the "resting ion channels", the "transient ion channels" play an increasingly important role.
Above the voltage threshold, the serial $RC$ circuit is the correct model.  Until the neuron's voltage reaches its threshold, a "parallel" and a "serial" resistor work simultaneously, this way implementing a mixed model (for experimental evidence, see sub-threshold excitations; for the correct mathematical handling, see Eq.~(\ref{eq:PID_Neuron})).

\subsubsection{Input resistance\label{sec:Physics-InputResistance}}
The equivalent circuit picture for the neuronal operation was abstracted originally~\cite{HodgkinHuxley:1952} from the real picture that distributed
resistors (the ion channels) are located in the wall of the membrane, and the system is in a balanced electrical state. 
Those distributed resistors were contracted to a single discrete resistor, that is switched in parallel with a discrete condenser, abstracted from the distributed parallel charge-storing surface elements.
Those two abstract discrete elements represented a parallel $RC$ circuit (apart from that the biological elements operate with slow currents). 
However, the later research~\cite{ActionPotentialGenerationNatrium:2008, AIS_Updated_Viewpoint:2018, AIS_NeuronalPolarity:2010} discovered that a much larger amount of ion channels, the \gls{AIS}
(abstracted as a discrete resistor array), is located at the
beginning of the axon. That means that the two-element equivalent circuit must be replaced by a three-element one 
to describe (more) correctly the neuronal configuration. 
However, biophysics has preserved the wrong circuit, although the additional resistance fundamentally changes the 
behavior of the electric operation of the neuron.

In the native resting operating mode, due to the different reversal potentials, a slight fluctuating local potential difference  exists across the ion channels in the membrane's wall, and so the nearby ions can pass across the membrane, observed as ion pumps (recall that \SI{5}{\micro\volt} change in the local potential causes  \SI{1}{\kilo\volt\per\meter} change in the electric field). The passed ion changes the local potential and the fluctuating potential causes a
slight resting current. The current remains local: no current can be measured on the \gls{AIS}, because the charge carriers
cannot reach it: they are "shunted" by nearby ion channels as they travel on the surface of the membrane.  Recall that the driving force depends on the chemical nature of the ions, so the same electric potential may drive ions in both directions. 

Since the charge transfer capacity of the individual channels is limited, as the amount of charge increases
(by natural or artificial current input),
the charge carriers can reach more distant ion channels,
and finally, the \gls{AIS} array; as the subthreshold 
excitation experiments demonstrate it.

When measuring the "input resistance" of neurons, one actually
adds a small external potential to the system (the conductance meter is not only an observer, it is also an active device). This way, changes the resulting potential, starts a small current across the membrane and changes the ion concentrations in the proximity of the membrane.  In this sense, one measures a mixture of the "input" and "output" resistance corresponding to the two models, given that the current is distributed between the one through the membrane and the other through the \gls{AIS}; plus the measuring electrode represents one more drain.

\subsubsection{Changing conductivity\label{sec:Physics-ChangingConductance}}
At the beginning, it was clearly told~\cite{COLE_CURTIS_IMPEDANCE:1939}, but forgotten:
"Since it is quite generally
believed that the depolarization of a nerve fiber membrane, during
excitation and propagation, involves an increased permeability to
ions. There have been many attempts to detect and measure this
change as an increase in the electrical conductivity. ...
In these cases, \textit{the measuring current was also the stimulating current}, and it was not possible
to analyze the changes satisfactorily."
Moreover, during the passage of an impulse, a "charge wave" (in another terminology: a "pressure wave") passes along the axon, and the
changed number of charge carriers in that cross-section, see Eq.(\ref{eq:DriftCurrent}), \textit{changes the current that creates the
	illusion that the conductance at the measured cross-section changed}.
In our terminology, the stabilizing feedback voltage provides an external force, and it also contributes an external current. The feedback amplifier's \textit{electron current} (converted to \textit{ion current}) is provided externally and is added to the closed system, without including the measurement device 
into the current budget of the closed system.
That interpretation of a closed system denies the fundamental symmetries of science. Moreover, it significantly contributes to the claim that "laws of science cannot describe life".

\emph{With wording
that "conductance changes", one states that charge carriers
appear/disappear/reappear; that is, the charge and mass conservation are not fulfilled}.
\emph{There is no voltage-dependent conductance}~\cite{KochVoltageDependentConductance:1999}.
Instead, the finite speed of ions and the wrong assumption that conductance
is a primary entity mislead physiological research. Claims such as "the current cannot disappear, it has to go somewhere"~\cite{KochBiophysics:1999}, page 9, are wrong: \textit{there is no conservation law for current, which is the derivative of charge with respect to time}. 
The correct physics background behind the early phenomenon~\cite{COLE_CURTIS_IMPEDANCE:1939} is that the number of charge carriers, at the cross section where the measurement is carried out, changes as a wave of ions with variable intensity moves slowly along the axon (they are apparently ``created'' for the static measurement arrangement). The changing conductance is a fallacy stemming from the idea of "fast current", which is valid for electrons but not for slow ions. 

In contrast, when the clamping voltage is switched off, the axon is still
filled with charge carriers. The resting potential reaches the end
at the membrane ``instantly'' (in this situation, the conduction mechanism is similar to that in metals). The driving force disappears, the
ion stream stops, and the ions discharge into the membrane. The lack and
the presence of ions in the axon when switching clamping on and off, respectively, produce the difference that "\emph{conductances [actually, currents]
increase with a delay when the axon is depolarized but fall with no
appreciable inflexion when it is repolarized}"~\cite{HodgkinHuxley:1952}.
The potential is equalized by the \gls{AIS}
current, producing a net exponential decay:
\begin{equation}
	I_{m}^{off}=I_{Wall}*e^{(-\frac{\alpha}{R_{m}C_{m}}*t)}\label{eq:I_Membrane_Off}
\end{equation}

\begin{figure}
	\includegraphics[width=0.95\textwidth]{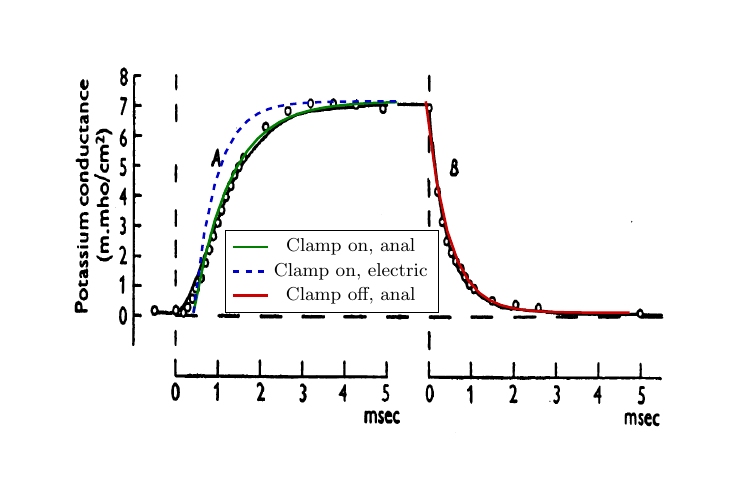}
	\caption{The measured current, instead of conductance (the black points and diagram lines in the figure are taken from Fig.~2 of~\cite{HodgkinHuxley:1952}), demonstrating that two different physical effects result in the rising and falling edge of the measured current.
		\label{fig:HHFig2Simple} }	
\end{figure}

During the regular operation of a neuronal membrane, after opening
the ion channels, a vast amount of ions flows into the intracellular
space from the extracellular space, somewhat similar to the effect of switching
a clamping voltage. The essential difference is that
the ions arrive through the axon to the joining point in clamping. In contrast, through the membrane's ion channels, they directly contribute to the current on the membrane’s entire surface. The membrane's size is finite, so with a finite
current speed, it takes time until the charges on the membrane's surface
arrive at the \gls{AIS}, as discussed for the
axonal current.

Fig.~\ref{fig:HHFig2Simple} shows the measured rising and falling edges
of the clamping experiment performed by \gls{HH}~\cite{HodgkinHuxley:1952}.
The continuous lines show the 
diagram lines fitted to the experimental data. Notice that the fitted polynomial significantly differs from  the correct~\cite{SodiumCurrentDelay:2006} exponential function around the time 0.5~msec. The dashed line shows the expected rising edge, calculated with the time constant of the falling edge.
The measured rising edge corresponds to the superposition of two processes, described by the same type of function form.
One is the charge-up process where the 
charge-up current is a saturating curve and alone would be described by the dashed line. However, the amplitude of the charge-up current is also a saturating curve with a different time constant; it describes the change in the number of
charge carriers in the axon; see Eq.(\ref{eq:StokesCurrent}).
Their time constants are very similar. 
\gls{HH} measured that their time constants differ far outside
their experimental error, which would not be possible if they were
the result of the charge-up and discharge of the same oscillator circuit.
However, their wrong idea about the "fast current" misled them.

%
That is, the \textit{slow} membrane current reaching the \gls{AIS} can be described by a product of two exponential curves, with different time constants given that they describe two physical processes. 
The first term describes how the current amplitude saturates
and the second how that the charge that that current delivers, discharges.
The \textit{slow} input currents formally are described by
the following analytic form valid for \textit{fast} currents:
\begin{equation}
	I_{in}=I_{o}*(1-\exp(-\frac{1}{\alpha}*t))*exp(-\frac{1}{\beta}*t)\label{eq:BioCurrentForm}
\end{equation}

\begin{figure}
	\includegraphics[width=0.75\textwidth]{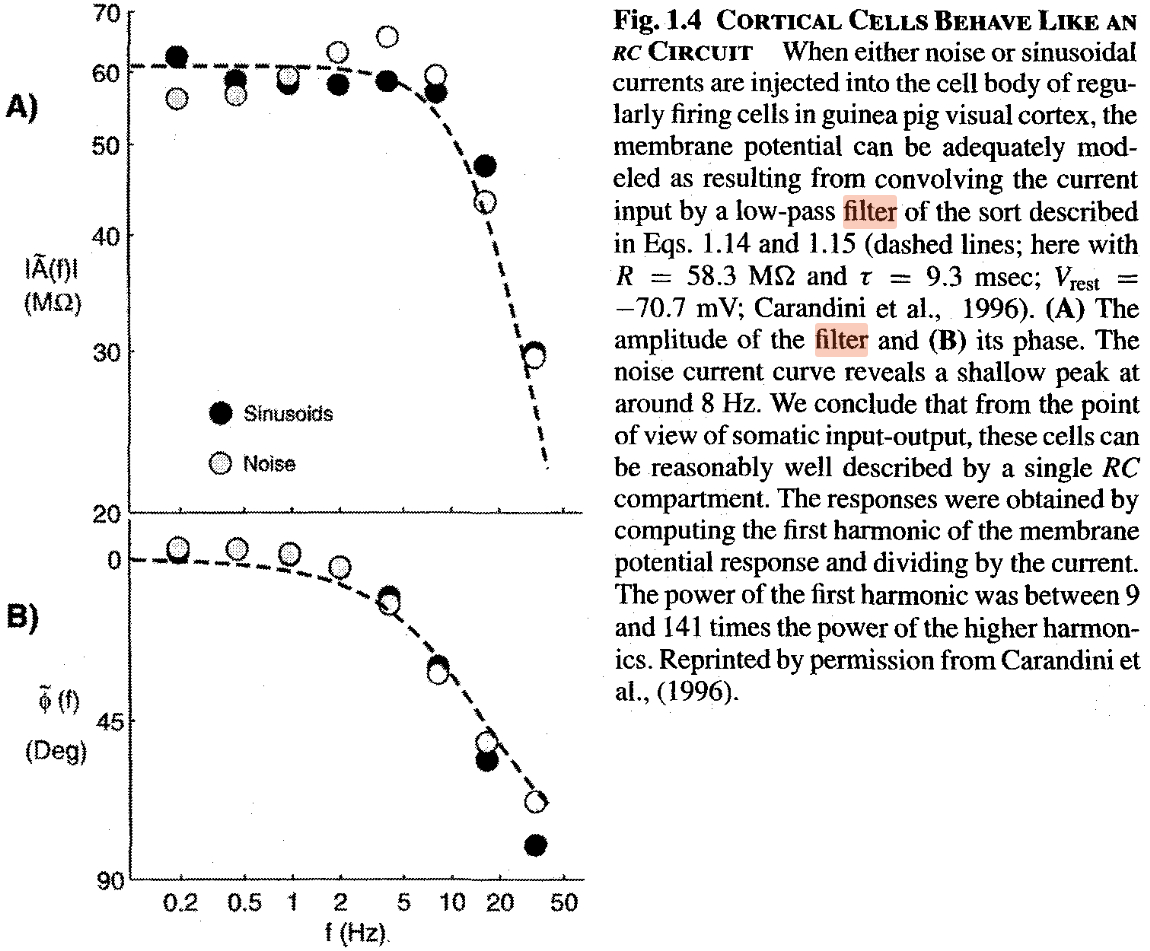}
	\caption{The "low pass filter" effect from a known textbook (Fig~1.4 of~\cite{KochBiophysics:1999}). The effect of the "slow current" is misinterpreted as "low pass filter.
		\label{fig:LowPassFilter} }	
\end{figure}

\subsubsection{Neuron as 'low pass filter'\label{sec:Physics-LawPassFilter}}

It is important to remember for alternating current experiments that 
the ions will move at different speeds
under the effect of $E(z,t)$, which is the function of
the local concentrations and the frequency of the alternating current.
The alternating current changes the current's direction and speed. It moves the ions
from one electrode toward the other, reproducing the rising and falling phase of the experiment displayed in Fig.~\ref{fig:HHFig2Simple} with reducing the distance between them as the frequency increases. At some value (from the typical length of an \gls{AP} it can be estimated to lie between one and ten millisecond, it can be estimated to be between hundred and ten KHz), the ions do not arrive at the electrodes. As displayed in Fig.~\ref{fig:LowPassFilter}, the current steeply breaks down when the frequency approaches that critical value. They "hesitate" between the electrodes.
The conduction speed sensitively depends 
on the temperature, and through it, the shape parameters~\cite{APTemperatureDependence:2012} of the \gls{AP} and especially the length of the "relative refractory" period~\cite{APTemperatureDependenceRefractory:2001} (although the causality is reversed).
The nature of the electrical signals differs: the sinusoidal 
current consists of a single frequency signal, while the 
noise contains also high-frequency components. The figures 
proves that "slow ionic current" moves in the neural circle, instead of that the circle is a low-pass filter (concluded from the wrong model).

\subsubsection{The 'outlaw' biology\label{sec:Physics-OutlawBiology}}

All conservation laws are correct
and valid for biology, only "the construction is different"~\cite{Schrodinger:1992} and the "test method" is wrong. 
Not mentioning that under some conditions,
the "system under test" also "produces" charge carriers by opening ion channels. The "construction" needs revisiting concepts of electricity
for living matter instead of projecting concepts valid for inanimate
matter only, to living matter. 
That wrong "grasp of concepts of electricity"  suggests that \textit{Ohm's Law is not valid for living systems}.
The correct statement is that \textit{Ohm's Law is not valid for arbitrarily chosen currents and voltages}.
Without that revisiting, by saying that 
living systems show \textit{non-Ohmic} behavior,  biology claims that it applies good laws inappropriately to a good system. 
The consequences are far-reaching: physiology uses magically
changing conductances in circuits with fixed-voltage batteries,
that blocks not only understanding the thermodynamic and mechanical effects (and introducing the corresponding theories~\cite{PerspectivesNerveSignalPropagation:2024,MechanicalPropertiesNerves:2025, CriticalElectricity:2018, MechanicalBrainPulses:2018, MechanicalCouplingNeuronalMembrane:2019,HeimburgPhysikOfNerves:2009,MechanicalWaves:2015, HEIMBURGReversibleHeatProduction:2021,LivingSystemPhysics:2021,ElectrochemistryMembrane:2022,ThermodynamicAPDrukarch:2022,NerveSignalAsWindow:2023}), but even consistently describing the electrical phenomena.
We are not speaking against using the concept of conductance, but rather against deriving its value incorrectly by including foreign current components, which obscures the actual electrical behavior of biological systems.
In addition, by introducing current feedback (which, by definition, compensates for any gradient in the system), physiology applies an opposite-phase control unit against the neuron’s native control unit and concludes that, after compensation, there is no gradient in the system; in other words, that \textit{life exists without needing a driving force}.

Neuroscience observes the thermodynamic and mechanical changes (for a comprehensive review, see~\cite{MechanicalPropertiesNerves:2025}) but cannot connect them to the electrical phenomena.
Lacking the required physical background, \textit{biology claims that physical laws cannot describe processes in living matter}, without providing a plausible and self-consistent theoretical model.

\section{Forming membrane's potential\label{sec:Physics-FormingPotential}}

In this section, we model the neuron as an electrical circuit that comprises an electrolyte condenser implemented by its membranes, surrounded by two electrolyte segments. The membrane comprises low-transmission-capacity ion channels in its wall and high-transmission-capacity channels toward its axon; it also uses slow ions rather than electrons.
The condenser produces a potential (that depends on the membrane's thickness and the concentrations in the two segments, as discussed above).  Furthermore, the two electrolyte segments produce Nernst voltages that are combined to yield a single voltage. 
According to Kirchhoff's Voltage Law, in a balanced state, the sum of those voltages between the intracellular and extracellular segments must be zero (see Fig.~\ref{fig:RestingPotential3}).
However, one must take into account the effects of the "slow current" (emulated by a
"fast current" with a plausible time course).
We omit biological regenerative processes (such as \gls{ATP} production via hydrolysis, ion pumps and channels functioning, and so on) and discuss only the most important physical primary processes.
In light of our findings, we must reconsider several key concepts in neurophysiology.

The electrical processes are inseparable from the corresponding thermodynamic processes~\cite{VeghNonOrdinaryLawsForLife:2025}, which we discuss by deriving an "equivalent thermodynamic electrical field" in section~\ref{sec:Physics-ElectrolyteInhomogeneity}.
The other minor effect is a low-intensity "passive flux" produced by the
"resting channels". They are different from the ion channels in the \gls{AIS}~\cite{AIS_NeuronalPolarity:2010, ActionPotentialGenerationNatrium:2008, BackpropagationAP:2012, AIS_Updated_Viewpoint:2018, AISStructureReview:2018}.
The "resting current" is about two orders of magnitude lower~\cite{EnergyNeuralCommunication:2021,NeuralEnergyConsumption:2017} than \gls{HH} assumed at the time of writing the book; correspondingly, its role is much smaller in the transient state, and it even drastically changes the electrical model used in producing an \gls{AP} which we discuss elsewhere~\cite{VeghDANCES:2026}.

\subsection{States\label{sec:PHYSICS_STATES}}

Figure~\ref{fig:NernstPlanckThermalWidth} shows, in the function of the membrane's thickness, the electrical field across the membrane due to electrical charges (it does not depend on the thickness of the membrane), for different physically plausible assumptions; furthermore, the thermal electrical field. The two electrical fields show significantly different dependencies on the neuron's parameters. Recall that although the parameters we used are biologically plausible, they are only estimations from non-dedicated, incomplete measurements. The qualitative conclusions, however, remain valid. 

\begin{figure}
	\includegraphics[width=.8\columnwidth]{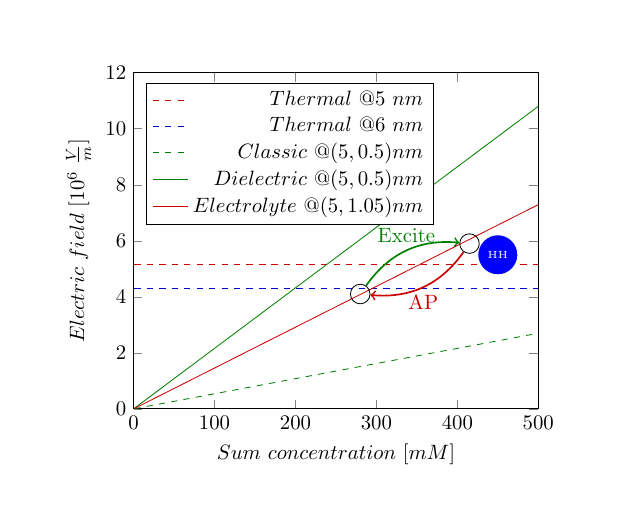}
	\caption{The electrical and the thermal "electrical fields" at different concentrations, see Eq.(\ref{eq:ElectricGradient}). The thermal electrical field varies with the thickness parameter and depends on the concentration; see Eq.(\ref{eq:NernstPlanckThermal2}). Balanced states are where 
		the sloping lines cross the horizontal ones. The dot marks where \gls{HH} provided measured data.
		\label{fig:NernstPlanckThermalConcentration}
	}
\end{figure}

The balanced state is set where the thermodynamic and electrical "electrical field" diagram lines cross each other; 
around $5\ nm$; the value we assumed in evaluating our equations.  
On the other hand, we can check the electrical field's dependence on the summed-up concentration of the ions in the segment (see Fig.~\ref{fig:NernstPlanckThermalConcentration})
(dehydration, for example, changes the setpoint through changing concentrations).
Here, the thermodynamic electrical field is constant but different at different width parameters. The field of the classic condenser does not provide a reasonably balanced state (a matching line). The dielectric diagram line seems to provide a realistic estimation 
when using biologically reasonable parameters. 
One can describe the neuron as a balanced system at the intersection of the sloping and the horizontal lines. In the resting state, only minor deviations (fluctuations) from this state exist
in both the electrical field and the chemical concentration.

In the transient state, if we assume that a charge transfer causes a significant potential change around $100\ mV$ in the layer of width 
$10\ nm$, it means a $10^7 \bigl[\frac{V}{m}\bigr]$ jump in the electrical field that is caused by a $100 \bigl[mM\bigr]$ jump in the concentration (for only a short time and only in the mentioned layer, but not in the bulk; see also Fig.~\ref{fig:RestingPotential4}). One can infer that a neuron can minimize its offset voltage (and so: its energy) by either decreasing its membrane's width, the sum concentration of the electrolyte, the electrical field, or by a process with a temporal course that changes their combination, thereby implementing a true multi-physics phenomenon. During issuing an \gls{AP}, the system finds its way back to the crossing point (the path in the phase space is not simple: concentrations on both sides of the membrane and the electrical field in the mentioned layers change, all having different speeds; electrical, thermodynamic, mechanical and so on, changes happen; the process is a relaxation (a waving), as it is seen in the shape of an \gls{AP}). 

We can imagine how the electrical process occurs in this 
controlled system (for visibility, the positions of the marks are not proportional to the mentioned numbers). The system is in a balanced state at a crossing point. The two circles connected by bent arrows represent that when an excitation begins at the point with lower
concentration and higher potential (the cytoplasm side and negative resting potential), ions rush into the surface layer, so the ion concentration increases, which increases the electric
field; this is seen that the membrane potential suddenly increases. (We note here that, due to ions' repulsion, the pressure also increases proportionally with the potential.)
The increased potential starts a (slow) current, 
and the intensity of the current through the channels in the
membrane's wall is much lower than the current of the drain (\gls{AIS})
so the system issues an \gls{AP} as it returns to its (balanced) starting point.
Given that it receives a large amount of $Na^+$ ions,
its primary effort is directed to get rid of them (the electrical gradient 
cannot distinguish the electrical charges by their chemical nature),
so the output current through the \gls{AIS} comprises mainly
$Na^+$ ions.
The system will move along a line, with the restriction that it is a movement in phase space (the change in concentration is not proportional to time), and the different points on the membrane's surface may have different potentials at different times (spatiotemporal behavior). Due to the capacitive current, the local potential may temporarily drop below the setpoint (as observed in physiology, this phenomenon is known as hyperpolarization). The time course of the processes is described by the laws of motion of biology~\cite{VeghNonOrdinaryLawsForLife:2025} (the time derivatives of the Nernst-Planck equation). The simplified model operates as described in section 4 of~\cite{VeghTechnomorphBiology:2025} and its computational details are provided  in~\cite{VeghNeuronAlgorithms:2025}.

\subsubsection{Resting state\label{sec:PHYSICS_RESTINGSTATE}}
In the resting state, the goal is to provide stability with slow, less intense currents. As~\cite{PrinciplesNeuralScience:2013} discusses, there are 'holes' (non-controlled ion channels) in the membrane where the counterforce is missing, so
the difference in the electrical and thermodynamic forces could accelerate ions until the Stokes-Einstein speed (see Eq.(\ref{eq:StokesEinsteinSpeeddV})) is reached.
Given that the deviation from the setpoint is slight, the current's intensity and speed are low. 
The slow ion transport delivers both charge and mass, so the gradients
permanently direct the process control variable toward the reference point. Given that the 'resting channels' are scattered on the surface of the membrane and it takes time for the gradients delivered by slow ions to reach the entrance of an ion channel, some slight variations 
in the actual value of the membrane's potential necessarily exist.
Given that the \gls{AIS} comprises non-gated ion channels with much higher density than those in the membrane, the variations and the control ("leaking") are marginal, and the current flows mainly through the ion channels in the membrane's walls, as observed in experimental physiology.

\begin{figure}
	\includegraphics[width=.8\columnwidth]{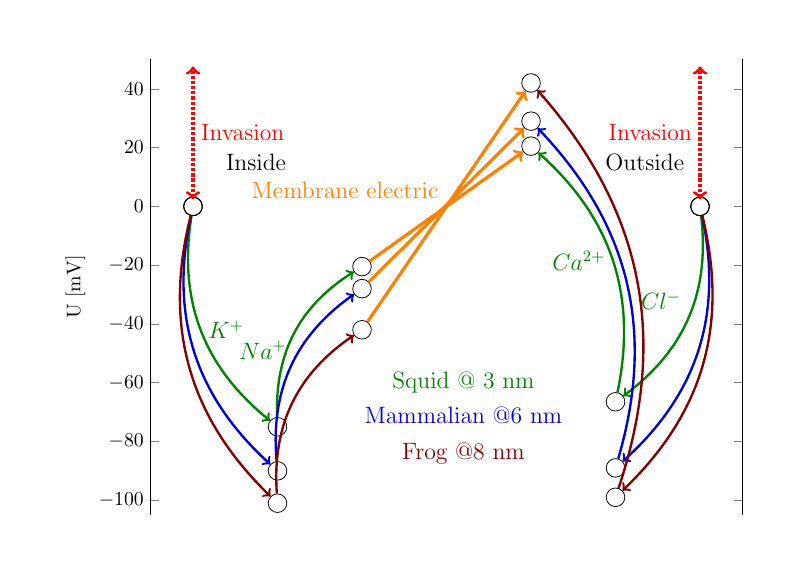}
	\caption{The mechanism for controlling neuronal operation is the same in all cases shown in Table~\ref{Tab:SummaryTable}. It appears that 
		the thicker membrane produces a higher membrane electrical potential, requiring a higher thermal potential to compensate. However, it can solve the task using a lower salt concentration, since the electrical membrane potential accelerates ion flow through the membrane channels, thereby speeding up the operation. The electrical fields for the different species have similar values, as shown in  Fig.~\ref{fig:NernstPlanckThermalWidth}.
		Furthermore, 
		the total concentration decreases as the membrane thickness and potential increase.  
		\label{fig:RestingPotential3}
	}
\end{figure}

In Fig.~\ref{fig:RestingPotential3}, the electrical voltage contribution is unidirectional. In contrast, the two-component thermodynamic contributions, which depend on the concentrations, are bidirectional.
If choosing the zero potential at the middle of the electrical potential, we can interpret that
the concentration combinations maintain their ionic concentrations (and so also thermodynamic voltage contributions) independently on both sides, starting from a zero potential level. They generate potentials autonomously in the segments proximal to the opposite membrane surfaces. The electrical charge on the membrane generates an additional voltage independently. However, that voltage must fit between the two thermodynamic voltages. Any deviation in potential between the intracellular and extracellular sides initiates a current that drives the concentrations and voltages toward a balanced state.
The process is complex: the sum of surface concentration of ions must be the same on both sides for the condenser, and the ratios on the intracellular and extracellular sides must be adjusted to satisfy the conditions 
\begin{align}
	U_{internal}^{K^+,Na^+} &= -U_{electric} &+ U_{internal}^{K^+} &+ U_{internal}^{Na^+}&(+U_{internal}^{invasion})\\
	U_{external}^{Cl^-,Ca^{2+}} &= U_{electric} &- U_{external}^{Cl^-} &- U_{external}^{Ca^{2+}}&(+U_{external}^{invasion})
\end{align}

In a balanced state, the left sides are zero: at the contact point between the electrolyte and the membrane,
there are no potential differences, while in a perturbed state, they are non-zero and so they represent driving forces for restoring the balance. (Recall that the ions are slow and that the local electrical and concentration gradients control the local potentials.)
From the derivation, it follows that
\begin{align}
	U_{electric} &= - 2*(U_{internal}^{K^+} &+ U_{internal}^{Na^+}) \\
	U_{electric} &= - 2*(U_{external}^{Cl^-} &+ U_{external}^{Ca^{2+}}) \\
	U_{electric} &=  U_{internal}^{K^+} &+ U_{internal}^{Na^+} 
	&= - (U_{external}^{Cl^-} &+ U_{external}^{Ca^{2+}})\quad
\end{align}

Our statement is the precise mathematical formulation of that
"Thus ions are subject to two forces
driving them across the membrane: (1) a \textit{chemical
driving force}, a function of the concentration gradient
across the membrane, and (2) an \textit{electrical driving force},
a function of the electrical potential difference across
the membrane."~\cite{PrinciplesNeuralScience:2013}, page~129.
Fig.~\ref{fig:RestingPotential3} quantitatively underpins that "action potential generation in nearly all types of neurons and muscle cells is accomplished through mechanisms similar to those first detailed in the squid giant axon in early research by \gls{HH}"~\cite{MembranePotentialMoleculesNetworks:2014}. \textit{In all species, the chemical and thermodynamic processes keep a balance in the cells of different construction in both resting
and transient states.}

According to the above derivation, we consider three characteristic segments in the neuron. In the middle, we have an electrostatically charged isolator plate with positive and negative charge carriers on its two surfaces and none inside, so it has a homogeneous electrical field. The magnitude of the field depends only on the total concentration of the corresponding ions in the segments. On both sides are ionic solutions of different compositions. If the membrane is permeable for the ions, they can move from one segment to the other as the resultant field dictates,
until the fields get balanced.

The membrane's electrical potential defines the resting potential, and the "electrical" and "thermodynamic" potentials work together to control neuronal operation.  
The anatomically defined "electrical" membrane potential, combined with the vast number of ions in the extracellular space, provides a stable concentration that serves as a solid reference point (a setpoint).
In the resting state, the membrane potential is balanced by the presence of always-open "resting ion  channels". When ions flow
into the membrane, the "resting ion  channels"
can keep the balance: the internal $K^+$ and $Na^+$ concentrations are adjusted as the $Na^+$ flows in.

Figure~\ref{fig:RestingPotential3} also shows that
in its native state, the neuron is electrically balanced: in the
internal and external segments, the resulting thermodynamic and electrical forces are equal.
Any foreign invasion into the neuron's life (chemical, mechanical, or electrical, including clamping) changes the left-side reference value to
an externally set value. The neuron must adjust its $Na^+$ and $K^+$ concentrations to reach a new balance while the invasion persists. That is, any invasion ("testing in the physical laboratory") changes the concentrations and voltages 
inside the cell. In addition, the internal biological processes can trigger internal processes that contribute (from the measurement's point of view) foreign currents or processes, see also section~\ref{sec:TransientState}. The feedback from the clamping methods introduces a compensating current. Even a simple conductance-measuring device is active: it applies a test voltage that generates a test current, which can sensitively affect the ionic composition of the neuron (recall that concentrations may differ by up to six orders of magnitude). Although modern electronic devices attempt to conceal this effect, they subtly influence biological operation.

From our result, it follows that for more chemical elements, a per-element coupled set of equations exists, plus the sum concentration
agreed electrical fields that are valid simultaneously and define the process variable (aka membrane potential) of the neural regulatory circuit.
In a more symmetrical form

\begin{align}
	0 &=\frac{U_{electric}^{Total}(c,\Delta z,d)}{2} + \sum_k U_{thermal}^{C_k}(d) + {invasion}
	\quad for\ all\ k;\ inside\ and\ outside \\
	c &=\sum_k C_k^{ext} =\sum_k C_k^{int}	\label{eq:coupling}
\end{align}
That is, we have an unidirectional electrical potential difference and oppositely directed thermal voltages. In the balanced state, the two potentials must be equal.
When changing, say, one concentration, the other concentration on the same side, and the surface charge density of the condenser, changes accordingly.  The changed electrical voltage 
re-adjusts the thermodynamic voltage and, consequently, the concentrations in the other segment.
There is no simple way to express, say,
$[Na^+]_i$ in the function of $[K^+]_i$ or vice versa, and similarly for the other ions. Everything moves.
The size matters: the amount of ions in the external segment is many orders of magnitude higher, and the gradients are inversely proportional to that amount.

As the figure suggests, an external invasion, typically an electrical voltage on the intracellular side, changes the balanced state, and due to the parameters being linked,
changes all concentrations. When moving the system out of
its balanced state, in any way, a driving force appears that moves the system towards finding a new balanced state. However, recall that the ions are slow, so \textit{the changes are not instant}.

Table~\ref{Tab:SummaryTable} shows our results, without fitting parameters, for three different
famous published cases against available experimental data (Table 2.1 in~\cite{JohnstonWuNeurophysiology:1995}). It looks like the calibration of the concentration 
of negative and positive ions needs scaling, and the measurement of $Ca^{2+}$
comprises issues (the $Ca^{2+}$ concentration in all cases is inaccurate and had to be corrected).

\renewcommand{\arraystretch}{1.3}
\begin{table}
	\caption{The summary data table for equilibrium (Experimental data from Table 2.1 in\cite{JohnstonWuNeurophysiology:1995} ) \label{Tab:SummaryTable}}
	\begin{tabular}{ |p{0.6cm}||p{0.7cm}|p{0.9cm}|p{3.0cm}| p{0.85cm}| p{0.85cm}| p{0.85cm}|p{0.6cm}|p{0.85cm}|}
		\hline
		\hline
		Ion& Inside &Outside &  $U_{therm}\frac{RT}{zF} log\bigl( \frac{[C]_{o}}{[C]_{i}}\bigr)$ &$U_{electr}^{theor}$&$U_{electr}^{exp}$&$U_{membr}$&Width&$E_{membr}$\\
		& [mM] & [mM] &  [mV] &  [mV] &[mV]&[mV]&[nm]&[MV/m]\\
		\hline
		&\multicolumn{6}{|c|}{Squid (\gls{HH}) \quad T=293\ $K^o$}&\textbf{3}&14.0\\
		\hline
		\em	$K^+$ & 400 & 20 & $-75.5= 58 log\bigl( \frac{20}{400}\bigr)$& \textbf{41.9}&\textbf{41.4}&-62.0&&\\
		$Na^+$ & 50 & 440 & $+54.8= 58 log\bigl( \frac{440}{50}\bigr)$&&&&&\\
		$Cl^-$ & 40 & 560 & $-66.5= 58 log\bigl( \frac{560}{40}\bigr)$&\textbf{41.9}&\textbf{42.0}&63.0&&\\
		$Ca^{2+}$ & $0.27^*$ & 10 & $+45.5= 29 log\bigl( \frac{10}{0.27}\bigr)$&&&&&\\
		&\multicolumn{8}{|l|}{$^*$ used for the published value 0.4} \\
		
		\hline
		&\multicolumn{6}{|c|}{Frog muscle (Conway) \quad T=293\ $K^o$}&\textbf{8}&9.61 \\
		\hline
		$K^+$ & 124 & 2.25 & $-101= 58 log\bigl( \frac{2.25}{124}\bigr)$&\textbf{88.9}&\textbf{83.6}&-125.4&&\\
		$Na^+$ & 10.4 & 109 & $+59.2= 58 log\bigl( \frac{109}{10.4}\bigr)$&&&&&\\
		$Cl^-$ & 1.5 & 77.5 & $-99.4= -58 log\bigl( \frac{77.5}{1.5}\bigr)$&\textbf{88.9}&\textbf{82.7}&124.1&&\\
		$Ca^{2+}$ & $0.021^*$ & 2.1 & $+58.0= 29 log\bigl( \frac{2.5}{0.25}\bigr)$&&&&&\\
		&\multicolumn{8}{|l|}{$^*$ used for the published value 4.9} \\
		
		\hline
		&\multicolumn{6}{|c|}{Typical mammalian cell \quad T=310\ $K^o$}&\textbf{6}&11.1 \\
		\hline
		$K^+$ & 140 & 5 & $-89.7= 62 log\bigl( \frac{5}{140}\bigr)$&\textbf{57.7}&\textbf{57.3}&-85.8&&\\
		$Na^+$ & 15 & 145 & $+61.1= 62 log\bigl( \frac{145}{15}\bigr)$&&&&&\\
		$Cl^-$ & 4 & 110 & $-89.2= -62 log\bigl( \frac{110}{4}\bigr)$&\textbf{57.7}&\textbf{56.9}&85.9&&\\
		$Ca^{2+}$ & $0.04^*$ & 5 & $+60.8= 31 log\bigl( \frac{5}{0.04}\bigr)$&&&&&\\
		&\multicolumn{8}{|l|}{$^*$ used for the published value 0.0001} \\
		
		\hline
		&\multicolumn{6}{|c|}{$Na^+$ ions only~\cite{OriginMembranePotential:2017}\quad T=310\ $K^o$}&6&10.0 \\
		\hline
		$Na^+$ & 12 &145 &$+66.6= 62 log\bigl( \frac{145}{12}\bigr)$&\textbf{60.7}&\textbf{66.6}&&&\\
		\hline
		\hline
	\end{tabular}
\end{table}

When calculating the values in Table~\ref{Tab:SummaryTable}, we used $\Delta z=1.05$ and employed the membrane thickness data from various publications to demonstrate that our theoretical approach is correct and that using the correct data can yield even the absolute values of the membrane potential.   
The higher-than-expected value of $\Delta z$ may suggest that the ions and the membrane's double layers in the electrolyte may also play a role. 
Dedicated, complete investigations can reveal the details.

\subsubsection{Transient state\label{sec:TransientState}}

\begin{figure}
	\includegraphics[width=.8\columnwidth]{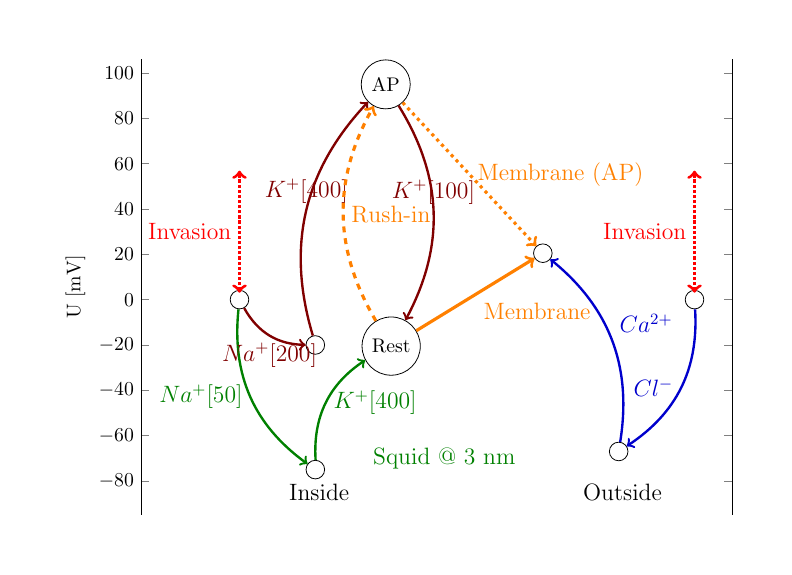}
	\caption{The mechanism of producing \gls{AP} (using numbers referring to the case of squid shown in Table~\ref{Tab:SummaryTable}). In the resting state, the inside thermal potentials follow the green paths.
	When the rush-in of $Na^+$ ions suddenly increases the internal concentration to $[200]$, the potential on the internal side of the membrane increases to $+90\ mV$. The neuron must decrease its concentrations to restore the resting potential, mainly by releasing an action potential through the \gls{AIS}.
	\label{fig:RestingPotential4}
	}
\end{figure}

In a special type of invasion, when suddenly a large amount of $Na^+$ rushes into the segment (i.e., the trigger is external, but the ions derive from a biological process),
the current throughput of the "resting ion channels" 
is not sufficient anymore: the internal $Na^+$ concentration in the proximal layer drastically increases, and so does the 
membrane's potential. At that point, the neuron is out of balance: although most of the $Na^+$ ions can
flow out through the \gls{AIS}, the system also changes its $K^+$ concentration through the limited capacity "resting ion channels" until the balanced state is regained. This latter effect must not be confused with the capacitive current, which (owing to its reversed direction) causes the phenomenon known as "hyperpolarization".

Compare the figure to Fig.~7.2 of~\cite{PrinciplesNeuralScience:2013}.
Notice the two-sided negative feedback effect: a change in the value of the 
"command potential" or "command concentration" (the "setpoint" in the language of control theory) changes the other value.
In the case of voltage invasion, the ion concentration ratio also changes. In this case, the feedback amplifier is represented by the 
delicate balance of the electrical and thermodynamic force fields.
In the case of clamping, another (electrical) feedback circuit is introduced
into the system, and they work against each other. The neuron attempts 
to restore its balance without external potential (it cannot distinguish the foreign potential from its membrane potential), and changes its concentrations accordingly.
A vital difference is the \textit{speed} of the feedback amplifier. The electrical one operates at electrical speed, while the thermodynamic one operates at a speed several orders of magnitude lower.
When switching (or changing) the external ("command") voltage,
the neuron must first get into its "steady state". It takes time.
"The command potential, which is selected
by the experimenter and can be of any desired amplitude
and waveform"~\cite{PrinciplesNeuralScience:2013}. However, if the measured data readings are not from a steady state,
the measurement is wrong: the different amplifier speeds (different current speeds) surely distort the measured value.

Figure~\ref{fig:RestingPotential4} shows the schematic course of potentials and concentrations during issuing an \gls{AP}. 
Follow the green path: the concentration ratio of 
$Na^+$ and $K^+$ ions define a thermodynamic contribution in the intracellular segment. Follow the 
blue path: the $Cl^-$ and $Ca^{2+}$ ions define a thermodynamic contribution in the extracellular segment.

On the left side, two positive ions are in the solution, and the membrane potential (due to charge separation) rises toward the outside segment; the ions cannot go "uphill". Similarly, 
the $Cl^-$ ions cannot go uphill (due to their negative charge, the same potential is also rising for them). This single membrane potential keeps all ions in their segments. However, the $Ca^{2+}$ ions can travel "downhill".
Calcium pumps are needed: although in minimal concentration, $Ca^{2+}$ ions flow into 
the intracellular segment, and they must be pumped out (BTW: this is likely also the reason why measuring $Ca^{2+}$ concentration is inaccurate).

As discussed above, 
the concentrations and voltages in the extracellular segment, with their vast amount of ions, remain unchanged
in resting and transient states.
Follow the red path: $Na^+$ ions rush into the intracellular segment.
They increase the concentration from $[50\ mM]$ to (say) $[200\ mM]$
(from point (Rest) to point (AP)). This sudden change increases the thermal contribution of $Na^+$ potential to $[-20\ mV]$, but 
because the total charge on the membrane significantly increases, the resulting voltage on the intracellular side of the membrane increases to $[90\ mV]$.
The neuron could decrease its potential to the resting value if it could decrease the $K^+$ concentration to $[100\ mM]$. However, the ion-delivery capacity of the "resting ion channels" is insufficient: they are sized to maintain the resting state.
(Furthermore, the electrical contribution appears instantly, while the thermodynamic one, due to the finite speed of ions, only with some delay~\cite{VeghNonOrdinaryLawsForLife:2025}.)
Follow the dashed orange path: the large amount of rush-in $Na^+$ ions drastically increases the surface concentration on the membrane, so the potential suddenly changes to $[90\ mV]$. Follow the dotted orange path: the electrical field across the membrane suddenly changes its direction from positive to negative (see the slopes of the solid and dotted orange lines) so that the excess positive ions can move "downhill" out of the intracellular space through the high-conductance ion channel array (\gls{AIS}) at the end of the neuron.
Its internal end is at the high membrane potential,
while the external end is at the low potential of the extracellular segment.

The setpoint on the extracellular side remains fixed.
On the intracellular side, at the beginning of the \gls{AP}, the actual voltage 
is above that of the setpoint on the extracellular side, so at the two ends of the \gls{AIS}, a positive driving force moves the ions toward the downstream neuron. 
Given that positive $Na^+$ ions leave the intracellular segment, its potential continuously decreases. At some point, 
\textit{the voltage drops below the setpoint of the extracellular side,
and the electrical field across the membrane reverses}. 
The driving force also decreases, and a low-intensity current
restores the balanced state to the previous setpoint.
When measuring at the \gls{AIS}, one observes that the
potential rises suddenly, then decreases below the setpoint (hyperpolarizes the membrane), then a decreasing current (flowing in the opposite direction due to the changed slope of the resulting potential, the changed electrical field across the membrane,
but comprising the originally rushed-in $Na^+$ ions) slowly restores the resting potential.
Here comes the finite speed of ions again. 
The ions can follow the potential changes with a delay
due to their finite speed, causing an apparent delay in the time course of the \gls{AP}'s current. 

Models in neuroscience (as reviewed in~\cite{BrainNetworkModels:2018})
almost entirely ignore these aspects. In our physical model, we see
that the measurable membrane potential and current change in the function
of the ions' speed, the concentration, and its time derivative.
Furthermore, all mentioned quantities depend on the effective potential. 
As Eq.~(\ref{eq:NernstPlanckThermal1}) shows, the "thermodynamic electrical field" increases
as the temperature increases, and raises the setpoint.  As Eq.(\ref{eq:StokesEinsteinSpeeddV})
shows that the current decreases as the temperature increases, in line with the experimental evidence~\cite{APTemperatureDependence:2012,APTemperatureDependence:1985}.
The amplitude of the action potential is decreased, given that the setpoint has increased, and its duration is reduced, given that the current decreases.

The thickness of the charged layer significantly affects the resulting electrical field (i.e., the neuron's 'setpoint'). Suppose biological fragments are formed and settle on the membrane's intracellular side due to the increased $\Delta z$. In that case, the resting potential (and the threshold potential) may change, potentially leading to neurological diseases.
For example, "in brain tissue from Alzheimer’s disease patients \dots the most common changes were a proximal shift or a lengthening of the AIS", so, “Structural alterations of the \gls{AIS} are likely to have an impact on normal neuronal activity”~\cite{AlzheimerDisease:2022}.
Lengthening means that the resistance of the 
\gls{AIS} increases that changes the parameters of the \gls{AP};
see Eq.(\ref{eq:AIS_Voltage}).
As~\cite{TemperatureOnAP:2022} observed, the setpoint changes with temperature (the "thermodynamic electrical field" changes with the temperature, while the "electrical electrical field" does not, so their crossing point changes), reflected in the dynamics of the \gls{AP}.

\subsection[Controlling states]{Controlling neuronal stages\label{Physics-ControlTheory}}

From control theory, it is known that the goal of a controlled system is to
govern the application of system inputs to drive the system to a desired state while minimizing delay and overshoot, steady-state error, and ensuring control stability.
The neuron implements a controller that monitors the controlled process variable (membrane voltage) and compares it with the reference or setpoint (resting potential).
Recall that the geometry and the composition and concentration of electrolytes define the setpoint. 
The difference between the process variable's actual and desired values, known as the error signal, represents the actual offset potential.
It is applied as feedback to generate a control action that brings the controlled process variable to its setpoint.

\subsubsection{Neuron as a PID controller\label{Physics-PIDController}}

Biology uses a simple \gls{PID} controller with proportional, integral, and derivative components. 
In simple words, it means that the output of the system (the output voltage measurable on the \gls{AIS}) depends on the input voltage (the present: input current if the resistance is constant),
the integrated input current(s) (the past), and their derivatives (the future). Notice that calculating the derivatives of the interdependent parameters needs care, as discussed in~\cite{MechanicalPropertiesNerves:2025}. The sum of those summands defines the resulting output voltage. 
The biological case is more complicated than the technical one because biology works with slow ions, and the components have
their temporal dependence (applying the laws created for fast currents requires emulation, as discussed in section~\ref{sec:Physics-FormingPotential}). Furthermore, the summands have several constituents, and the neurons handle those current-related constituents autonomously.
That is, unlike what is assumed in the classic physiology,
the output voltage is not necessarily a simple
function of the mentioned input variables.
Furthermore, it is not sufficient to consider the "present" of the system.
The classic models all omit the derivatives' contributions (a direct consequence of using 'clamping': in that "frozen" state, the "future" cannot be interpreted)
except that of the capacitive current, which is the passive consequence of the input currents.
Furthermore, the 'Integrate and Fire'-type models
consider the integrated currents only in the 'charge-up' ('Computing') period of operation, and they entirely omit that different physical processes are going on in different stages of operation.
That is, \textit{even the best classic models can not describe neuronal operation completely and correctly}.

The gradients are used to adjust the process variable through their positive and negative contributions (corresponding to rising and falling edges), and the different speeds of the thermodynamic and electrical interactions minimize the delay (i.e., provide the maximum operational speed, vital for survival).
The steady-state error is minimized by setting the process variable to the reference point using long-term stable parameters (geometry and overall concentration). The low-intensity current through the always-open resting ion channels provides dynamic stability in the steady state.

We considered that the neuron has a stable base state. On the one side, this resting state must be dynamically stabilized for little perturbations using as little energy as possible
(and to provide a mechanism when the cell grows, divides, or ages). On the other side, it must be able to restore the state after rough perturbations as quickly as possible, causing short-time transients (when restoring the
membrane's potential after issuing a spike).
In both the resting and transient states, the system attempts to return to its balanced state, though the mechanisms required differ. 

In its excited state, the system aims to provide an intense output that informs the downstream neurons.
This overshoot is initiated by a large number of voltage-gated ion channels (distinct from the non-gated ion channels used to maintain the resting potential) distributed in the neuron's membrane.
The overshoot current flows through those persistently open ion channels, which are concentrated in \gls{AIS} (which has about two orders of magnitude higher channel density than the wall).
The intense slow current produces a condenser-like behavior (capacitive current); the phenomena called "polarization" and "hyperpolarization" (not polarization; instead, a movement of completely separated charge in the surface layer of the electrolyte) of the membrane provide the necessary positive and negative error signals for the controller in the transient state by moving the actual potential value of the membrane above and below the resting potential value. The potentials and the electrical field's magnitudes 
depend on the concentration and the geometry (the finite size of the membrane). The ions' chemical nature comes into play only if the polarizability can differ 
for different molecules.

\begin{figure}
	\includegraphics[width=0.8\columnwidth]{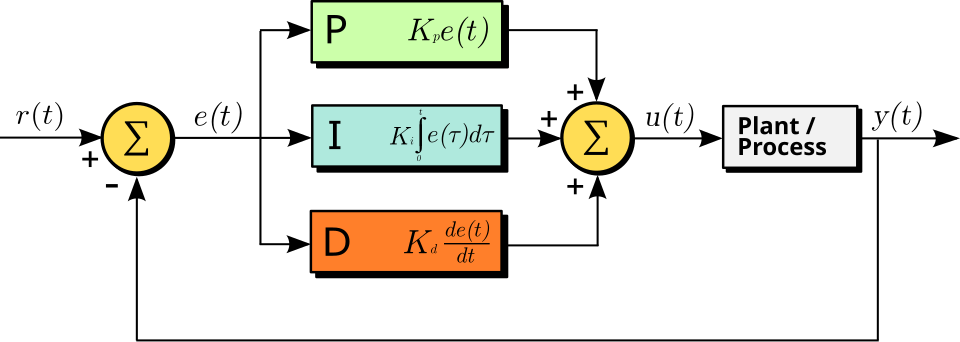}
	\caption{A block diagram of a PID controller in a feedback loop. r(t) is the desired process variable (PV) or setpoint (SP), and y(t) is the measured PV. (Wikipedia)
		\label{fig:PID_Controller}
	}
\end{figure}

	The time course of the overall control function $u(t)$ of the general \gls{PID} controller in function of the error variable $e(t)$ in Fig.~\ref{fig:PID_Controller} is
	described by
	\begin{equation}
		u(t)= K_p\biggl( e(t)
		+ \frac{1}{T_i} \int_0^t e(\tau) d\tau
		+ T_d \frac{de(t)}{dt}\biggr)
		\label{eq:PID}
	\end{equation}
	where $T_i=\frac{K_p}{K_i}$ is the time integration constant, $T_d=\frac{K_d}{K_p}$
	is the derivative time constant (notice the independent time constants).
	That is, in the world of a neuron, the "theory of everything" is
	\begin{equation}
		V_M^{OUT}(t)= K_p\biggl( \underbrace{\overbrace{V_M^{IN}(t)}^{proportional}}_{clamping,\ synaptic}
			+ \overbrace{\underbrace{\frac{1}{T_i} \int_0^t V_{M}^{IN}(\tau) d\tau}_{membrane's\ wall}}^{integral\ (parallel\ RC)}
		+ \overbrace{\underbrace{T_d \frac{dV_M^{IN}(t)}{dt}}}^{derivative\ (serial\ RC)}_{AIS,\ synaptic}\,\biggr)
		\label{eq:PID_Neuron}
	\end{equation}
The first term represents the constant effect of the external world (including synaptic inputs, clamping, and excitatory effects on brain tissue).
	The second term represents the time-averaged contribution, such as 
	charging up the membrane, including clamping current. The voltage is mainly due to current through ion channels in the membrane's wall; it represents a parallel $RC$ circuit. The third term represents
	the sum of the gradients due to all effects (the external world, the internal processes, and the outflow through the \gls{AIS}); it represents a serial $RC$ circuit.
The underbraces of the equation identify which components of the neuron
contribute the terms of the \gls{PID} equation.
The presence of some terms depends on the neuron's actual stage; furthermore, the $R$ values in the parallel and serial $RC$ circuit differ
(resulting in different time constants).

One can reformulate the \gls{HH}'s famous Eq.~(1) in~\cite{HodgkinHuxley:1952} to the form of the \gls{PID} equation
	\begin{equation}
		\underbrace{I \times R_{AIS}}_{V_M^{OUT}} = 
		\overbrace{\underbrace{I_i\times  R_{AIS} }_{ionic}}^{proportional}
		+\overbrace{
				\frac{1}{\underbrace{C_M\times R_{AIS}}_{T_i}} \int_0^t \xcancel{(I_{bio}-I_{clamp})(\tau)} d\tau
		}^{integral,\ clamping\ cancels\ this}
		+\overbrace{\underbrace{C_M\times R_{M}}_{T_d}\frac{dV_M}{dt}}^{derivative,\ capacitive}
		\label{eq:HH_PID}
	\end{equation}
The proportional term comprises the ionic (and external) currents.
At the time when Eq.~(\ref{eq:HH_PID}) was set up, \gls{AIS} was not yet known,
so \gls{HH} assumed that in the proportional term, the resistance is identical to the membrane's resistance implemented by the ion channels in the wall (leading to the ideas of "leakage current" and "resting potential", misleading physiological research.)
The experimental conditions canceled the integral term entirely: the negative feedback from clamping precisely counterbalances the neuron's internally generated current, so the current integrates to zero.
(BTW: the "integrate and fire" type models artificially revive the forgotten
integral term.)
The derivative term assumes that all currents are constants,
so the voltage changes only due to the capacitive current.
It entirely forgets that gradients are needed to operate an
electrical system (leading to the ideas of constant-voltage batteries
and magically operated resistors in the equivalent circuits). 
Although it was known from the beginning that the \gls{AP}s comprise relatively steep rising and falling edges, it was forgotten that a non-constant current contributes a gradient to the derivative term.
\gls{HH}'s formalism describes only the 'present' (as can be expected
when freezing the state), has no predictive power, nor can it give an account of neuronal memories.

\gls{HH} considered only the proportional term, plus in the derivative term, the voltage gradient due to the condenser (but not the capacitive current itself, directly leading to the wrong ad-hoc hypothesis of the presence of $K^+$ for explaining hyperpolarization). Given that clamping obscures the presence of
gradient-like changes in the system, they missed that \textit{the input currents also produce a voltage gradient and that the output current through the \gls{AIS} also produces a gradient}. Furthermore, the equation shows that external currents (e.g., the rising and falling edges of step functions) also cause significant changes in the output voltage.
Similarly, a "foreign" invasion (changing mechanically the positions of charges by pressure, ultrasound, magnetic pulse, changing the concentration in one of the segments, and so on) generates a change in the position of charges on the neuronal membrane. This way, it generates a voltage gradient (along with pressure and other gradients), which may trigger neuronal spikes.
As discussed in section~\ref{sec:AP-PhysicalProcess},
subthreshold excitations provide direct experimental evidence that the description of the resting state includes both serial and parallel contributions.

From a control theoretical point of view, the condenser geometry and the ion concentrations set
the 'always the same' value (the setpoint; see also Figure~\ref{fig:NernstPlanckThermalWidth}) and the currents serve to keep or restore
that value.
After introducing a finite-width membrane into an electrolyte,
a potential difference between the two membrane surfaces is created; see Eq.~(\ref{eq:UGapTotal400}).
When adjusting the membrane's potential, we must consider its ground and excited states separately,
given that the perturbations in these two cases are vastly different. Nature employs various mechanisms to maintain the ground state and recover it after generating a spike. Although nature's tools are remarkably similar (channels and pumps; here, for discussing 'downhill' channels, we must consider the charge-up of different components), the significant difference in current intensity warrants a separate discussion.

When comparing Eq.~(\ref{eq:RC_Circuit_Output}) describing the operation of a simple serial $RC$ oscillator taken from the theory of electricity,  Eq.~(\ref{eq:HH_PID}) describing \gls{HH}'s differential equation
(having \gls{PID} in mind),  and the complete Eq.~(\ref{eq:PID}) describing \gls{PID} in general, one can see that they describe the 
same process in different (experimental and mathematical) approximations.
As experience shows, Eq.~(\ref{eq:RC_Circuit_Output}) describing 
the 'net electric' model provides a sufficiently accurate description
of the \gls{AP}, proving that the dominating effect comes from the derivative term and that the thermodynamic term is linearly proportional to the
electrical one. Furthermore, as we emphasized by discussing the 
physical processes and the mathematical terms, to some measure, 
the "parallel $RC$ circuit" is also present in the process, although
its current amplitude is by two orders of magnitude 
lower than that of the "serial $RC$ circuit". Furthermore, the time constants of the two circuits are also largely different, making the effect of the parallel circuit unnoticeable. 
The slow current also plays a role here: the ion currents are 'local'. 
Given that the channels in the membrane's wall are always open, the current on its way towards the \gls{AIS} may flow out
through a nearby ion channel, so only a fragment of the 'resting current' reaches the \gls{AIS}.
Eq.~(\ref{eq:PID})
provides a high-accuracy description of the process, but for most 
practical applications using Eq.~(\ref{eq:RC_Circuit_Output})
provides sufficiently good results, while Eq.~(\ref{eq:HH_PID}) is based
on an incorrect 'physical model' and requires a series of incorrect ad-hoc hypotheses to describe observations.
Note that the forces have both electrical and thermodynamic components, but they are proportional to each other. That is, although the absolute values of the coefficients in the equations are not accurate, their effects are.
Surely, the description of the processes
is complicated, but appropriate approximations give sufficiently
precise and computationally reasonable results.

Notice, that the equations describing a \gls{PID} circle
are valid for "fast" currents, so applying those formulas for
neurons need adaptation. Furthermore, in the two states (and due to the finite speed of the current: the intermediate state)
different terms of  Eq.(\ref{eq:PID_Neuron}) dominate.

\subsubsection{Robustness of the control circle}

An exciting thought experiment is discussed in~\cite{OriginMembranePotential:2017}. "First, we consider a hypothetical membrane that is permeable only to $Na^+$ ions. Suppose that $[Na^+]_o$, the outside or extracellular sodium concentration, is $145\ mM$, and $[Na^+]_i$ is 12 mM. Moreover, suppose that initially, there is no membrane potential. The diffusion gradient for $Na^+$ favors $Na^+$ entry into the cell, and the initial $Na^+$ influx carries a charge that builds up on the inside of the cell. This produces a potential (recall that separation of charge produces a potential) across the membrane that impedes further $Na^+$ ion movement because the positive charges repel the positively charged $Na^+$ ion."

The paper correctly explains that in the case of $Na^+$ only, the potential is set up
to $66.6\ mV$. (We quietly add that plus a layer of the corresponding $Cl^-$ ions on the other segment, in a similar concentration ratio, builds up; so also, the attraction between positive and negative charges contributes to the effect.) 
In the absence of our argumentation, it cannot explain \textit{why} 
that potential builds up; only why that concentration ratio
establishes: the Nernst law alone does not explain the effect. Furthermore, the statement is valid only for
a well-defined semipermeable membrane.
We can add that the same concentrations develop if initially identical concentrations
are present on both sides. 
The finite width of the membrane creates an electrical potential, a consequence of charge separation, as stated. 
If there are no ions initially inside, then no charge separation occurs, and therefore, no ion transport begins due to the membrane's electrical potential. However, the thermal driving force exists, and the started ion transport quickly builds the electrical layer, which effectively helps to establish the resting potential. 

This robustness also explains why, during replication (cell division), cells also inherit their operating ability and resting potential; the existing membrane cell continues to grow in the same way. Its voltage does not depend on its size, so it remains constant, and the voltage defined in this way adjusts the concentrations accordingly. Moreover, it explains why, during evolution, cells first formed: a couple of lipids came together to form an imperfect membrane with holes, establishing cellular electricity. As Fig.~\ref{fig:RestingPotential3} depicts, the concentrations may adapt to the living conditions (whether living in seawater), but the general operating principles remain the same. The seawater environment 
enables the use of thinner membranes and lower membrane voltages; however, the same electrical field is used in all cases.

Given that neurons share the external (global) concentrations (defined by the vast amount of ions in the bulk), they provide a firmly fixed offset value to the always-the-same electrical potential. This way, the operations of the individual neurons do not interfere. Furthermore, the internal (local) concentrations can adjust themselves even following a very rough charge perturbation
while issuing an \gls{AP}; moreover, when the living organism is growing.

\subsection{Goldman-Hodgkin-Katz potential\label{sec:Goldman-Hodgkin-Katz}}

It has already been stated, based on experimental evidence, that "the membrane permeability to the ions has nothing to do with the potential generation and the ions' adsorption on the membrane surface generates the membrane potential"; for a review, see~\cite{MembranePotentialPermeability:2024}.
Furthermore, the model is mathematically inconsistent in some important cases~\cite{RevisitingGHK:2025}; includes three arbitrarily chosen ions (an insufficient set) and an unmeasurable parameter "mobility".
The idea itself is nonsense: in a balanced state, no ion transport happens, so even without permeability, the balanced state persists;  the resting potential has nothing to do with either ions' mobility or permeability or with ion absorption~\cite{Hodgkin-HuxleyAdsorption:2021}.
Furthermore, there is no idea in \gls{GHK} about whether the 'setpoint' (why that specific concentration or potential difference) is present and why the same potential is reset after rough perturbations such as issuing an \gls{AP} or replicating a cell.
The causality is reversed: the potential is a static concept and is set electrically, and
the two gradients form the experienced concentrations from the available solvent molecules.
For the time course of gradient formation, mobility and permeability play a role, but not in determining concentrations or the resting potential.

In our clear physical picture, the thermodynamic forces on one side
and on the other side of the membrane are summed. They counterbalance each other; furthermore, jointly the effect of the neuronal condenser, see  Fig.~\ref{fig:RestingPotential3}.
As we emphasized, the concentration gradients are ion-specific.
Furthermore, as discussed in connection with Eq.(\ref{eq:Nernst1}), the Nernst equation comprises a per-ion indefinite constant (a potential difference).
To calculate a linear combination of terms comprising an arbitrary constant is nonsense,
and so is adding absolute concentrations on the different sides of the membrane, or changing the base of the logarithm used in a calculation to match
the experimental value. It is not more than number magic. The potential is described by coupled equations as discussed in connection with Eq.~(\ref{eq:coupling}).

The $Ca^{2+}$ ions do not fit into the \gls{GHK} picture. 
One of the reasons why \gls{GHK} cannot be good is that $[Ca^{2+}]$
is omitted. Biologically, it is hard to believe that $Ca^{2+}$ does not participate in the game of life (especially since biology sees the need for $Ca^{2+}$ pumps). Physically, a single concentration on one side of the membrane cannot maintain balance; as the different ion-exchange processes shown in Table~\ref{Tab:SummaryTable} and Fig.~\ref{fig:RestingPotential3} demonstrate.
When adding a new ion to the solution, the sum concentration
increases, and so the electrical force increases, forcing the previous 
elements to find new concentrations on both sides.
The appearance of a new chemical element indirectly changes the concentrations of the others (a good example is the role of the negligible amount of $Ca^{2+}$).

\section{The Action Potential\label{sec:ActionPotential}}

As discussed, a smaller number of ion channels is distributed over the surface of the membrane, and the overwhelming majority of ion channels is concentrated in the \gls{AIS}.
The always-open channels in the membrane's wall are "on the 
way" for the ions entering in the resting state, so, essentially, no ion reaches the \gls{AIS}: the resulting (fluctuating) voltage remains around the resting potential. The resting current
flows through the resting channels: the low conductance 
is sufficient for keeping the balance. In the transient state,
after that a large amount of ions arrives to the membrane,
the potential is about \SI{100}{\milli\volt} above the resting potential;
to restore the resting state as quickly as possible, the high conductance of the \gls{AIS} is required. This way, 
in a good approximation, the neuron switches from the 
parallel $RC$ circuit model to the serial $RC$ circuit model.

\subsection{The physical process\label{sec:AP-PhysicalProcess}}

In the resting state, we have a condenser that has some low-conductance resistors (the resting ion channels) and a very low intensity (nearly constant) distributed current flows, under the effect of a slightly varying (nearly constant) membrane voltage. Due to the membrane potential, a current through the always-open channels flows on the dendritic surface
and the resting channels' conductivity represents a drain with 
sufficient transmission. The resting current practically does not 
reach the \gls{AIS}: the ion channels along the current's path
practically "shunt" the current.
This is a static state that classic neuroscience assumes: some static current in and some static (leakage) current out; no significant gradient.
\textit{In this state}, an almost correct model is a condenser with a
\textit{parallelly} switched resistor; an integrator-type  $RC$ circuit, see Table~\ref{tab:Electric_RCOscillator_Circuits}
(although when approaching the threshold voltage, the \gls{AIS} plays some role. The smaller gradient changes, such as subthreshold excitations, produce "mini-\gls{AP} waves"~\cite{NeuralEnergyConsumption:2017}, providing direct experimental proof that in this state the parallel $RC$ circuit is not correct).
Some perturbations must be counterbalanced, but their
amount does not exceed a predefined threshold.

In the transient (perturbed) state, the process variable
(the membrane's potential) exceeds the threshold. That event acts
as a fast trigger signal. The amount of charged particles (and so:
the membrane's voltage) suddenly increases, the created gradient
drives a current.
When $Na^+$ ions rush into the intracellular layer, they roughly increase the overall concentration and the potential in that thin layer.
All other ions, including $K^+$, also feel a driving force. The targeted membrane potential is set electrically,
and the driving gradients may behave unexpectedly during the transient period.
The actual voltage gradient may temporarily reverse the direction of the chemical gradients.

Our results align with the observation (see caption of 11.22 in~\cite{MolecularBiology:2002}), that \textit{the significant processes
occur in a thin layer of the electrolyte proximal to the membrane surface}.
The amount of unbalanced ions is in the range of $10^7$,
and so is the amount of rush-in ions. In addition, those 
ions on the high-concentration side rush into the 
low-concentration side and cause a significant change in the 
membrane potential (and concentration). Their absolute amount is small compared to the total number of ions in the cell,
but it is significant compared to the number of unbalanced ions in that layer.
However, the layer itself can also be modeled
as having just a few ions under their mutual repulsion on the surface
or in a few atomic layers on top of each other, depending on the concentration.

Suppose that at the beginning of an \gls{AP} a large amount of $Na^+$ ions are transferred from the extracellular to the intracellular side. In that case,  the $K^+/Na^+$ concentrations change from $145/15$ to $45/115$ and the 
corresponding thermodynamic voltage contribution changes from $+61\ [mV]$ to $-25\ [mV]$ (i.e., altogether a $86\ [mV]$ sudden increase in membrane potential). For the $Na^+$ ions, the resultant potential changes from $+4\ [mV]$ to $-83\ [mV]$. That means when the \gls{AP} begins, the "Na-K pump" stops (if the resulting potential provides the driving force for the exchange pump. The only way for the neuron to remove the excess  $Na^+$ ions is to generate a current toward the \gls{AIS} (where the other end of the ion channels remained at the extracellular potential), until the $Na^+$-specific driving force disappears. The $K^+$-specific driving force, changes from $-90\ [mV]$ to  $-176\ [mV]$, i.e., the $K^+$
intake gets more intensive, misleading researchers into believing that it causes
the observed hyperpolarization. However, as  Fig.~3 in~\cite{NeuralEnergyConsumption:2017} demonstrates, it occurs instantly, not with a delay; furthermore, the "leakage current" and the stimulus current are negligible. The low extracellular $K^+$ concentration plus the 
low number of channels in the membrane's wall
do not enable a significant increase in the intracellular $K^+$. 
(Given that complex changes occur, including changes in the electrical potential that also change the concentrations, different waves start; the statement is not strictly valid.)

%
%
%
As the current flows, the thermodynamic force decreases, and at some point the resultant force changes sign; so does the current direction. This is seen as the "condenser current" (and mis-identified as an intense $K^+$ current). The ions move slowly, and due to the enormous electrostatic repulsion, their "electrostatic fluid" must be continuous. A special "damped wave" is formed (with all secondary mechanical, optical, etc. effects; see section~\ref{sec:DampedOscillation}) and the condenser current reverses. Electricity interprets the phenomenon as "capacitive current" in the equivalent circuit, and thermodynamics interprets it as 
changed chemical concentration (or the negative amplitude of the damped elastic oscillation). Our discussion adds that electrical repulsion generates a pressure wave, or shock wave, within the neuron's closed volume.

\begin{figure}
	\includegraphics[,trim={.5cm -0.cm 0cm 0cm},
	scale=2.0]{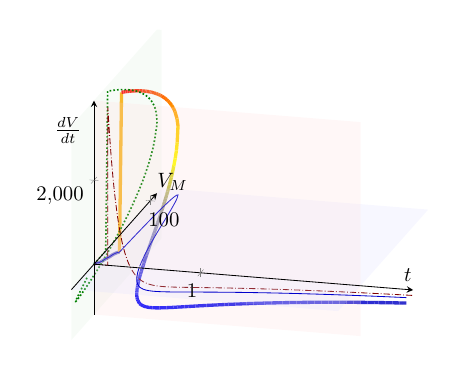}
	\caption{A summary of producing neuronal \gls{AP} as a 3-D figure. The thick, solid 3-D line shows how the color-coded voltage gradient drives the membrane potential over time.
		The three dashed diagram lines are the projections of the 
		spatial curve to the boundary planes of the cube. The experimental physiology delivers those 2-D plots; the figure explains how those parameters are interrelated. 
		The red line shows the voltage gradient
		produced by the rush-in of the $Na^+$ ions, see also 
		the middle inset in Fig.~\ref{fig:PhysicalProcessesMembrane}.
		It produces the blue diagram line  \gls{AP}, see also 
		the bottom inset in Fig.~\ref{fig:PhysicalProcessesMembrane}.
		The interdependence of the gradient and the voltage it produces
		leads to Fig.~\ref{fig:Bean_V_dV}, left side, that allows to 
		calculate the thermodynamic characteristics, right side,
		of neuronal operation.		
		\label{fig:SummaryPicture} 	}
\end{figure}

The neuronal \gls{AP} generation can be entirely described  
by the 3-D line in 
	Fig.~\ref{fig:SummaryPicture}. The three dashed diagram lines are drawn in 
	three perpendicular planes; see the figure's caption and the discussion below for the details.

\subsection{The oscillator model\label{sec:OscillatorModel}}

Despite the early warning that '\textit{it was not possible to separate the change into
	resistance and capacity components}' \cite{COLE_CURTIS_IMPEDANCE:1939},
a commonly accepted truism was that neurons, in some sense, behave
as electrical oscillators.
\gls{HH} introduced the idea explicitly that the electrically equivalent circuit
of a neuron is an $RC$ oscillator.
We must also add the invention from about three decades later: "Neurons ensure the directional propagation of signals throughout the nervous system.
The functional asymmetry of neurons is supported by cellular compartmentation: the cell body and dendrites (somatodendritic compartment) receive synaptic inputs,
and the axon propagates the action potentials that trigger synaptic release toward target cells.
\textit{Between the cell body and the axon sits a unique compartment called the axon initial segment}.
The \gls{AIS} was first described 50 years ago [i.e., nearly two decades after \gls{HH} published their study], and its molecular composition
and organization have been progressively elucidated during the following decades. \dots.
Recent years have also brought crucial insights into the functions of the \gls{AIS}:
how \textit{ion channels at its surface generate and shape the action potential}."~\cite{AIS_Updated_Viewpoint:2018}
We provide the physics and mathematics of how \gls{AIS} shapes
the action potential (or more precisely, we show what an important role it plays in forming \gls{AP}).

Inventing \gls{AIS} changed the viewpoint of neuroscience~\cite{AIS_Updated_Viewpoint:2018}, a half century later, after setting up the commonly accepted model by \gls{HH}.
"The \gls{AIS} is located at the proximal axon and is the site of action potential initiation. This
reflects the high density of ion channels found at the \gls{AIS}.
... The summation of
synaptic inputs gives rise to action potentials at the
\gls{AIS}, a 20--60 $\mu m$ long domain
located at the proximal axon/soma interface that has
a high density of voltage-gated ion channels."	
As discussed in \cite{AISStructureReview:2018}, see also their Figure 1, 
the structure of the \gls{AIS}
is known to the most minor details.
As the illuminating investigations in 2008
\cite{ActionPotentialGenerationNatrium:2008}
revealed, the 
\gls{AIS}
has very dense ion channels. That is, from an electrical point
of view, those parallelized channels can be abstracted as a  \textit{discrete  conductance}
(or resistance) between the membrane and the axon.
The membrane itself can be abstracted as a \textit{distributed condenser} with no resistance.
This model is in contrast with the viewpoint of biophysics, that the membrane plus \gls{AIS}
is considered a distributed element, where the capacitor and condenser cannot be separated. The current model is the result of projecting the physical processes in the resting state to the transient state.
Notice the important point:
"Neurons are also anatomically polarized, as they can be
subdivided into a somatodendritic input domain and an axonal output domain"~\cite{AIS_NeuronalPolarity:2010};
providing direct evidence that (unlike in 
\gls{HH}'s
model) \textit{the input and output currents (and voltage time derivatives) are independent}.

\begin{table}
	\caption{\href{href="https://www.electronics-tutorials.ws/rc/rc-integrator.html} {RC circuit types}\label{tab:Electric_RCOscillator_Circuits}}
	\centering
	\begin{tabular}{cc}
		\hline
		\hline
		The RC Integrator
		\label{RCIntegratorCircuit}                                                                                                                                   & The RC differentiator                                                                                                                                  \label{RCDifferentiatorCircuit}                                                                                                                                  \\
		\hline
		$
		V_{out}^{Integrator}=\frac{1}{RC}\int_0^t V_{in}{dt}\label{eq:RC_Integrator}
		$
		& $V_{out}^{Differentiator}=RC\frac{dV_{in}}{dt} $ \\
		\hline
		Low Pass Filter&\textbf{High Pass Filter}\\
		\hline
Resting state&\textbf{Transient state}\\
		\hline
		\includegraphics[width=.4\textwidth]{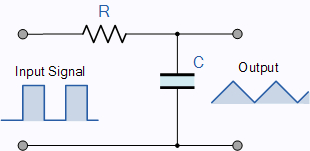} & 
		\includegraphics[width=.5\textwidth]{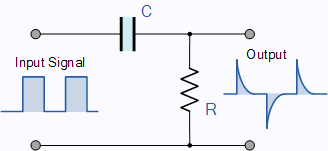}\\
			\hline
(Mostly) resting state& (Mostly) transient state\\
\hline
Ion channels in the membrane&
Ion channels in the AIS\\
		\hline
	\end{tabular}
\end{table}

In our model, the 
\gls{AIS} gets part of the membrane,
and this leads to crucial changes. (BTW, the name is slightly misleading:
the \gls{AIS} is part of the neuronal oscillator, and it forwards a traveling
potential wave to the axon instead of belonging to it.)
'Although by definition a neuron must have an
axon to assemble an \gls{AIS}, the relationship between \gls{AIS}
assembly and axon specification in vivo has not been
determined yet'~\cite{AIS_NeuronalPolarity:2010}.
Anyhow, a neuron can be abstracted (in the disciplinary view of electricity) as comprising a resistor-like and a capacitor-like component. However, in the resting and the transient states, they are combined in different ways.

In the resting state, we can model neurons as a \textit{parallel} $RC$ circuit (see the "integrator" in Table~\ref{tab:Electric_RCOscillator_Circuits}), where "random" currents flow in and out. Practically,
the output voltage is constant (the resting potential): the
ion channels and pumps are distributed over the membrane's surface,
and due to the low speed of currents, at low local voltage gradients,
the \gls{AIS} is not involved. Only the synaptic inputs cause a
measurable gradient on the \gls{AIS}. In this native state, there is no gradient; in this sense, it is similar to the case of clamped operation, where the gradient is (apparently) suppressed by the feedback.
\gls{HH} did not see any structural elements on the membrane,
so logically, they modeled it as a distributed resistor and capacitor,
which really has resemblance with a \textit{parallel $RC$ oscillator.}
However, they made a wrong choice of the circuit type, and their choice (likely due to inertia)
was repeated in good textbooks such as  (\cite{ JohnstonWuNeurophysiology:1995}
Figure 3.1 or \cite{KochBiophysics:1999} Figure 1.1), and
it is a commonly accepted fallacy even today~\cite{NeuralDynamicsGertsner:2014}.

In the transient state,  the 
high-transmission ion channels in the \gls{AIS} (high conductance $R$) replace the low-transmission ion channels (low conductance $R$) in the membrane's wall, and the rush-in current flows out entirely on the \gls{AIS}
(for the model, in this state, we omit the low-transmission ion drain).  
Those two elements form a \textit{serial} $RC$ circuit (see the "differentiator" in Table~\ref{tab:Electric_RCOscillator_Circuits}) that acts as a damped oscillator (see section~\ref{sec:DampedOscillation}) and produces a current pulse, known as \gls{AP}. Neuroscience mistakenly claims it is a parallel $RC$ circuit; that statement is valid only in the resting state.
However, the number and intensity of the 'resting' ion channels in the membrane's wall are insignificant for producing \gls{AP}, so  we use the approximation that only the 'transient ion channels' in the \gls{AIS} 
are active when generating an \gls{AP}.   

\begin{figure}[]
	\includegraphics[scale=1]{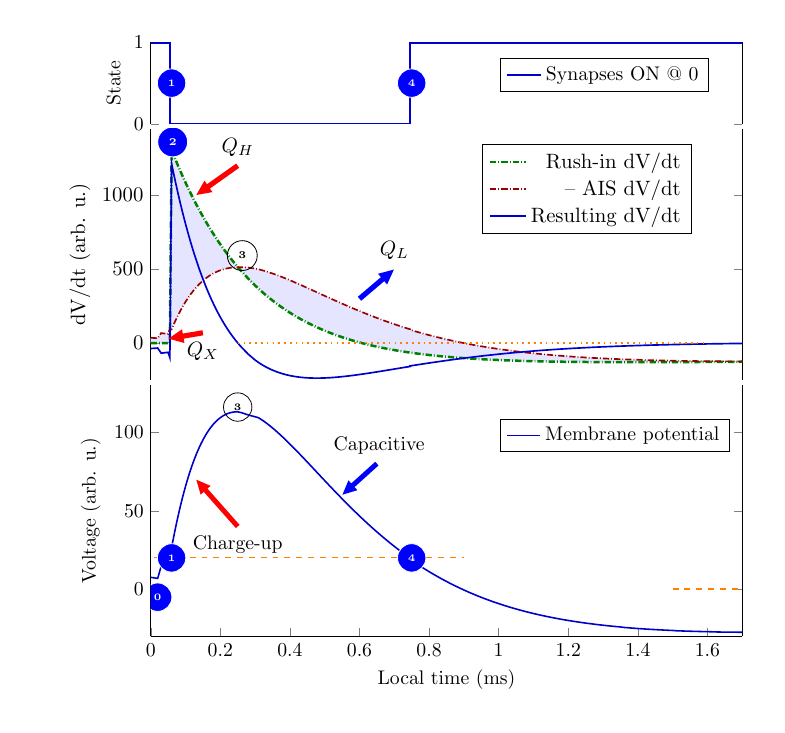}
	
	\caption{The physical processes describing the membrane's operation\label{fig:PhysicalProcessesMembrane}. The rush-in $Na^+$ ions instantly increase the membrane's charge, and the membrane's capacity discharges (producing an exponentially decaying voltage derivative). The ions are created at different positions on the membrane, so they take different times to reach the \gls{AIS}, where the current produces a peak in the voltage derivative. The resulting voltage derivative, the sum of the two derivatives (the \gls{AIS} current is outward), drives the oscillator. Its integration produces the membrane potential. When the membrane potential crosses the threshold, it switches the synaptic currents on or off.
	The upper two diagram lines have no corresponding ones 
	in \gls{HH}-based theories.}
	
\end{figure}

In electrical view, the process can be modeled as follows: a sudden voltage gradient  (see Figure~\ref{fig:PhysicalProcessesMembrane} middle inset) appears on the membrane (as a discrete capacitor), and a current flows out (see Figure~\ref{fig:PhysicalProcessesMembrane} bottom inset) through the serially connected ion channel array (as a discrete resistor). The neuron forms a simple \textit{serially }(\textit{not parallelly}, as assumed by Hodgkin and Huxley~\cite{HodgkinHuxley:1952} and mistakenly claimed by neurophysiology) connected $RC$ oscillator.
For the physical and mathematical description, see section~\ref{sec:MathematicsAP}, furthermore, ~section~4 in~\cite{VeghTechnomorphBiology:2025}; for its algorithmic specifics~\cite{VeghNeuronAlgorithms:2025}, for further details~\cite{VeghDANCES:2026}.
Notice that the difference between the rush-in and the resultant gradients
emphasizes the role of the finite resources: without the \gls{AIS},
no turn-back of the resultant gradient would occur.

The fundamental difference between the parallel and serial $RC$ circuits is that \textit{the parallel circuit cannot produce output voltage with opposite sign}. 
As discussed, the capacitive current, which by definition changes its direction and so 
generates an opposite voltage on the resistor represented by the \gls{AIS}, perfectly describes the so-called "hyperpolarization". It is not the effect of a $K^+$ current: the resting ion channels do not have sufficient conductance.  
Not knowing about the \gls{AIS} leads to assuming that the resting ion channels work also as transient channels (in other words, assuming a parallel $RC$ circuit instead of the serial one, furthermore, that
the output current flows through the membrane instead of the axon). Some $K^+$ current certainly exists; for the magnitude of currents during \gls{AP}, see~\cite{NeuralEnergyConsumption:2017}.

The shape of the output waveform depends on the ratio of the pulse width to the $RC$ time constant. When $RC$ is much larger than the pulse width, the output waveform resembles the input signal, even with a square wave input. (In the case of the neuronal oscillator, the shape of the front side of the spike is similar to the one derived for the rush-in current, while the back side is very prolonged.)

\subsection{Mathematics of the Action Potential\label{sec:MathematicsAP}}

\begin{figure}
	\includegraphics
	{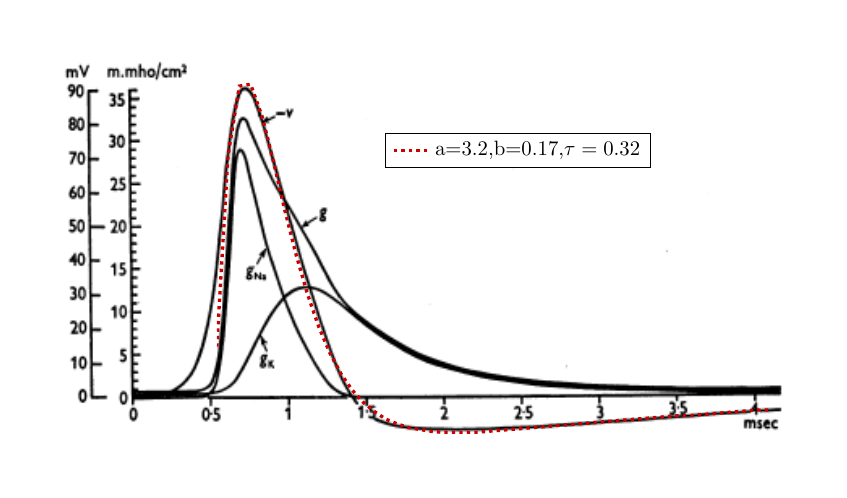}
	\caption{Comparing the calculated \gls{AP} to the 
		empirical function fitted by Hodgkin-Huxley
		(Fig.~17 of ~~\cite{HodgkinHuxley:1952}).
		\label{fig:AP_to_HH}
	}
\end{figure}

Although the implementation of a neuronal circuit works with ionic
currents and has a charge transmission mechanism drastically different
from the one based on free electrons used in metals, the operation
can align (to some measure) with the instant currents of physical
electronics (as we suggest, the effect of the slow current and finite
size can be reasonably imitated with current generators generating
a special current shape, see section~\ref{sec:SlowCurrent}). The input currents in a biological circuit
are a large rush-in current through the membrane in an extremely short
time, and much smaller gated synaptic currents arriving at different
times through the synapses. Those currents generate voltage gradients,
and those gradients direct the operation of the $RC$ oscillator
(Notice that due to the multi-physical effects, the parameters $R$ and $C$ may change in the function of time and the actual voltage).
As it is well known from the theory of electrical circuits, the output
voltage measured on the output resistor is as follows:

\begin{equation}
	V_{out}^{Differentiator}=RC\frac{dV_{in}}{dt}\label{eq:RC_Circuit_Output}
\end{equation}

\noindent where the input voltage is as follows:

\begin{equation}
	\frac{d}{dt}V_{in}=\sum\frac{d}{dt}V_{IN}^{Component}-\frac{d}{dt}V_{OUT}^{AIS}\label{eq:RC_Circuit_Input}
\end{equation}

\noindent that is, the (temporally gated) sum of the input gradient
that the currents generate, plus the gradient of the output current
through the \gls{AIS}~\citep{AISStructureReview:2018,AIS_Updated_Viewpoint:2018}.
The latter term can be described as follows (see the discussion around also Eq.(\ref{eq:PID}):
\begin{equation}
	\frac{d}{dt}V_{OUT}^{AIS}={\frac{1}{C}}\frac{V_{internal}-V_{external}}{R_{AIS}}\label{eq:AIS_Voltage}
\end{equation}
\noindent Fig.~\ref{fig:AP_to_HH} compares the calculated function shape
calculated by Eq.~(\ref{eq:AIS_Voltage}) to the function
fitted by \gls{HH} to their experimental observations.

The input currents are slow (we typically imitate them using a current generator with a special form); see section~\ref{sec:SlowCurrent}.
In the simplest case, the resulting voltage derivative comprises only
the rush-in contribution described by Equation~(\ref{eq:PSPderivative})
and the \gls{AIS} contribution described by Equation~(\ref{eq:AIS_Voltage}).
The gating implements a dynamically changing temporal window and dynamically
changing synaptic weights. The appearance of the first arriving synaptic
gradient starts the ``Computing'' stage. The rush-in gradient starts
the ``Delivering'' stage~\cite{VeghNeuronAlgorithms:2025}. The charge collected in that temporal window is as follows: 
\begin{equation}
	Q(t)=\int_{t_{0,i}}^{{t_{thr}}}dt\underbrace{\sum_{i}}_{{t_{0,i}\le t\le t_{thr}}}I_{syn,i}(t)
\end{equation}

The total charge is integrated in a mathematically entirely unusual
way: this is \textit{NOT} a simple "integrate and fire": the collected charge defines the upper bound of the integration time. Through adjusting the integration time window, the neuron selects which
one(s) out of its upstream neurons can contribute to the result; the
individual contributions are the result of a bargain between the sender and the
receiver. The upstream neuron decides when the time window begins, but the neuron decides when a contributing current terminates. \emph{The beginning
	of the time window is set by the arrival of a spike from one of the
	upstream neurons, and it is closed by the neuron when the charge integration
	results in a voltage exceeding the threshold.}
The speculations~\cite{MechanicalPropertiesNerves:2025} about the spatial length of the spike are correct, but its "payload length" is defined by the receiving neuron. The computation (a
weighted summation) is performed on the fly, excluding the obsolete
operands (the
idea of \gls{AIMC}~\citep{AnalogInMemoryComputing:2024}
almost precisely reproduces the charge summation except for timing). The
only operation that a neuron can perform is that unique integration.
Its result is the reciprocal of the weighted sum, and it is represented
by the time when a spike is sent out (\emph{when} that charge on the
membrane's fixed capacity leads to reaching the threshold). Notice
that the time is defined on the neuron's local time scale and is interpreted
similarly by the downstream neurons: the data transmission time matters.

\subsection{Damped oscillation\label{sec:DampedOscillation}}
\begin{figure}
	\includegraphics[width=.8\columnwidth]{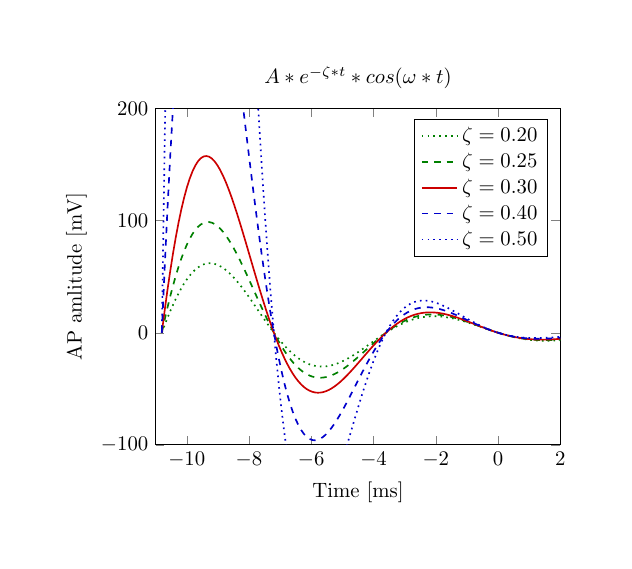}
	\caption{The analytical solution of the motion equation of the underdamped oscillator for some damping parameters. 
		\label{fig:DampedOscillationSolution}
	}
\end{figure}

According to Newton's 2nd Law, a body with mass $m$, under the effect  of a  "spring force" proportional to its position $x$, and a 
linear damping force proportional to its velocity $v$ 
moves according to the equation 
\begin{equation}
	\underbrace{m*\times\frac{d^2\,x}{{dt^2}}}_{moving\ force} + \underbrace{c\times \frac{dx}{dt}}_{viscous\ damping} + \underbrace{k\times x}_{spring\ force} = 0 \label{eq:DampedOscillation}
\end{equation}
The equation directly corresponds to our Eq.~(\ref{eq:IonicForces}),
given that neither constraint force nor external force is present.
Recall that according to the Stokes-Einstein relation (see Eq.~(\ref{eq:StokesSpeed})), the viscosity force is proportional to the speed of the body in the fluid. The damped oscillation has an analytic solution
(see Fig.~\ref{fig:DampedOscillationSolution} for the corresponding graph)
\begin{equation}
	x = e^{-\gamma* t}*a*\cos{(\omega*t-\alpha)}\label{eq:DampedOscillationSolution}
\end{equation}

A remarkable resemblance with the shape of an \gls{AP} is evident,
but there exist significant differences from the 'net' physical case.
Among others, the mass $m$ changes continuously; furthermore, it is distributed in the volume of the neuron, so one cannot expect a perfect description from such a simple motion equation. However, it is perfect for demonstrating that such a system can produce the symptoms of relaxation 
and hyperpolarization that physiology observes.
(The electrostatic interaction is the strongest and fastest interaction in nature, stronger than the other interactions that have a role in the process, so we have reasons to think that Eq. (\ref{eq:RC_Circuit_Output}) provides the best description.)

The "moving force" has two major components. The sudden rush-in of
$Na^+$ ions has (at least) a double effect. 
One effect is that they press the surface of the membrane that
elastically increases its diameter~\cite{NeuronalDeformation:2020}.
 Due to their mutual repulsion,
the ions experience a strong electrical force toward the membrane. It represents an impulse $J=F\,\Delta t$ (N\, s), that enables estimating the time and energy of the 
change the rush-in causes and explains that the energy needed to 
generate an \gls{AP} is suddenly produced electrically, 
and stored temporarily as elastic energy (the neuron produces the required energy in the background, when \gls{ATP} produces ions by hydrolysis, as described above). 
The case is practically identical to
the deformation of an elastic plate induced by the hydrostatic pressure of a water column. At the beginning, the aluminum plate is suddenly subjected to hydrostatic pressure, and it reaches equilibrium after initial oscillations. The diagram line of the process is shown in the figure
\cite{HydrostaticWaterColumn:2021}, and the method of simulation is discussed in \cite{Hydrostatic_SPHinXsys:2021}. Compare the diagram line to the
gradient in Fig.~\ref{fig:PhysicalProcessesMembrane}, middle inset.

At the same time, 
it offers a plausible way of unifying the electrical and thermodynamic 
attempts to describe neuronal operation; furthermore,
a pictorial understanding of why the elastic change produces 
opposite amplitudes in the size of the membrane and the voltage in its
surface layer. Similarly, it explains that in the first phase of generating an \gls{AP} the neuron invests (performs a work $Q_H$ against the membrane's elastic force), while in the second phase the electic membrane performs work $Q_L$ on pressing out the ions physiology observed in an \gls{AP}; for the details see section~\ref{sec:Physics-Thermodynamics}. These two changes are inseparable, and their combined effect can explain the observed consequences.
The compression and expansion give way to explaining the
observed mechanical, optical, thermal, etc. changes.

Given that, as we explained, the energy operating the neuron 
is temporarily stored in the form of elastic energy, it is not surprising
that its time course is in resemblance to a damped elastic oscillation. 
We can consider that the neuron forms a damped oscillator (the Stokes-Einstein force represents a viscous fluid, the $Na^+$ represents 
a $\delta$-function like excitation) with time constant about $\zeta=0.35$.

Their other effect is that the appearance of the ions drastically increases 
the potential in the surface layer of the neuron membrane. During regular operation, a very intense current pulse is created:
"A small flow of ions carries sufficient charge to cause a large change in the membrane potential.
The ions that give rise to the membrane potential lie in a thin ($< 1\ nm$) surface layer close to the membrane"~\cite{MolecularBiology:2002}
(see the dotted line in Fig.~\ref{fig:The-membrane's-extra-gradient}; notice that the horizontal scale on the figure spans a few $nm$).
For this, a considerable number of ion channels are used.
When the trigger signal is received, the gated ion channels in the membrane's wall open quickly and only briefly. The rushed-in ions create a large offset potential in the layer proximate to the membrane's surface (from an electrical point of view, a condenser) that drives an intense current through the always-open ion channel array \gls{AIS} (from an electrical point of view, a resistor) at the beginning of the axon. The slow current on the membrane's surface mimics a condenser: in the picture of classical electricity, the charges stay inside the distributed condenser while they travel on its surface. In classical electronics, the condenser is sizeless.

\section{Thermodynamics of Action Potential\label{sec:Physics-Thermodynamics}}

As discussed above, by finding the connection
between apparently separated disciplines,
one can derive from quantities belonging to the discipline 
electricity quantities belonging to thermodynamics; see section~\ref{sec:Physics-MagicConnection}.
In this section, we derive the description of the
thermodynamic Carnot cycle from measured electrophysiological 
data, essentially from the 2-D diagram lines shown
in Fig.~\ref{fig:SummaryPicture}.

\subsection{Deriving thermodynamic description\label{sec:Physics-DerivingThermodynamics}}
\begin{figure}[b]
	\begin{tabular}{cc}
		\includegraphics[height=8cm,trim={1.1cm 1.0cm 1cm 0cm },clip]{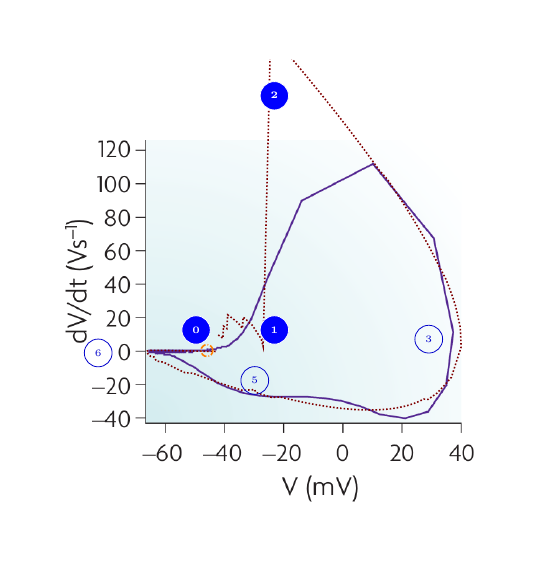}
		&
		\includegraphics[,trim={1.5cm -0.26cm 0cm 0cm},scale=1]{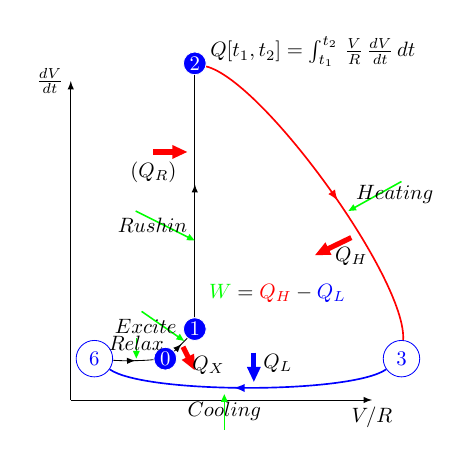}
	\end{tabular}
	\caption{Thermodynamic consequences of the processes shown in Figure~\ref{fig:PhysicalProcessesMembrane} during the generation of an Action Potential. Left: Comparing the measured (broken blue) and theoretical (red dotted) voltage gradient vs voltage phase diagrams~\cite{BeanActionPotential:2007}. 
		Right: Reinterpreting the phase loop in terms of thermodynamics: the “Carnot cycle” (theoretical thermodynamic cycle) of a neuron \label{fig:Carnot} \label{fig:Bean_V_dV}	}
\end{figure}

As the middle inset in Fig.~\ref{fig:PhysicalProcessesMembrane} shows, the sum of the rush-in and \gls{AIS} gradients (plus more gradients if any) controls the generation of the \gls{AP}. During excitation, see Fig.~\ref{fig:Bean_V_dV}, $Q_X$ energy is fed into the neuron. The rush-in of the $Na^+$ ions injects more $Q_R$ energy, which can be used to generate an \gls{AP}.
It is a kind of potential energy, which is converted in the first 
phase of $AP$ (while the voltage gradient is positive) to the kinetic energy $Q_H$ of \gls{AP}.
When the membrane potential reaches its peak, the gradient reverses, decreasing the potential.  
In the second phase of \gls{AP} (while the voltage gradient is negative)
the neuron harvests $Q_L$ energy.
At the beginning of the transient state, the size of the neuron (due to the increased electrostatic repulsion) slightly increases~\cite{NeuronalDeformation:2020} and stores the energy mainly in the form of elastic energy. The resultant of the electrical, elastic, and thermodynamic forces moves the ions toward the \gls{AIS}; see Fig.~\ref{fig:Physics-MembraneForces}. 
 Although part of the energy is dissipated while the ions exit through the \gls{AIS} with their Stokes-Einstein speed, the rest of the energy (the elastic part) is recovered.
 The concentration and potential values change as discussed in connection with Fig.~\ref{fig:RestingPotential4}. The membrane voltage controls synapses as shown in the top inset in Fig.~\ref{fig:PhysicalProcessesMembrane}.
 
The ‘phase loop’ (how the voltage gradient and the membrane voltage correlate; a parametric curve that uses time $t$ as parameter) shown on the left side of Fig.~\ref{fig:Bean_V_dV} represents a strong test of the theoretical line shape: the values of the potential and its gradient are simultaneously compared to the experimentally measured ones. Comparing the experimental and theoretical shapes supports the theory that describes the mutual dependence of voltage and its gradient. Furthermore, it shows that omitting the gradient from the classic theory misled the experiment designers, and the measurement design was wrong: the time interval between recording data points was too long. 
Where the gradient is high and changes quickly (see Fig.~\ref{fig:PhysicalProcessesMembrane}, middle inset), the blue experimental diagram line is roughly broken. The changes in the values are too significant, causing the integration to distort the value and resulting in an inaccurate broken line. In the rush-in region, the theoretical line is even outside the figure's range. Where the gradient change is slow, the measurement points are sufficiently dense, and the theory perfectly describes the experimental data. Neither of the mentioned competing (or other classical) theories can interpret the phenomenon and produce any similar dependence.

\begin{figure}
	\includegraphics[width=\textwidth]
	{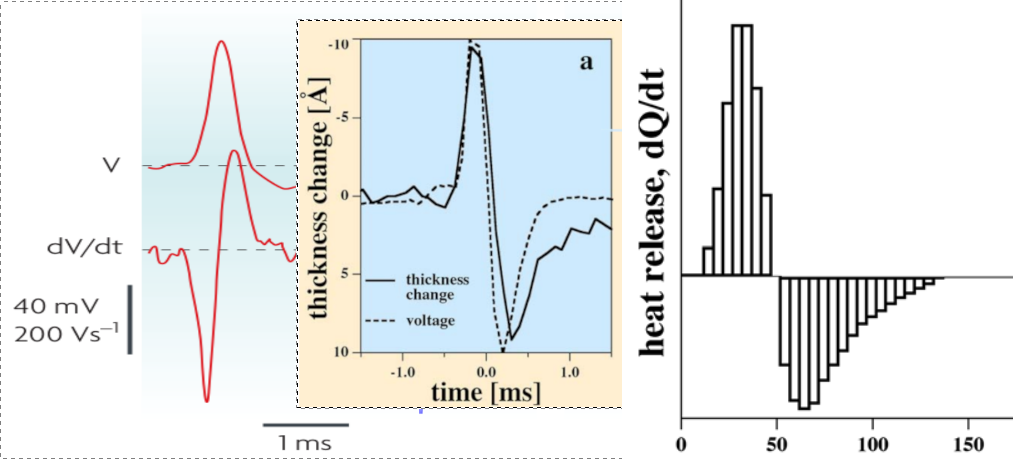}
	\caption{The red diagram lines show the membrane voltage
		and its gradient during producing an action potential~\cite{BeanActionPotential:2007}, Fig. 4. The black diagram lines show the size-related and heat-related changes during producing an action potential~\cite{HeimburgPhysikOfNerves:2009}, Fig. 1.
		and~\cite{ThermalBiophysics:2007}, Fig. 18.5. }
	The time course of the electrical gradient~\cite{BeanActionPotential:2007} perfectly matches that of the mechanical and thermal changes~\cite{HeimburgPhysikOfNerves:2009}; however, classic theory cannot explain the connection between them~\cite{VeghNonOrdinaryLawsForLife:2025}.
	The experimental $\frac{dV}{dt}$ uses an opposite sign convention.
	\label{fig:HeimburgElascticity}
\end{figure}

An exciting aspect is that the phase plot in Fig.~\ref{fig:Bean_V_dV}, left side, shows a remarkable resemblance to a specialized thermodynamic Carnot cycle in Fig.~\ref{fig:Bean_V_dV}, right side; and inspires us to interpret the parameters of the "neuronal Carnot machine" (the bent arcs serve only for illustration; they do not follow the path on the left side). 
As discussed in~\cite{HeimburgPhysikOfNerves:2009}, a neuron emits and absorbs heat (energy) in the first and second phases, respectively, of an \gls{AP}; see Fig.~\ref{fig:HeimburgElascticity}.
The experience, alone, defies the classical \gls{HH} mechanism: the leakage current can only emit heat, but not absorb it.
\textit{Laws describing currents of electrons must not be used directly to currents of ions. Neglecting the repelling force between the charge carriers of the ion current, which can be omitted in the case of electron current, leads to entirely wrong conclusions, from the wrong oscillator model to losing the connection between the electrical and mechanical changes, to losing the connection between inanimate and living nature.}
Our theoretical model, unlike the others, can explain that observation quantitatively.

We rescale the horizontal axis by the resistance $R$ of the \gls{AIS} (actually, the horizontal axis is transformed to $I_{AIS}$); thus, the plot becomes a neuronal energy density diagram.
Actually, it corresponds to a usual $P-V$ thermodynamic diagram, 
and we expect that the temperature changes in the same analogy
(we keep the convention of using the letter $V$ to denote both potential and volume).
By integrating the parametric $t$ values of any two points on the diagram ‘circle’ line, we receive the work performed by the neuron.

We consider that the environment in the period ‘Excite’ invests energy $Q_X$ through the neuron’s synapses (increases the membrane’s voltage to the threshold) between [0,1] to open the valves.
 Opening the valves enables the neuron receiving in period ‘Rushin’ energy $Q_R$ by letting in $Na^+$ ions, accelerated by the field across the membrane. Suddenly, the ion concentration and the membrane potential reach their maximum values (given that the current is slow,
 the sudden change generates a smeared and delayed current on the \gls{AIS}, see Fig.~\ref{fig:PhysicalProcessesMembrane}, middle inset). Given that the process is instant, the potential $V$ does not change (no current flows), so the integral [1,2] evaluates to zero (an ‘iso-voltage’ transition). Now the neuron has a high potential to prepare its (already encoded) message (the \gls{AP}).
In terms of electricity, in the first phase (from the threshold potential to the maximum value of the \gls{AP}, where $\frac{dV}{dt}$ is positive), the neuronal power cycle produces a positive heat $Q_H$, given that $\frac{dV}{dt}$ is always positive. 
In the second phase (from the top value of \gls{AP} to where the driving force $\frac{dV}{dt}$ becomes zero again), the neuronal power cycle produces a negative heat $Q_L$, given that $\frac{dV}{dt}$ is always negative; in line with the measurement results shown in Fig.~\ref{fig:HeimburgElascticity}~\cite{HeimburgPhysikOfNerves:2009}. The integral [2,3] provides the “heating” $Q_H$, the energy invested. Similarly, the integral [3,6] provides the “cooling” $Q_L$, representing the energy recovered (assuming no artificial current). 

In terms of thermodynamics, the ions stop in the immediate vicinity of 
the membrane, i.e., initially, the pressure created by their mutual repulsion is present at the inner surface of the sphere (it is a mechanical shock wave). In the positive phase of the damped oscillation, it compresses the fluid in the sphere. In the second phase, in the negative 
phase of oscillation, it decompresses the fluid (sucks back the fluid).
Given that the density of ions in the volume is constant
(the ions' speed is too low to follow those sudden changes),
the electrical and thermodynamic quantities faithfully follow each other.
Measuring electrical potential and mechanical pressure yields measurements of the same physical effect within the same discipline.

In the “Relax” phase, $\frac{dV}{dt}$  is (almost) zero, so the corresponding integral evaluates to zero (‘iso-gradient’ transition). In this phase, the neuron must restore its resting state: by using \gls{ATP}, it creates new ions. The difference in the energies of $Q_H$ and $Q_L$ is the net energy used to forward an \gls{AP}; while $Q_X$ is the energy used to perform the computation. The energy $Q_X$ is an approximate value: as shown in Fig.~\ref{fig:PhysicalProcessesMembrane}, at the time of re-opening the synaptic inputs, the membrane can be above or below the resting potential, which actually means a neuron-level memory. In this sense, adjacent neural spikes can borrow energy from each other (which explains why the spikes in a burst behave differently). The energy $Q_R$ is produced by the \gls{ATP} hydrolysis in parallel with the 'primary processes' described above, and the produced ions are acquired on the condenser plates, thus restoring the resting potential and providing potential energy for the forthcoming spikes. 

From Fig.~\ref{fig:Carnot}, after recalibration (with $R_{AIS}=10^7$ Ohm), one 
can estimate (by approximating roughly the area under the arcs by triangles, in units of $10^{-7} [J]$), $Q_H =5.2, Q_L=1.8, W=3.4, Q_X=0.2$. That is, according to the "phase plot"~\cite{BeanActionPotential:2007}, the neuron operates with an efficiency around 65\%, and approximately 3\% of the energy is derived from the upstream neurons. The values (without using data from dedicated measurements and making precise numerical integration) agree excellently with the measured~\cite{NeuralEnergyConsumption:2017} values $W=2.5$ and 74\%; furthermore, the $Q_X$:total ratio $1:35$, derived by measuring the direct energy consumption~\cite{EnergyNeuralCommunication:2021}.
\textit{Another interesting proof that the two disciplines see the same physical effect, that the electrical parameters can measure the thermodynamic efficiency.} The invested energy can be safely estimated from the work the rush-in charge performs, given that the energy is almost entirely stored as elastic potential energy due to the enormous elasticity modulus.


\subsection{Neural entropy\label{sec:HeatEntropy}}
The theoretical diagram line in Fig.~\ref{fig:Carnot}, based on the presented correct physical model, might serve as a good starting point for interpreting \textit{entropy} for neuronal operation, again in analogy with the Carnot cycle; although the charge of the ions (the macroscopic current) complicates the process. Furthermore, material convection happens in
different ill-defined volume parts of the membrane in different directions. Initially, in the resting state, the ions are in a (relatively) disordered state. After the rush-in, in the dynamic layer, the ions have a highly ordered macroscopic speed (see Stokes-Einstein speed, the gradually changing gradients, and the finite speed) component toward the \gls{AIS}. In the layer, the maximum ordered state is reached when the voltage drop on the \gls{AIS} reaches its peak value (the current is slow). Then, the ions slow down, and at zero voltage gradient, they reach again a minimally ordered state. After that, due to the 'capacitive current', the current reverses its direction, and the backward current again represents an ordered state.  
Perhaps the latter-mentioned negative contribution is the one referred to as ‘negative entropy’ when describing life in terms of thermodynamic processes? We must be aware that, in the background, energy-producing processes are at work, involving the ordered transport of material. Considering the neuron, based only on the described primary processes, would be a mistake.

The above discussion might help in understanding the \textit{physical entropy}
in the neuron; although, as discussed above, there are issues with calculating partial derivatives (and so: deriving entropy from the Gibbs energy of the system) and with the applicability of Boltzmann assumptions to ensembles of ions. However, the \textit{information entropy} is different. 
As our discussion underpins, the "information" arrives at the neuronal inputs as charge carriers, and the time course of the input currents, in cooperation with the neural environment and the neuron itself, carries the information in some form. 
The information is distributed in space and time~\cite{VeghNeuralShannon:2022, VeghChannelCapacity:2023}, so we can be sure that using the pulse count as a measure of neural information is wrong. We must distinguish the \textit{carrier of the information}, the ion, and the \textit{information content} that its temporal appearance carries. After having the correct physical model of operation, we have a chance to gain more knowledge about how the
information is encoded in the spatially and temporally distributed
observable signals; furthermore, we have greater hope of understanding neural information processing, including the brain's entropy handling.

\section{Summary}

The paper reviewed the competing disciplinary electrical and thermodynamic theories
for describing neuronal operation and described their synthesis, the non-disciplinary unified
theory that resolves all former controversies between the two.
The restrictive disciplinary approach on both sides causes the apparent incompatibility between them. The interplay of those two disciplines
is vital in all aspects of neuronal operation.
Using a cross-disciplinary approach 
(and revisiting the fundamental laws of science for describing
living matter) one can perfectly describe neuronal operation.
The neuron represents a well-designed, simple control circuit with separate ground and excited state mechanisms, 
where the construction defines the setpoint  
and the control is implemented by regulating the ion concentrations 
on the two sides of the semipermeable membrane. The geometry and the overall concentrations of the cellular fluids define the nominal values of the electrical and thermal gradients. The \gls{AP} is 
part of the operation of the control circuit.
The low and largely variable speed of ionic currents,
in which the discrete features of the carrier dominate,
creates local ionic gradients and movement in the electrolyte
in the segments of the neuron.

Slow currents deliver ions through the permanently open "resting" and the gated "transient" ion channels, respectively, which means they affect the 
gradients until the equilibrium state is kept or restored.
By introducing the theoretical idea that biological currents are slow, we underpinned the experimental observation by Hodgkin and Huxley~\cite{MembranePotentialMoleculesNetworks:2014} that
"the changes
in ionic permeability depend on the movement of some component of the
membrane which behaves as though it had a large charge or dipole moment".
Using the principles of classical electricity, we calculated
that neurons handle charge as electricity describes (electrolyte) condensers. 
Given the presence of electrolytes on both sides of the membrane, the correct values for the resting potential and the time course of the action potential have been derived numerically from first principles, without empirical functions.
We defied the hypothesis suggested by \gls{HH} that some
"leakage current" produces the resting potential; this way
solved the decades-old mystery of "heat production" by neurons and
the controversy of the experienced and theoretically expected 
energy consumption of neural computing.
We fixed the mistakes
\gls{HH} claimed in the physical picture behind their empirical
description and mathematical equations, and introduced the correct
physical model by providing its detailed mathematical description.
By understanding that the biological "construction"
defines the value of the 
resting potential, we defied the decades-old fallacy that 
a mystic concerted operation of ion channels and 
a linear combination of poorly-defined permeability values (the \gls{GHK} equation) define the resting potential.

We could theoretically underpin 
that "action potential generation in nearly all types of neurons and muscle cells is accomplished through mechanisms similar to those first detailed in the squid giant axon in early research by \gls{HH}"~\cite{PrinciplesNeuralScience:2013}.
However, not because "this unequal distribution of ions is maintained by ionic pumps and exchangers". Instead, electrostatic charging is due to charge separation at the membrane and to polarization in the electrolyte surrounding it.
The mechanically separated segments produce two charged layers in the
electrolyte proximal to the membrane. Those layers have a decisive impact on the neuron's electrical operation.

The operation can be described in terms of electricity and thermodynamics, although the 'non-ordinary' laws~\cite{VeghNonOrdinaryLawsForLife:2025} of science must be considered.
We demonstrated the need for a cross-disciplinary approach by calculating the thermodynamic parameters 
from experimental data that can be measured electrically. 
We showed that, due to the disciplinary properties, one cannot derive the thermodynamic entropy using the usual methods of information processing.

\section*{Competing interests}
The authors declare that they have no competing interests

\subsection*{Funding}
The research did not receive any support.

\begin{appendices}

\section{Dichotomies in neuron modeling\label{sec:Physics-Fundamental}}

Paper~\cite{ThermodynamicAPDrukarch:2022}, which, in the context of this section, tends to be the precursor of our paper, sheds light on some remarkable dichotomies.
Scrutinizing them can result in "a sound basis for unification of the physics of nerve impulses".
By deeply agreeing with the fact that they have a "potential impact on our understanding of (the physical nature of) neuronal signaling", we list and uncover more dichotomies; furthermore, we uncover their fundamental,
partly philosophical reasons.

\subsection{Science methods\label{sec:Physics-ScienceMethods}}

For centuries, science has developed its methods for deriving abstract concepts
by reducing the features of a real object to an abstract one that cannot be reduced further, such as mass and charge. It derived laws for the forces acting on those abstract objects, such as Newton's universal law of gravitation and Coulomb's law of electricity. Then one could apply Newton's laws of motion. Experience shows that the generated forces, independently of their origin, can be summed, and one can apply 
the laws of motion by using the resultant force.
Ions are special from the point of view of reducing the 'material point'
to one single abstraction: \textit{the ions are charge and mass simultaneously, without a further possibility of reduction}.
Notice that in the electrical abstraction, no mass is present,
so one can use the equations assuming 'instant interaction',
which in biology led to non-physical explanations of the observations (such as 'delayed current').
In the mechanical/thermodynamic abstraction, mass is to be moved, 
making the finite-speed interaction evident. The same physical phenomenon,
the interaction (or movement) of ions is described using an 'infinite' electrical speed and 
a million times less mass propagation speed, respectively, which leads
to an unresolvable discrepancy, given that physics is not prepared to handle
different speeds in the same interaction event~\cite{VeghNon-ordinaryLaws:2025}.
This item is listed in Table~1~in~\cite{ThermodynamicAPDrukarch:2022}
as the "Electrical vs Mechanical" dichotomy.

The low speed of ions in electrolytes introduces further problems.
The farther parts of the cell will see any change in the local value of the state variables with a delay.
The material transport represents simultaneously mass and charge,
so the transport itself changes the gradients. This process keeps the entire volume of the electrolyte in (more or less) continuous change (that makes life). Furthermore, biological objects inside the cell
can absorb ions and charge up. With their potential, they alter local gradients, accelerating or decelerating ions.  
Not to mention that biological objects can be active in the sense that 
(depending on the environmental conditions) they can let ions from one separated volume part into the other.

The closed volume of the biological objects is also
a problem of finite resources. In biology, we cannot use the
fundamental assumption of physics that, although a field acts on the 
ion in question, the ion does not affect the field (that the other ions generate).  The transferred ions decrease the field in the volume they departed from
and increase it in the volume where they arrived. Due to the 
field-dependent speed within the electrolyte, we must consider the autonomous
change between the microvolumes where the ion traverses. 

These processes are what E.~Schrödinger coined as "\textit{the construction is different from anything we have yet tested in the physical laboratory}"~\cite{Schrodinger:1992}. Consequently, measurements must be designed and carried out with care; the routine
methods used for measuring objects from inanimate nature,
cannot surely be applied to living objects.

The interdependent behavior of charge and mass has an interesting
mathematical consequence as well. 
The definition of a partial derivative of a function of several variables is its derivative with respect to one of those variables,
\textit{with the others held constant}. In the case of ions, changing the function with respect to mass or charge means simultaneously changing the other; that is, the partial derivatives depend on each other.
In other words, one cannot calculate the partial derivatives; only the total derivative.
As a consequence, equations that calculate the partial derivatives of concentration and electrical potential with respect to time are either incorrect or approximations. Classical mathematics cannot be applied.
Consequently, in the case of ions and electrolytes, one must not use the well-established concepts (enthalpy, entropy, etc) of thermodynamics in unchanged form.

Although science is aware that the apparently continuous matter in nature cannot be divided infinitely, even though it knows that the abstracted discrete
'material points' (as A.~Einstein coined) have well-defined discrete values,
it occurs rarely that both views must be applied in describing a single
phenomenon. The continuous and discrete approaches (also called macroscopic and microscopic views) seem to be independent of each other.
Connecting those views was one of the tasks performed by thermodynamics for particles with no long-range interactions. However, there is no similar discipline for electricity,
although the behavior of charge carriers is similar to that of neutral discrete 
particles. That effect is not evident for electrons, but significant for ions.
This item is listed in Table~1~in~\cite{ThermodynamicAPDrukarch:2022}
as the "Macroscopic vs Microscopic" dichotomy.

Thermodynamics provides a framework for handling ions and determining their thermodynamic properties. However,
as good thermodynamic textbooks (including the one on the thermodynamics of the membrane~\cite{ThermalBiophysics:2007}) emphasize, thermodynamics
derives its concepts for non-interacting particles, so one cannot expect 
its validity for ionic solutions~\cite{VeghNon-ordinaryLaws:2025}
(Boltzmann assumed that, in the absence of long-range interaction between the particles, the sizes of cells in the phase space do not change).
In addition, he required  the presence of a vast number of particles
in a homogeneous, isotropic, infinite volume,
which is typically not the case in biology.

The dichotomy about reversibility is 
closely related to the exclusively electrical nature of the
\gls{HH} theory. In the framework of that
theory, Ohmic currents ﬂow through resistors that irreversibly dissipate heat due to
friction, no matter in which direction ($W=I^2\times R$) the ion currents ﬂow. 
By introducing the concept of
'delayed current' and the hypothesis that some hidden power controls the operation
of neurons by altering their conductance, physiology gave rise to the fallacy that the sciences and life sciences are almost exclusive fields.
Furthermore, their (unintended) model provokes questions (for a review
see \cite{HH_Potential_Controversies_2017}) whether it is a model
at all, and what controversies it delivers.  
As Hodgkin wrote~\cite{HodgkinConduction:1964}: "Hill and his colleagues found~\cite{HeatProductionNeuron:1958} that an initial phase [of the action potential] was followed by one of
heat absorption. [...] a net cooling on open-circuit was totally unexpected
and has so far received no satisfactory explanation."
Since that invention, experimental evidence from the missing "leakage current"~\cite{EnergyNeuralCommunication:2021} to the conversion between
elastic and kinetic energies~\cite{MechanicalWaves:2015} to demonstrating 
the pressure-wave-like behavior of \gls{AP} witnesses that 
at least part of the phenomenon is of thermodynamic origin;
that is, it requires a cross-disciplinary explanation.
For discovering the reciprocal relations in thermodynamics~\cite{OnsagerExperimental:1959}, also in electrolytes, Lars Onsager was awarded the 1968 Nobel Prize in Chemistry.
Those relations, together with the theoretical understanding,
manifesting in the cross-disciplinary Nernst-Planck relation,
paved the way to a combined theory.
However, no thermodynamic explanation has been given so far.
Our discussion provides, in line with the experimental results,
a complex electrical/thermodynamic description of neuronal operation.

Those dichotomies are rooted in deeper layers of science.
Classical physics is based on the Newtonian idea that space and time are absolute, so everything happens simultaneously. Consequently, when their objects
interact, it must be instantaneous; in other words, their
interaction speed is infinitely large. Furthermore, electromagnetic
waves with the same high (logically, infinitely high) speed inform the observer. This
self-consistent abstraction enables us to provide a "nice"
mathematical description of nature in various phenomena: the classical
science. In the first year of college, we learned that the idea resulted
in "nice" reciprocal square dependencies, Kepler's and Coulomb's
Laws. We discussed that the macroscopic phenomenon "current" is
implemented at the microscopic level by transferring (in different
forms) discrete charges; furthermore, that solids show a macroscopic
behavior "resistance" against forwarding microscopic charges. We also learned that \textit{without charge (and, without atomic charge carriers), neither potential nor current exists}. We did not learn, however,
that thermodynamic forces can move the ions,
given that they have inseparable mass.

Physics notoriously suffers from a lack of handling
{different simultaneous interactions}; facing such a case leads to misunderstandings, debates, and causality problems.
Such a famous case is the entanglement speed.
In that time, E. Schrödinger introduced his famous law of motion in quantum mechanics entirely analogously to how I. Newton introduced his Laws of Motion.
Similar to the Newtonian 'absolute time', the quantum mechanical interaction is supposed to be 'instant'
(this is the price for having 'nice' equations in classical and quantum mechanics),
i.e., its speed is supposed to be infinitely high.
However, at that time, it was already known that the speed of electrical interaction (propagation of electromagnetic waves) is finite.
So if an object has quantum-mechanical interaction (aka entanglement) and electrical interaction simultaneously, the corresponding forces act simultaneously but reach the other object at different times.
The entanglement arrives instantly; the electromagnetic effect arrives at a time we can calculate from the interaction speed and the spatial distance between the objects.
This effect leads to causality problems: the two interactions of photons entangled earlier in an exploded supernova should be measured at two different times, meaning a "spooky remote interaction" as A. Einstein coined,
and leads to contradictions such as the Einstein–Podolsky–Rosen paradox.
Actually, the issue stems from the improper handling of mixing interaction speeds:
the Schrödinger-equation introduces the infinitely large interaction speed,
while the
\gls{EM}
interaction has a finite speed.

\textit{The confusion and question marks in connection with describing life by science mostly arise from the interpretation of
	notion 'speed' in physics.}
When discussing the underlying physical laws for biology, we
go back to the fundamental physical concepts instead of taking over the
approximations and abstractions
used in the \textit{classical physics for non-biological matter} and
less complex interactions.
As we emphasized many times, we construct laws and conclusions based
on simplified abstractions about nature in all fields of science.
The notions and laws depend
on the circle of phenomena we know and want to describe.
The Newtonian and Einsteinian worlds are
basically distinguished by considering \textit{speed dependence} that actually means \textit{explicit time dependence}.
Interesting consequences are that in the Einsteinian world,
the mass is not constant, time and space are not absolute, and so on.
We can be prepared for some similar counter-intuitive experiences in physiology: "we must be prepared to find it working in a manner that cannot be reduced to the \textbf{ordinary} laws of physics"\cite{Schrodinger:1992}.

Science uses 'instant' in the sense that one interaction
is much faster than the process under study; we consider the faster interaction as instant.
\index{instant interaction}
The approach of classical science is based on the oversimplified
approximation that the interaction speed is \emph{always} much higher
than the speed of changes it causes and that the processes can \emph{always}
be described by a single stage. In our approach, for biology, we put together a
\textit{series of stages} to describe the observed complex phenomena, where the stages provide input and output
for each other, involve more than one interaction speed, and use  per-stage validity. We simplify the approximations by omitting the less
significant interactions and introduce ideas for accounting for the different
interaction speeds. This way, we reduce the problem to a case that
science can describe mathematically. \emph{This procedure is fundamentally different
	from applying some mathematical equations derived for an abstracted
	case of inanimate science to a complex biological phenomenon without validating
	that we use the appropriate formalism}.

\subsection{Complete measurement\label{sec:Physics-Complete}}

In physics, measurement means the quantization of something 
relevant to the process under study.  A 'complete' measurement
measures all relevant quantities. \textit{The different disciplines of physics restrict the measured quantities to the ones which are,
in general, that the discipline studies.
The remaining quantities remain outside the scope of the discipline.}
The fundamental physical quantities of mechanics are length, mass, and time, which form the basis for defining all other quantities like velocity, force, and energy in that field.
The fundamental physical quantities in thermodynamics are Temperature, Energy (Internal Energy, Heat, Work), and Entropy, which characterize systems at equilibrium and describe energy transformations; Pressure and Volume are also key variables. 
Those of electricity include electric charge (Coulomb), the basis of all electric phenomena; electric current (Ampere), the rate of charge flow; voltage (Volt), the potential difference driving current; resistance (Ohm), opposition to current; power (Watt), the rate of energy transfer; and energy (Joule), the capacity to do work. 
As seen, there is little overlap of the studied quantities. 
\textit{A disciplinary study (such as the electrical and thermodynamic ones in physiology) provides an incomplete set of measured quantities}. None of the disciplines alone can describe electrolytes, since they consider different and incomplete sets of physical quantities.
As discussed in connection with the Onsager relations, a disciplinary (incomplete) measurement does not discover that an unexpected change (a miracle) in the value of one of the quantities belonging to another discipline is accompanied by a change in the value
of the studied quantity. This change is clearly the case with the measurements performed in the spirit of \gls{HH}: the values the
electrical instruments provide are accompanied by the values
of mechanical/thermodynamic quantities which are not measured, partly due to the obvious 
measuring difficulties, and because they are outside the scope of the discipline. 
One cannot reduce ion-related phenomena to a single abstraction as thermodynamics does with mass or electricity does with charge.
If one attempts (see the mentioned theoretical descriptions),
one experiences that some quantity (the charge or the mass)
changes in an uncontrolled way.

\subsection{Physics and biology\label{sec:Physics-Biology}}

A common fallacy in biology is that 
{physics} cannot underpin the operation of living matter, citing E.~Schrödinger. However, the claim falsifies his opinion by omitting the most essential word, 'ordinary'. 
Schrödinger wanted to emphasize the opposite: there is no new force (no unknown new interaction, as biophysics attempts to introduce "protein mechanism"), only that studying living matter requires different testing methods (and we add: different uses of physics' concepts) in the physical laboratory. He suggested answering the question
"Is life based on the laws of physics?” affirmatively, but expected to discover the appropriate forms of physical laws describing the 'non-ordinary' (in our reading: non-disciplinary) behavior of living matter. No doubt, the basic concepts and terms must be interpreted precisely for living matter,
much beyond the level we used to at the college level.
However, after that reinterpretation, we can interpret features of living matter, although we need a more careful, cross-disciplinary analysis to do so. 
We must use appropriate abstractions and approximations for the phenomena,
depending on the level required for the given cooperation of objects and interactions.
Furthermore, we discuss some of the relevant terms and notions of physics,
differentiating which approximation is appropriate only for physics (mainly electricity),
and,  instead, which approximation should be used for biology.
As we discuss, \textit{biophysics translated the corresponding major terminus technicus words
	from the theory and practice of physics' major disciplines, mainly from electricity, which were
	worked out for homogeneous, isotropic, structureless metals,
	and for strictly pair-wise interactions with a single (actually, 'instant') interaction speed; to the structured,
	\index{interaction!attributes}
	non-homogeneous, non-isotropic, material mixtures and for multiple interaction speeds}.
Those notions do not always have unchanged meaning, and how much they do,
depends on the actual conditions. The precise meaning needs a case-by-case analysis.

The physical models consider infinitely large volumes,
surfaces, distances; furthermore, and most importantly, instant interactions. Is the cell large enough to
consider it infinitely large (at least on the scale using ions' size); that is, to apply laws of science
derived for the abstraction 'infinitely large'? 
When working with charge, we know that charge is quantized, while the macroscopic quantities voltage and current are 
continuous (derivable). Do cells contain a sufficient 
number of charge carriers to apply macroscopic notions?
Do the thousands of times smaller ion channels transfer enough charge
to speak about a well-defined current?
When a couple of ions are transferred through an ion channel,
do they significantly change the potential that accelerates them?

Science could serve as a firm base for all its disciplines.
As we discuss, \textit{its disciplines use abstractions based on limited-validity approximations} based on the same first principles.
However, \textit{the approximations differ between biology and physics}.
In physics, some processes we observe are fast enough that we can treat them as essentially state transitions.
In some cases, the approach can be --more or less-- successful. For the slower, well-observable processes,
we have the 
laws of motion
that describe how processes occur under the influence of a driving force.
We also found that nature is not necessarily linear (in the sense that it depends only on the mentioned quantities, not on their derivatives), which we can describe with "nice" mathematical formulas.
A century ago, A.~Einstein discovered that the approximations I.~Newton introduced two centuries earlier are not sufficiently accurate for describing the movement of bodies at high speeds. In other words, a new paradigm, the constancy of the speed of light, must have been introduced
that caused a revolution in physics and led to the birth of "modern physics". 

\textit{Life, including the brain's operation, is dynamic.} 
As Schrödinger formulated, the "construction of living matter" differs
from the one science used to test in its labs.
The scientific abstraction based on "states" (i.e., on instant changes)
fails for the case of biology, where "processes" happen (i.e., the changes
are obviously much slower).
The commonly used measuring methods, such as {clamping, patching, and freezing},  reduce the life to states. 
The related theories describe states with perturbation~\cite {PerturbationNeuralComputation:2002}. 
On the one hand, this technology fixes the cell in a well-defined, static state, enabling us to observe a static anatomic picture of the cell. On the other hand,
it eliminates the dynamic processes from the theory, i.e., \textit{hides forever the essence of the life that the cell exists in a continuous change governed by laws of motion}.
Those methods stop the processes under study, thereby depriving them of their dynamicity and preventing measurement.
It was forgotten that using feedback for stabilizing an autonomously
working electrical system means introducing foreign currents,
and this way falsifying its operation. 

Furthermore, it is hazardous to introduce technically (and incorrectly) derived
and misinterpreted macroscopic features and interpret them as fundamental
electric notions. In general, instead of understanding and developing the proper scientific basis for the operation, they say that science cannot describe it. The idea of {conductance}
has been introduced
to neurophysiology almost a century ago. It was taken from physics, where the
notion was derived for metals (conductors, instead of electrolytes). Since then, its original interpretation
has been forgotten, and today (in contrast with physics), it has become
a primary entity for describing electrical characteristics of biological
cells. We explain how the right physics background enables us to
discover wrong physical models and 
misinterpreted notions of physics in neurophysiology.
Furthermore, the proper interpretation opens the way to the correct interpretation
of neuronal information. We set up an abstract electrical/thermodynamic model of neuronal
operation.

We derived the needed 'non-ordinary laws'~\cite{VeghNon-ordinaryLaws:2025}, which are derived by using the same first principles as the 'ordinary laws', but are abstracted for the
approximations valid for living matter.
As we discuss, those 'ordinary' laws were derived for strictly pair-wise interactions at very high speeds.
In biology, we can observe interactions at a million times lower speed, in inhomogeneous, non-isotropic, structured material.
\textit{Biology does not have the conditions for which physics derived its ordinary laws.}
By using the appropriate approximations for the 
biological cases, we can derive the required 'non-ordinary' laws of physics,
which laws are more complex to derive and use; furthermore, we must use several
stages (with the approximations changing from stage to stage) instead of 
one single stage, as in the case of the 'ordinary' laws. 
However, \textit{all laws follow the same principles}.

Biology, and predominantly neuronal operation, produces examples where using wrong omissions
in complex processes results in absolutely wrong results. In those cases,
some initial resemblance between our theoretical predictions and our phenomena exists. However, the success
in simple cases provides no guarantee that the model was appropriate: "the success of the equations is no evidence in
favour of the mechanism"~\cite{HodgkinHuxley:1952}. 
\textit{Finally,
all laws are approximations, and the accuracy of verifying their predictions is limited.}
Several theories can describe the same phenomenon with the required accuracy.
We also show in the section about the 
{finite interaction speeds}
that the most well-known laws (from Newton, Coulomb, Kirchoff, etc.) are
also approximations. They have their range of validity, although it is often
forgotten.

One such neuralgic point of omissions and approximations is the vastly different
{interaction speeds}. Furthermore, where the speed is considered at all, \textit{the same speed is assumed for all interactions}.
The laws are abstract also in the sense that, say, the objects in the laws of physics have either mass or
electric charge, but not both. It is the researcher's task to decide
which combination of laws
can be applied to the given condition. For example, one can assume in most cases
that the speeds sum up linearly, except at very high speeds.
Biology provides excellent case studies where different interactions
shape the phenomenon, and special care must be exercised.
We give a short review of 
{history and kinds of interaction speeds}.

Another point is that science started with the assumption that
the non-living matter is continuous, although it was early
discovered that there are the smallest pieces of that matter.
When we reached that size, we experienced that different subsets
of science laws describe that matter and the atoms they contain.
Establishing relations between those subsets is one of the most challenging tasks. Again, we used abstractions that the 
continuous matter is infinitely large and that the isolated
atoms are infinitely far from each other and from the external world. We also experienced the semi-infinite cases, and studied
the behavior of surfaces and interfaces, which, again, is different from both that of the atoms and their large masses. Given that biological objects span the microscopic to macroscopic size range and are surrounded by surfaces, we must be prepared for the fact that no simple rules describe their behavior.

Neuronal operation is at the boundary, where sometimes, in the same phenomenon,
one interaction can be interpreted at the macroscopic level, another
must already be interpreted at the microscopic level.
Furthermore, a series of stages (instead of a single state) and
processes (instead of stages) describes the subject under study.
Given the vital role of 
charge and current in neuronal operation, we provide their precise interpretations.
Furthermore, we must consider that the processes happen in a finite volume,
"within the spatial boundary".

\end{appendices}


\end{document}